\documentclass[3p]{elsarticle}

\usepackage{amssymb}
\usepackage{mathrsfs,amsmath}
\usepackage{amsfonts}
\usepackage{subfigure}
\usepackage{color}
\usepackage{graphicx}
\usepackage{tabu}
\usepackage{undertilde}
\usepackage{lineno,hyperref}
\newcommand{\captionabove}[2][]{%
    \vskip-\abovecaptionskip
    \vskip+\belowcaptionskip
    \ifx\@nnil#1\@nnil
        \caption{#2}%
        
    \else
        \caption[#1]{#2}%
    \fi
    \vskip+\abovecaptionskip
    \vskip-\belowcaptionskip
}
\newtheorem{theorem}{Theorem}

\newtheorem{remark}[theorem]{Remark}

\makeatletter
\def\ps@pprintTitle{%
 \let\@oddhead\@empty
 \let\@evenhead\@empty
 \def\@oddfoot{}%
 \let\@evenfoot\@oddfoot}
\makeatother

\begin{document}

\begin{frontmatter}

\title{A Deep Material Network for Multiscale Topology Learning and Accelerated Nonlinear Modeling of Heterogeneous Materials}
\address[Zeliangaddress]{Livermore Software Technology Corporation (LSTC), Livermore, CA 94551, USA}
\address[Koishiaddress]{Koishi Laboratory, The Yokohama Rubber Co., LTD., Kanagawa 254-8601, Japan}
\cortext[mycorrespondingauthor]{Corresponding author.}
\author[Zeliangaddress]{Zeliang Liu\corref{mycorrespondingauthor}}
\ead{zlliu@lstc.com}
\author[Zeliangaddress]{C.T. Wu}
\author[Koishiaddress]{M. Koishi}
\begin{abstract}

In this paper, a new data-driven multiscale material modeling method, which we refer to as deep material network, is developed based on mechanistic homogenization theory of representative volume element (RVE) and advanced machine learning techniques.  We propose to use a collection of connected mechanistic building blocks with analytical homogenization solutions which avoids the loss of essential physics in generic neural networks, and this concept is demonstrated for 2-dimensional RVE problems and network depth up to 7. Based on linear elastic RVE data from offline direct numerical simulations, the material network can be effectively trained using stochastic gradient descent with backpropagation algorithm, further enhanced by model compression methods. Importantly, the trained network is valid for any local material laws without the need for additional calibration or micromechanics assumption. Its extrapolations to unknown material and loading spaces for a wide range of problems are validated through numerical experiments, including linear elasticity with high contrast of phase properties, nonlinear history-dependent plasticity and finite-strain hyperelasticity under large deformations. 

By discovering a proper topological representation of RVE with fewer degrees of freedom, this intelligent material model is believed to open new possibilities of high-fidelity efficient concurrent simulations for a large-scale heterogeneous structure. It also provides a mechanistic understanding of structure-property relations across material length scales and enables the development of parameterized microstructural database for material design and manufacturing.

\end{abstract}

\end{frontmatter}
\section{Introduction}
In the past decade, multi-scale simulation methods have demonstrated significant advantages for computational mechanics due to their ability to consider microscopic heterogeneities inside a material. The macroscopic material properties are strongly affected by the morphology and evolution of the microstructures, especially under extreme events with large material deformations and geometric nonlinearities. While direct simulations for large-scale heterogeneous structures are extremely expensive and uncommon in industrial applications, the representative volume element (RVE) techniques based on homogenization theory \cite{hill1965self,feyel2000a,geers2010multi} are a type of hierarchical multi-scale simulation methods offering the numerical constitutive closure relationship at the macroscopic point.

In terms of RVE analysis and homogenization, direct numerical simulations (DNS), such as finite element method (FEM) \cite{feyel2000a,belytschko2008a}, meshfree methods \cite{wu2012three,wu2016immersed}, and fast Fourier transform (FFT)-based methods \cite{moulinec1998a,de2017finite}, offer high accuracy at the expense of high computational costs.  A myriad of model reduction techniques have been introduced for predicting the effective mechanical properties in a manner that balances computational cost and accuracy. Analytical micromechanics methods \cite{eshelby1957determination,hashin1963variational,hill1965self,mura1987micromechanics,liu2015statistical,liu2016extended} can be regarded as one type of reduced-order models with high efficiency. However, due to a loss of detailed physics in the microscale, they normally lose accuracy or require extensive model calibrations when irregular complex morphologies, nonlinear history-dependent properties or large deformations are presented. For heterogeneous hyperelastic materials, manifold-learning methods like isomap are used for nonlinear dimensionality reduction of microscopic strain fields \cite{bhattacharjee2016nonlinear}. 
The model reduction of history-dependent plastic materials can be more complex and challenging. Two examples are non-uniform transformation field analysis (NTFA)  \cite{michel2003a,michel2016model} and variants of the principle component analysis \cite{jolliffe2002a} or proper orthogonal decomposition (POD) \cite{yvonnet2007a,kerfriden2013a,oliver2017reduced}.  However, they usually require extensive \textit{a priori} simulations for interpolating nonlinear responses, and their extrapolation capability for new material inputs is usually limited, Recently, the self-consistent clustering analysis (SCA) \cite{liu2016self,liu2018microstructural} has demonstrated a powerful trade-off between accuracy and efficiency in predicting small-strain elasto-plastic behavior though clustering techniques, and it only requires linear elastic simulations in the offline stage.

Meanwhile, current advanced machine learning models (e.g. artificial neural networks and deep learning) have achieved great successes in broad areas of computer engineering, such as computer vision, gaming, and natural language processing \cite{hinton2012deep,lecun2015deep, Goodfellow-et-al-2016, silver2017mastering}. Although these techniques are able to construct models for complex input-output relations, their applications to mechanics of materials are still limited.  Methods like tensor product approximation and neural networks have been employed to directly construct the overall strain energy density surface of the RVE \cite{le2015computational,yvonnet2013computational,bessa2017,ibanez2017data}. Artificial neural networks have also been used to approximate constitutive behavior by fitting the stress-strain relation directly \cite{ghaboussi1991knowledge,unger2008coupling,wang2018multiscale}.  However, these techniques are usually problem-dependent and may suffer from "the danger of extrapolation" beyond the original sampling space, e.g. different material laws and loading paths. Issues like material history dependency, physical invariance and conservation laws are not naturally resolved, mainly due to the loss of physics in the current machine learning models.

The paper introduces a novel multiscale material modeling method called deep material network, which represents the DNS model of RVE by a hierarchical topological structure with mechanistic building blocks, and is able to predict nonlinear material behaviors both accurately and efficiently. In the proposed material network, the above-mentioned limitations of various reduced order methods are addressed simultaneously by meeting three fundamental goals : 1) avoiding an extensive offline sampling stage (e.g. POD, NTFA and  generic neural network) and only requiring linear elastic RVE analysis; 2) eliminating the need for extra calibration of the constitutive laws (e.g. NTFA, methods based on isomap) or micro-mechanical assumption of homogenization (e.g. micromechanics methods, self-consistent scheme in SCA); 3) discovering an efficient reduced representation of RVE without the loss of physics and danger of extrapolation (e.g. generic neural network). Due to its intrinsic hierarchy structure,  its computational time is proportional to the number of degrees of freedom in the system. After one time offline training, the optimized material network creates a microstructural database of the RVE by virtue of its unique capability of extrapolation to unknown material and loading spaces, which is useful for multiscale concurrent simulation and material design.

In Section \ref{sec:theory}, the theory of material network, including the building block and network architect, are explained for RVEs under 2-dimensional (2D) plane strain condition. In Section \ref{sec:learning},  machine learning approaches, such as stochastic gradient descent (SGD) with backpropagation algorithm and model compression algorithms, are developed for training the material network. Extrapolations to general RVE problems with material and geometric nonlinearities are discussed in Section \ref{sec:online}. Finally, applications to several challenging problems are addressed in Section \ref{sec:results}, including linear elasticity with high contrast of phase properties, nonlinear history-dependent plasticity and finite-strain hyperelasticity under large deformations. Concluding remarks are given in Section \ref{sec:conclusion}.

\section{Theory of material network}\label{sec:theory}
The basic concept of material network is to use a collection of connected simple building blocks to describe complex RVE responses, similar to the one of artificial neural network where neurons are connected to define an arbitrary function. The building block is chosen to be a simple structure with analytical homogenization solutions, and the architect of the material network represents the path of the homogenization process from each individual phase to the overall macroscopic material. With advanced mechanistic machine learning approaches introduced in Section \ref{sec:learning}, the trained material network provides a possibility to obtain a simplified representation of the RVE with heterogeneous microstructures.

\subsection{The physically based building block} \label{sec:buildingblock}
The theory of material network is first developed for a two-phase linearly elastic building block in 2-dimensional (2D) plane strain condition. Under small-strain assumption, the strain and stress measures are the infinitesimal strain $\utilde{\boldsymbol{\varepsilon}}$ and the Cauchy stress $\utilde{\boldsymbol{\sigma}}$, which are related by the fourth-order compliance tensor $\utilde{{\textbf{D}}}$. The overall stress-strain relation of the building block can be expressed as
\begin{equation}
\utilde{\bar{\boldsymbol{\varepsilon}}}=\utilde{\bar{\textbf{D}}}:\utilde{\bar{\boldsymbol{\sigma}}}.
\end{equation}
For materials 1 and 2, we have
\begin{equation}
\utilde{\boldsymbol{\varepsilon}}^1=\utilde{\textbf{D}}^1:\utilde{\boldsymbol{\sigma}}^1, \quad \utilde{\boldsymbol{\varepsilon}}^2=\utilde{\textbf{D}}^2:\utilde{\boldsymbol{\sigma}}^2 \quad\text{and}\quad f_1+f_2=1,
\end{equation}
where $f_1$ and $f_2$ are the volume fractions of materials 1 and 2, respectively. In Mandel notation, the stress and strain can be written as
\begin{equation}
\boldsymbol{\sigma}=\{\utilde{\sigma}_{11},\utilde{\sigma}_{22},\sqrt{2}\utilde{\sigma}_{12}\}^T=\{\sigma_{1},\sigma_{2},\sigma_{3}\}^T 
\end{equation}
and
\begin{equation*}
 \boldsymbol{\varepsilon}=\{\utilde{\varepsilon}_{11},\utilde{\varepsilon}_{22},\sqrt{2}\utilde{\varepsilon}_{12}\}^T=\{\varepsilon_{1},\varepsilon_{2},\varepsilon_{3}\}^T,
\end{equation*}
where the subscript 3 denotes the shear direction. The compliance matrix can be written as
\begin{equation}
 \boldsymbol{\varepsilon}=\textbf{D}\boldsymbol{\sigma}, \quad
 \textbf{D}=
\begin{Bmatrix}
D_{11}&D_{12}&D_{13}\\
&D_{22}&D_{23}\\
\text{sym}&&D_{33}\\
\end{Bmatrix}.
\end{equation}

\begin{figure}[!htb]
	\centering
	\includegraphics[clip=true,trim = 4cm 7cm 4cm 6cm, width = 0.8\textwidth]{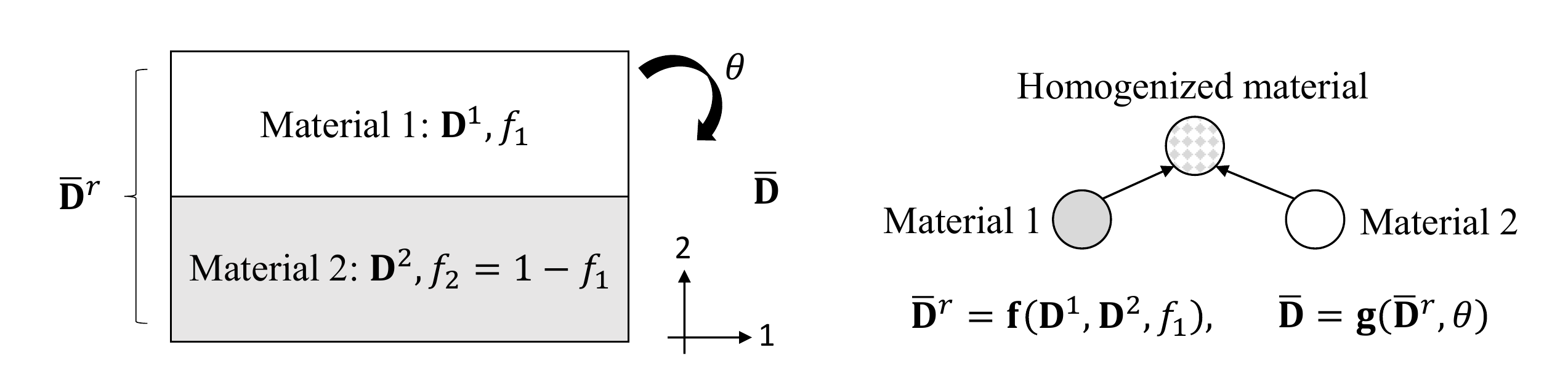}
	\caption{Illustration of the two-layer building block. The compliance matrix after the homogenization operation is $\bar{\textbf{D}}^r$, and the one after the rotation operation is denoted by $\bar{\textbf{D}}$.}
	\label{fig:twolayer}
\end{figure}
Similar to artificial neural networks, we want the building block to be simple and easy for the analysis. Here a two-layer structure is proposed. As shown in Figure \ref{fig:twolayer}. there are two operations in the building block: 1) homogenization which gives $\bar{\textbf{D}}^r$ and 2) rotation of the two-layer structure which gives $\bar{\textbf{D}}$.

The homogenized compliance matrix before the rotation $\bar{\textbf{D}}^r$ can be expressed as a function of the microscopic compliance matrix of each constituent and its morphological descriptors, in this case, the volume fractions $f_1$, with $f_2=1-f_1$. Mathematically, it can be written as
\begin{equation}\label{eq:twolayer}
\bar{\textbf{D}}^r=\textbf{f}\left(\textbf{D}^1,\textbf{D}^2,f_1\right).
\end{equation}

Analytical homogenized results are available for this simple two-layer structure, derived based on the equilibrium condition
\begin{equation}
\quad \sigma_2^1 = \sigma_2^2, \quad \sigma_3^1 = \sigma_3^2.
\end{equation}
and kinematic constraint
\begin{equation}
\varepsilon_1^1 = \varepsilon_1^2.
\end{equation}
Analytical expressions of the components in the homogenized compliance matrix $\bar{\textbf{D}}^r$ are
\begin{equation}
\bar{D}_{11}^r=\dfrac{1}{\Gamma}(D_{11}^1D_{11}^2),
\end{equation}
\begin{equation*}
\bar{D}_{12}^r=\dfrac{1}{\Gamma}(f_1D_{12}^1D_{11}^2+f_2D_{12}^2D_{11}^1),
\end{equation*}
\begin{equation*}
\bar{D}_{13}^r=\dfrac{1}{\Gamma}(f_1D_{13}^1D_{11}^2+f_2D_{13}^2D_{11}^1),
\end{equation*}
\begin{equation*}
\bar{D}_{22}^r=f_1D_{22}^1+f_2D_{22}^2-\dfrac{1}{\Gamma}f_1f_2(D_{12}^1-D_{12}^2)^2,
\end{equation*}
\begin{equation*}
\bar{D}_{23}^r=f_1D_{23}^1+f_2D_{23}^2-\dfrac{1}{\Gamma}f_1f_2(D_{13}^1-D_{13}^2)(D_{12}^1-D_{12}^2),
\end{equation*}
\begin{equation*}
\bar{D}_{33}^r=f_1D_{33}^1+f_2D_{33}^2-\dfrac{1}{\Gamma}f_1f_2(D_{13}^1-D_{13}^2)^2,
\end{equation*}

where
\begin{equation*}
\Gamma=f_1 D_{11}^2+f_2 D_{11}^1 \quad\text{and}\quad f_2=1-f_1
\end{equation*}

After the homogenization operation, the two-layer structure is rotated and the new compliance matrix $\bar{\textbf{D}}$ is passed to a child node of the next material building block in the upper level. As shown in Fig \ref{fig:twolayer}, the rotation angle is denoted as $\theta$. The matrix $\textbf{R}$ defines the rotation of a second-order tensor through the angle $\theta$ under Mandel notation,
\begin{equation}\label{eq:rotatem}
\textbf{R}(\theta)=
\begin{Bmatrix}
\cos^2\theta&\sin^2 \theta&\sqrt{2}\sin\theta\cos\theta\\
\sin^2 \theta&\cos^2\theta&-\sqrt{2}\sin\theta\cos\theta\\
-\sqrt{2}\sin\theta\cos\theta&\sqrt{2}\sin\theta\cos\theta&\cos^2\theta-\sin^2\theta\\
\end{Bmatrix},
\end{equation}
and it satisfies
\begin{equation}
\textbf{R}^{-1}(\theta)=\textbf{R}(-\theta).
\end{equation}
After rotation, the new compliance matrix can be expressed as
\begin{equation}\label{eq:rotate}
\bar{\textbf{D}}=\textbf{g}(\bar{\textbf{D}}^r,\theta)=\textbf{R}(-\theta)\bar{\textbf{D}}^r\textbf{R}(\theta).
\end{equation}
Combining the homogenization and rotation operations in Eq. (\ref{eq:twolayer}) and (\ref{eq:rotate}) yields the completed homogenization function of the two-layer building block,

\begin{equation}\label{neuron}
\bar{\textbf{D}}=\textbf{g}\left[\textbf{f}\left(\textbf{D}^1,\textbf{D}^2,f_1\right),\theta\right].
\end{equation}

\begin{remark}
It is equivalent to deriving all the analytical homogenization and rotation functions based on the stiffness tensor $\textbf{C}$, as we will show in Section \ref{sec:online} and \ref{sec:a1}. However, the reason why we use the compliance tensor $\textbf{D}$ here is that the expressions of the analytical solution based on $\textbf{D}$ are neater for this small-strain 2D building block in plane strain condition.
\end{remark}

\begin{figure}[!htb]
	\centering
	\includegraphics[clip=true,trim = 4cm 5cm 4cm 4cm, width = 0.8\textwidth]{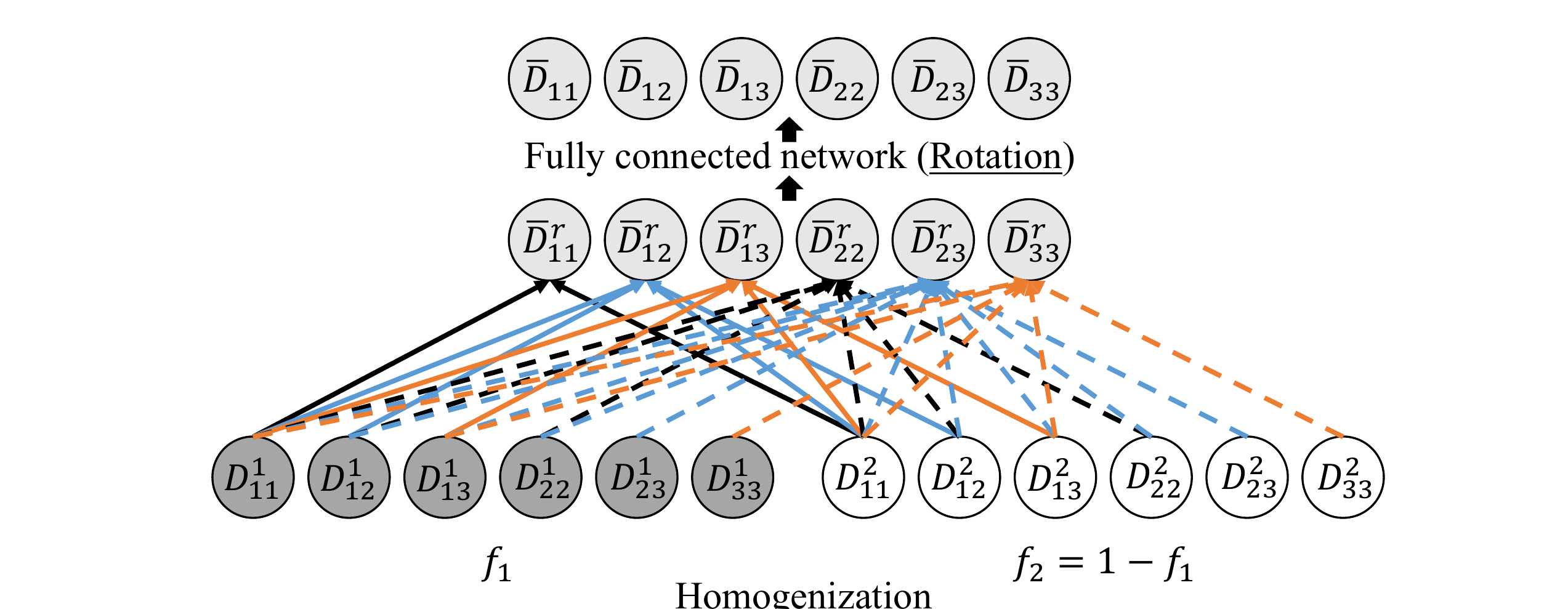}
	\caption{Detailed data flows of the components in the compliance matrices inside a building block.}
	\label{fig:detailednetwork}
\end{figure}

Note that analytical forms for building blocks with multiple layers could also be derived for multi-phase material, however, the material network would become more complex and more partial derivatives would be evolved in the learning process. For simplicity, this paper will focus on two-phase RVEs, and the two-layer structure, as well as the corresponding binary-tree architects, will be mainly discussed. In practice, the compliance matrix are stored in a vectorized form with 6 independent variables. The data flow for one building block after the vectorization is shown in Fig. \ref{fig:detailednetwork}. 

Derivatives of $\bar{\textbf{D}}$ and $\bar{\textbf{D}}^r$ will be used in the stochastic gradient descent with backpropagation algorithm introduced in Section \ref{sec:sgd}. Specifically, derivatives of $\bar{\textbf{D}}$ with respect to the rotation angle $\theta$ are
\begin{equation}\label{eq:dtheta}
\dfrac{\partial\bar{\textbf{D}}}{\partial \theta}=-\textbf{R}'(-\theta)\bar{\textbf{D}}^r\textbf{R}(\theta)+\textbf{R}(-\theta)\bar{\textbf{D}}^r\textbf{R}'(\theta)
\end{equation}
with
\begin{equation}
\textbf{R}'(\theta)=
\begin{Bmatrix}
-\sin 2\theta&\sin 2\theta&\sqrt{2
}\cos 2\theta\\
\sin 2 \theta&-\sin 2\theta&-\sqrt{2}\cos 2\theta\\
-\sqrt{2}\cos 2\theta&\sqrt{2}\cos 2\theta&-2\sin 2\theta\\
\end{Bmatrix}.
\end{equation}
In terms of the derivatives of $\bar{\textbf{D}}$ with respect to the components in $\bar{\textbf{D}}^r$, we have
\begin{equation}\label{eq:a}
\dfrac{\partial \bar{D}_{ij}}{\partial\bar{D}^r_{kl}}=R_{ik}(-\theta)R_{lj}(\theta).
\end{equation}

Derivatives of $\bar{\textbf{D}}^r$ with respect to the volume fraction $f_1$ ($f_2=1-f_1$) are
\begin{equation}\label{eq:f}
\dfrac{\partial \bar{D}_{11}^r}{\partial f_1}=\dfrac{1}{\Gamma}\left(D_{11}^1-D_{11}^2\right)\bar{D}_{11}^r,
\end{equation}
\begin{equation*}
\dfrac{\partial \bar{D}_{12}^r}{\partial f_1}=\dfrac{1}{\Gamma}\left[\left(D_{11}^1-D_{11}^2\right)\bar{D}_{12}^r+D_{12}^1{D}_{11}^2-D_{12}^2{D}_{11}^1\right],
\end{equation*}
\begin{equation*}
\dfrac{\partial \bar{D}_{13}^r}{\partial f_1}=\dfrac{1}{\Gamma}\left[\left(D_{11}^1-D_{11}^2\right)\bar{D}_{13}^r+D_{13}^1{D}_{11}^2-D_{13}^2{D}_{11}^1\right],
\end{equation*}
\begin{equation*}
\dfrac{\partial \bar{D}_{22}^r}{\partial f_1}=D_{22}^1-D_{22}^2+\dfrac{1}{\Gamma^2}\left[(f_1)^2D_{11}^2-(f_2)^2D_{11}^1\right](D_{12}^1-D_{12}^2)^2,
\end{equation*}
\begin{equation*}
\dfrac{\partial \bar{D}_{23}^r}{\partial f_1}=D_{23}^1-D_{23}^2+\dfrac{1}{\Gamma^2}\left[(f_1)^2D_{11}^2-(f_2)^2D_{11}^1\right](D_{13}^1-D_{13}^2)(D_{12}^1-D_{12}^2),
\end{equation*}
\begin{equation*}
\dfrac{\partial \bar{D}_{33}^r}{\partial f_1}=D_{33}^1-D_{33}^2+\dfrac{1}{\Gamma^2}\left[(f_1)^2D_{11}^2-(f_2)^2D_{11}^1\right](D_{13}^1-D_{13}^2)^2.
\end{equation*} 

Moreover, derivatives of $\bar{\textbf{D}}^r$ with respect to the components in ${\textbf{D}}^1$ are
\begin{equation}\label{eq:d}
\dfrac{\partial\bar{\textbf{D}}^r}{\partial D_{11}^1}=\dfrac{1}{\Gamma}
\begin{Bmatrix}
{f_1D_{11}^2D_{11}^2}/{\Gamma} & f_2(-\bar{D}_{12}^r+D_{12}^2)&{f_2(-\bar{D}_{13}^r+D_{13}^2)}\\
&{f_1(f_2)^2(D_{12}^1-D_{12}^2)^2}/{\Gamma}&{f_1(f_2)^2(D_{13}^1-D_{13}^2)(D_{12}^1-D_{12}^2)}/{\Gamma}\\
\text{sym}&&{f_1(f_2)^2(D_{13}^1-D_{13}^2)^2}/{\Gamma}\\
\end{Bmatrix},
\end{equation}
\begin{equation*}
\dfrac{\partial\bar{\textbf{D}}^r}{\partial D_{12}^1}=
\begin{Bmatrix}
0 & f_1D_{11}^2/\Gamma & 0\\
&-2f_1f_2(D_{12}^1-D_{12}^2)/{\Gamma}&-f_1f_2(D_{13}^1-D_{13}^2)/{\Gamma}\\
\text{sym}&&0\\
\end{Bmatrix},
\end{equation*}
\begin{equation*}
\dfrac{\partial\bar{\textbf{D}}^r}{\partial D_{13}^1}=
\begin{Bmatrix}
0 & 0 & f_1D_{11}^2/\Gamma\\
&0&-f_1f_2(D_{12}^1-D_{12}^2)/{\Gamma}\\
\text{sym}&&-2f_1f_2(D_{13}^1-D_{13}^2)/{\Gamma}\\
\end{Bmatrix},
\end{equation*}
\begin{equation*}
\dfrac{\partial\bar{\textbf{D}}^r}{\partial D_{22}^1}=
\begin{Bmatrix}
0 & 0 & 0\\
&f_1&0\\
\text{sym}&&0\\
\end{Bmatrix},\quad
\dfrac{\partial\bar{\textbf{D}}^r}{\partial D_{23}^1}=
\begin{Bmatrix}
0 & 0 & 0\\
&0&f_1\\
\text{sym}&&0\\
\end{Bmatrix},\quad
\dfrac{\partial\bar{\textbf{D}}^r}{\partial D_{33}^1}=
\begin{Bmatrix}
0 & 0 & 0\\
&0&0\\
\text{sym}&&f_1\\
\end{Bmatrix}.
\end{equation*}
Similarly, forms of $\partial\bar{\textbf{D}}^r/\partial{\textbf{D}}^2$ can be obtained by switching the phase index in Eq. (\ref{eq:d}).

In Section \ref{sec:onlinehyper}, we will extend the theory of material network to general finite-strain problems, such as heterogeneous hyperelastic RVEs. Furthermore, the concept of two-layer building block can be applied to 3-dimensional (3D) problems, which will be explored in our future work. It should be noted that the existence of an analytical homogenization function is crucial for optimizing the fitting parameters in the material network using gradient-based methods. More details on the training of material network can be found in Section \ref{sec:learning}.

\subsection{Architects of material network}\label{sec:architects}
\begin{figure}[!htb]
	\centering
	\includegraphics[clip=true,trim = 4cm 4cm 4cm 3.5cm, width = 0.8\textwidth]{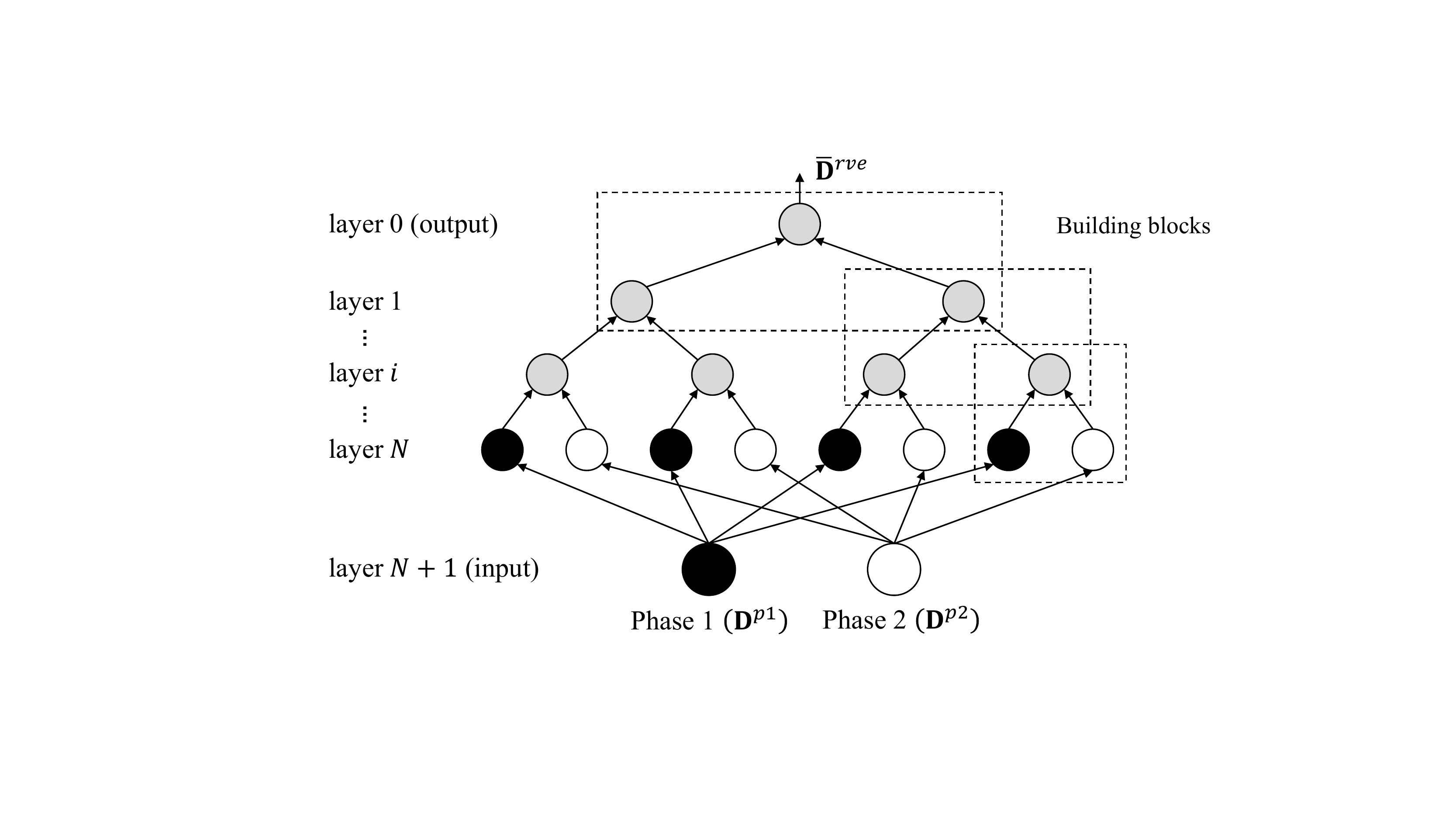}
	\caption{Illustration of material network with depth N=3. The nodes in each dashed box form a building block.}
	\label{fig:network}
\end{figure}

As shown in Fig. \ref{fig:network}, a material network with a binary-tree structure is proposed. The depth of the layer in the tree structure is denoted by $N$, and $i$ is the layer index. Layer 0 represents the output layer equivalent to the macroscopic homogenized material, and layer $(N+1)$ contains the input from each individual microscopic constituent. Layer $N$ is regarded as the bottom layer. Since a two-layer structure is considered as the basic homogenization building block, each node has 2 child nodes, and there are $2^i$ nodes at layer $i$ for $i\in[0,N]$.

\begin{figure}[!htb]
	\centering
	\includegraphics[clip=true,trim = 4cm 6cm 4cm 5cm, width = 0.8\textwidth]{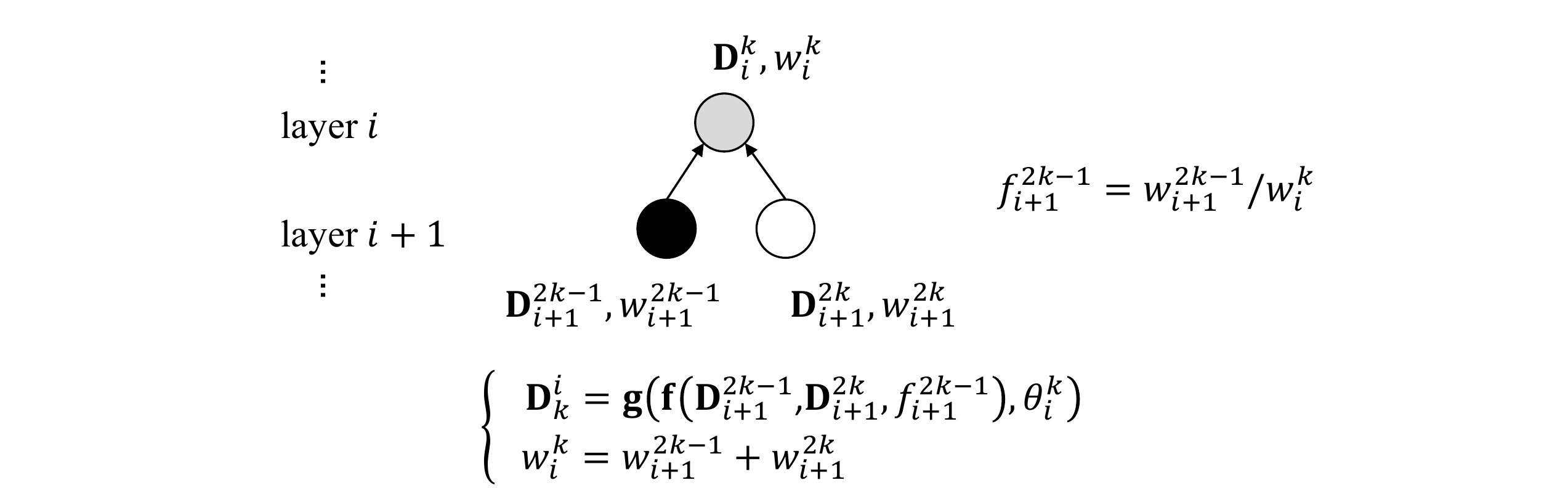}
	\caption{Data flow in the building block for the $k$-th node at layer $i$.  The volume fraction $f_{i+1}^{2k-1}$ is defined upon the weights.}
	\label{fig:block}
\end{figure}

To explain the data flows inside the material network, a homogenization building block for the $k$-th node at layer $i$ is provided in Fig. \ref{fig:block}. Correspondingly, the indices of its two child nodes at layer $i+1$ are $2k-1$ and $2k$. 

Other than the compliance matrix at the each node, a so-called weighting function $w$ is added into the data flow, which is used for tracking the volume fractions in each building block. When feed forward, the weights of the two child nodes are summed up and passed to their parent node,
\begin{equation}
w_i^k=w_{i+1}^{2k-1}+w_{i+1}^{2k}.
\end{equation}
By performing the summation recursively, the weight $w_i^k$ can be expressed as a summation of weights of its descendant nodes at the bottom layer $N$,
\begin{equation}\label{eq:wsum}
w_i^k=\sum_{j=2^{N-i}(k-1)+1}^{2^{N-i}k}w_N^j.
\end{equation}
Therefore, only the weights at bottom layer $N$ are independent parameters. The partial derivative of $w_i^k$ with respect to $w^j_N$ can be written as
\begin{equation}
\dfrac{\partial w_i^k }{\partial w^j_N}=
\begin{cases}
1 \quad\quad\text{if } k=\lceil{j/2^{N-i}}\rceil\\
0\quad\quad \text{otherwise}
\end{cases}.
\end{equation} 

The weights at layer $N$ are activated through the rectified linear unit (ReLU) \cite{glorot2011deep}, which have been widely used in deep learning. For $j\in[1,2^N]$, we have
\begin{equation}\label{eq:relu}
w_N^{j}(z^j_N)=Re(z^j_N)=\max(z^j_N,0),
\end{equation}
The activations $z^j_N$ are defined at the bottom layer and determine all the weighting functions $w$ in the network. Therefore, $z^j_N$ are treated as the fitting parameters.
The derivative of the ReLU is
\begin{equation}
Re'(z_N^j)=
\begin{cases}
1 \quad\text{if }  z_N^j>0\\
0\quad \text{otherwise}
\end{cases}.
\end{equation}
Once a unit is deactivated ($z^j_N<0$ and $Re(z^j_N)=0$), its gradient vanishes and the unit will never be activated again during the training process. This is good for automatically simplifying the material network and increasing the training speed, as long as the learning rate is set appropriately.

Back to the building block in Fig. \ref{fig:block}, the volume fraction of the first child nodes in the corresponding two-layer structure can be written as
\begin{equation}\label{eq:vf}
f_{i+1}^{2k-1}=w_{i+1}^{2k-1}/w_i^k.
\end{equation}
The derivative of $f_{i+1}^{2k-1}$ with respect to $w^j_N$ at the bottom layer is
\begin{equation}\label{eq:fzn}
\dfrac{\partial f_{i+1}^{2k-1}}{\partial w^j_N}=\dfrac{1}{w_i^k}\left(\dfrac{\partial w_{i+1}^{2k-1}}{\partial w^j_N}-f_{i+1}^{2k-1}\dfrac{\partial w_{i}^{k}}{\partial w^j_N}\right)
\end{equation} 
The compliance matrices $\textbf{D}^r$ and $\textbf{D}$ at the $k$-th node in layer $i$ become
\begin{equation}
{\textbf{D}^r}^k_i=\textbf{f}\left(\textbf{D}^{2k-1}_{i+1},\textbf{D}^{2k}_{i+1},f_{i+1}^{2k-1}\right)
\end{equation}
and
\begin{equation}\label{eq:oneblock}
\textbf{D}^k_i=\textbf{g}\left[\textbf{f}\left(\textbf{D}^{2k-1}_{i+1},\textbf{D}^{2k}_{i+1},f_{i+1}^{2k-1}\right),\theta_i^k\right].
\end{equation}

By combining Eq. (\ref{eq:wsum}), (\ref{eq:relu}), (\ref{eq:vf}) and (\ref{eq:oneblock}) the homogenized compliance matrix of the RVE $\bar{\textbf{D}}^{rve}$ can be written as a function of the compliance matrix from each material phase ($\textbf{D}^{p1}$ and $\textbf{D}^{p2}$) and the fitting parameters ($z$ and $\theta$), 
\begin{equation}\label{eq:final}
\underbrace{\bar{\textbf{D}}^{rve}}_\text{Outputs}=\textbf{D}_0^1=\textbf{h}(\underbrace{\textbf{D}^{p1},\textbf{D}^{p2}}_\text{Inputs}, \overbrace{z^{j=1,2,..,2^N}_N,\theta_{i=0,1,...,N}^{k=1,2,...,2^i}}^\text{Fitting parameters}),
\end{equation}
The number of layers $N$ can be regarded as a hyper-parameter of the material network, and there are totally $(3\times 2^N-1)$ fitting parameters in a material network with depth $N$ for a 2D problem. The optimum choice of $N$ needed to be studied through numerical experiments. A shallow material network with a small $N$ may not be sufficient for capturing the RVE behavior, while  a deep network with a large $N$ will increase the training time, as well as the computational cost for online extrapolation. 

\begin{remark}
	Other than $z^j_N$ in the bottom layer, it is also possible to directly use the volume fraction of the nodes at each layer as the fitting parameters. However, according to our tests, this would cause the vanishing gradient problem of volume fractions in the deep layers and might also result in unwanted early deactivation of sub-tree structures, whereas, the training process based on $z^j_N$ is more continuous and stable.
\end{remark}

\section{Machine learning of deep material network}\label{sec:learning}
\subsection{Cost function and dataset for training}\label{sec:training}
\begin{figure}[!htb]
	\centering
	\includegraphics[clip=true,trim = 4cm 7cm 4cm 5cm, width = 0.8\textwidth]{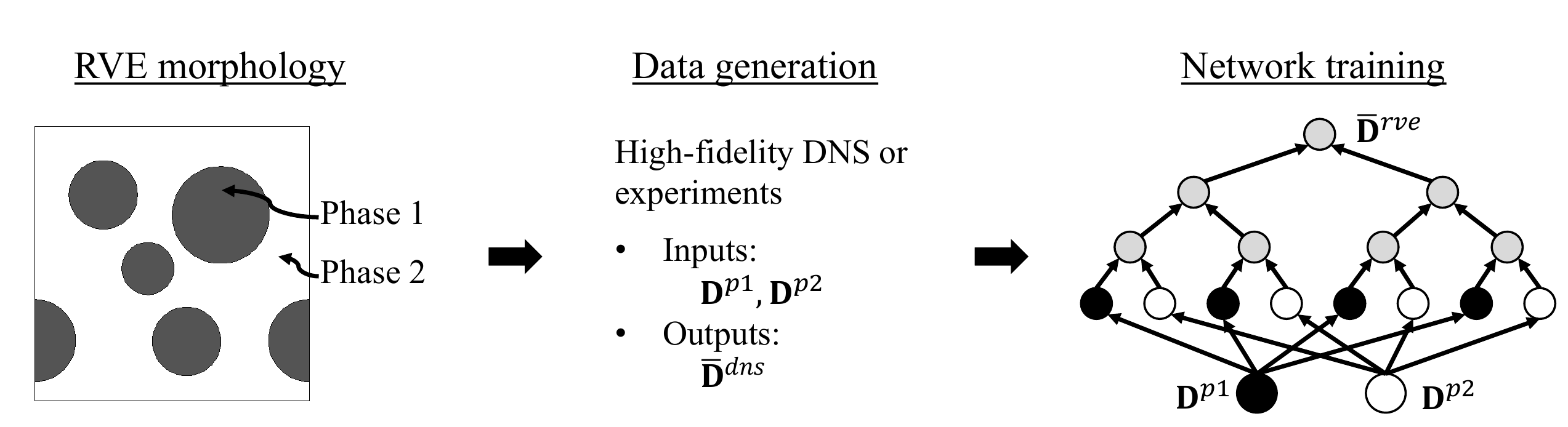}
	\caption{Training procedure of material network. The training data may be generated from high-fidelity DNS and experiments.}
	\label{fig:train}
\end{figure}

The procedure for training the material network is illustrated in Fig. \ref{fig:train}.  The dataset for training is generated through high-fidelity DNS of an RVE, or experiments (this can be challenging due to limited experimental tests). The inputs, outputs and fitting parameters for this training/optimization problem are listed below, with the number of variables given in the parentheses,

Inputs (12):
\begin{equation*}
\textbf{D}^{p1}\quad\text{and}\quad \textbf{D}^{p2}.
\end{equation*}
\par Outputs (6):
\begin{equation*}
\bar{\textbf{D}}^{dns}.
\end{equation*}
\par Fitting parameters ($3\times 2^N-1$):
\begin{equation*}
z^{j=1,2,..,2^N}_N,\theta_{i=0,1,...,N}^{k=1,2,...,2^i}.
\end{equation*}

When generating the input data, the components in $\textbf{D}^{p1}$ and $ \textbf{D}^{p2}$ cannot be assigned arbitrarily. For a physical material, situations resulting in negative strain energy density should be avoided. For the ease of data generation, the two material phases are considered to be orthotropical elastic,
\begin{equation}
\textbf{D}^{p1}=
\begin{Bmatrix}
1/E_{11}^{p1}&-\nu_{12}^{p1}/E_{22}^{p1}&\\
&1/E_{22}^{p1}&\\
&&1/(2G_{12}^{p1})\\
\end{Bmatrix}\quad\text{and}\quad
\textbf{D}^{p2}=
\begin{Bmatrix}
1/E_{11}^{p2}&-\nu_{12}^{p2}/E_{22}^{p2}&\\
&1/E_{22}^{p2}&\\
&&1/(2G_{12}^{p2})\\
\end{Bmatrix}.
\end{equation}
To remove the redundancy due to the scaling effect, we have
\begin{equation}
E_{11}^{p1}E_{22}^{p1}=1, \quad \log_{10}(E_{11}^{p2}E_{22}^{p2})\in U[-4, 4].
\end{equation}
The other inputs are selected randomly as
\begin{equation*}
\log_{10}(E_{22}^{p1}/E_{11}^{p1})\in U[-1, 1], \quad \log_{10}(E_{22}^{p2}/E_{11}^{p2})\in U[-1, 1],
\end{equation*}
\begin{equation*}
\dfrac{G_{12}^{p1}}{\sqrt{E_{22}^{p1}E_{11}^{p1}}} \in U[0.25, 0.5],\quad \dfrac{G_{12}^{p2}}{\sqrt{E_{22}^{p2}E_{11}^{p2}}} \in U[0.25, 0.5],
\end{equation*}
where $U$ stands for uniform distribution. Poisson ratios are selected to guarantee the compliance matrices are positive definite for the strain energy density to be positive, 
\begin{equation*}
\dfrac{\nu_{12}^{p1}}{\sqrt{E_{22}^{p1}/E_{11}^{p1}}}\in U[0.3,0.7], \dfrac{\nu_{12}^{p2}}{\sqrt{E_{22}^{p2}/E_{11}^{p2}}}\in U[0.3,0.7].
\end{equation*}
Design of experiments based on the Latin hypercube is performed to generate the input space.

To quantify how well we are training the network, a cost function based on the mean square errors (MSE) is proposed:
\begin{equation}\label{eq:cost}
C_0(z,\theta)=\dfrac{1}{2N_s}\sum_sC_s(z,\theta)=\dfrac{1}{2N_s}\sum_s {||\bar{\textbf{D}}^{dns}_s-\textbf{h}(\textbf{D}^{p1}_s,\textbf{D}^{p2}_s,z,\theta)||^2}/{||\bar{\textbf{D}}^{dns}_s||^2}.
\end{equation}
Here, $z$ and $\theta$ are the parameters to be fitted, $s$ is the index of the sample (or data point), and $N_s$ is the total number of training samples. Note that the cost function is normalized by the squared norm of $\bar{\textbf{D}}^{dns}_s$ to remove the scaling effect.
The operator $||...||$ denotes the matrix norm. For a general second-order matrix $\textbf{B}$ of dimension $m\times m$, the matrix norm is defined as
\begin{equation}\label{eq:matrixnorm}
||\textbf{B}||=\sqrt{\sum_{i=1}^m\sum_{j=1}^mb_{ij}^2}=\sqrt{\text{trace}(\textbf{B}^T\textbf{B})},
\end{equation}
and it is also called the Frobenius norm of matrix $\textbf{B}$. Since the trace of a matrix product is invariant under cyclic permutation, it can be proved that the norm of a compliance matrix under Mandel notation is invariant under arbitrary rotation. By using Eq. (\ref{eq:rotate}) and the symmetry of the compliance matrix, we have
\begin{equation}
||\textbf{D}||=\sqrt{\text{trace}(\textbf{D}\textbf{D})}=\sqrt{\text{trace}(\textbf{R}^{-1}{\textbf{D}}^r\textbf{R}\textbf{R}^{-1}{\textbf{D}}^r\textbf{R})}=\sqrt{\text{trace}(\textbf{R}\textbf{R}^{-1}{\textbf{D}}^r{\textbf{D}}^r)}=||\textbf{D}^r||.
\end{equation}
This guarantees that the cost function defined in Eq. (\ref{eq:cost}), as well as the optimized fitting parameters, is independent of the choice of the coordinate system.

To constrain the magnitude of activations $z_N^j$ and make the optimization problem well-posed, an additional term is appended to the cost function $C_0$,
\begin{equation}
C(z,\theta,\lambda)=C_0(z,\theta)+\lambda L(z),
\end{equation}
where $L(z)$ is defined as
\begin{equation}
L(z) = \left(\sum_j Re(z_N^j)-\xi\right)^2.
\end{equation}
The hyper-parameter $\xi$ determines the magnitude of $z_N^j$.
Although the hyper-parameter $\lambda$ will not alter the optimum fitting parameters for a given training dataset, in practice, it should be set appropriately to expedite the gradient descent algorithm in the training process.

\subsection{Stochastic gradient descent and backpropagation algorithm}\label{sec:sgd}
To minimize the cost function, the stochastic gradient descent (SGD) with backpropagation algorithm will be adopted. The gradient vector of the cost function can be written as
\begin{equation}
\nabla C=\left(\dfrac{\partial C}{\partial z_N^j},\dfrac{\partial C}{\partial \theta^k_i}\right)=\left(\dfrac{\partial C_0}{\partial z_N^j}+\lambda\dfrac{\partial L}{\partial z_N^j},\dfrac{\partial C_0}{\partial \theta^k_i}\right)
\end{equation}
with
\begin{equation*}
j=1,2,...,2^N; \quad i=0,1,...,N; \quad k=1,2,...,2^i.
\end{equation*}
Due to the simplicity of $L$, we will only focus on deriving the gradients of the cost function $C_0$ in this section.

In the context of training the neural networks, backpropagation is commonly used to adjust the weight of neurons by calculating the gradient of the cost function. Sometime, this technique is also called backward propagation of errors, because the error is calculated at the output and distributed backward through the network layers. Similar technique can be utilized to train the material network. The heart of backpropagation is the chain rule in calculus, by which the time for computing those partial differentiates can be greatly reduced and less memory is needed. The derivation and implementations of backpropagation for training the material network will be discussed below.

As shown in Fig. \ref{fig:detailednetwork}, each independent component of compliance matrix is regarded as a neuron in the network. After the vectorization, the components in the compliance matrices ${\textbf{D}}$ for all the nodes at layer $i$ are denoted by $d_i^n$, with the index $n$ ranging from $1$ to $6\times2^i$. Meanwhile, we use $a_i^n$ to represent the components in $\textbf{D}^r$ for all the nodes at layer $i$.  In the algorithm, we define the error $\delta_i^n$ of neuron $n$ in layer $i$ by
\begin{equation}
\delta_i^n = \dfrac{\partial C_0}{\partial d_i^n},
\end{equation}
and the error $\alpha_i^n$ by
\begin{equation}
\alpha_i^n = \dfrac{\partial C_0}{\partial a_i^n}.
\end{equation}

In the output layer, the components of $\delta_0$ are given by
\begin{equation}\label{eq:dlayer0}
\delta_0^n=\dfrac{\partial C_0}{\partial d_0^n}.
\end{equation}
Since $C_0$ takes a quadratic form in Eq. (\ref{eq:cost}), $\delta_0^n$ can be easily computed. For $i\geq1$, the error $\alpha^k_{i-1}$ and $\delta_i^n$ can be computed by
\begin{equation}\label{eq:alayeri1}
\alpha^k_{i-1}=\dfrac{\partial C_0}{\partial d_{i-1}^j}\dfrac{\partial d_{i-1}^j}{\partial a_{i-1}^k}=\delta_{i-1}^j\dfrac{\partial d_{i-1}^j}{\partial a_{i-1}^k},
\end{equation}
\begin{equation}\label{eq:dlayeri}
\delta_i^n=\dfrac{\partial C_0}{\partial a_{i-1}^k}\dfrac{\partial a_{i-1}^k}{\partial d_{i}^n}=\alpha_{i-1}^k\dfrac{\partial a_{i-1}^k}{\partial d_{i}^n}.
\end{equation}
All the derivatives have been previously defined. The expressions of ${\partial d_{i-1}^j}/{\partial a_{i-1}^k}$ and ${\partial a_{i-1}^k}/{\partial d_{i}^n}$ can be found in Eq. (\ref{eq:a}) and (\ref{eq:d}), respectively.

By combining Eq. (\ref{eq:dlayer0}), (\ref{eq:alayeri1}) and (\ref{eq:dlayeri}), we can compute the error $\delta_i$ and $\alpha_i$ at any layer in the network. Within one step, the error of cost function propagates backward from the output layer $i=0$ to the bottom layer $i=N$. After the backpropagation of error for compliance matrices, we can compute the rate of change of the cost function with respect to a rotation angle $\theta_i^k$ at layer $i$ by
\begin{equation}\label{eq:C0theta}
\dfrac{\partial C_0}{\partial \theta^k_i}=\delta_i^j\dfrac{\partial d_i^j}{\partial \theta^k_i}.
\end{equation}
where ${\partial d_i^j}/{\partial \theta^k_i}$ can be found in Eq. (\ref{eq:dtheta}). Meanwhile, the rate of change of the cost function with respect to an activation $z_N^j$ in the bottom layer can be computed by
\begin{equation}\label{eq:C0z}
\dfrac{\partial C_0}{\partial z_N^j}=\left(\alpha_{i-1}^l\dfrac{\partial a_{i-1}^l}{\partial f_i^m}\dfrac{\partial f_i^m}{\partial w_N^j}\right)\circ Re'(z_N^j),
\end{equation}
where $\circ$ denotes the Hadamard (element-wise) product. Analytical forms of ${\partial a_{i-1}^l}/{\partial f_i^m}$ are provided in Eq. (\ref{eq:f}), and ${\partial f_i^m}/{\partial w_N^j}$ is defined in Eq. (\ref{eq:fzn}).

Finally, we conclude the backpropagation algorithm for finding gradients of the cost function $C_0$ with respect to the fitting parameters in the material network.
\begin{enumerate}
\item \textbf{Input}: Set the initial values for $z_N$ and $\theta_i$ $(i=0,1,...,N)$
\item \textbf{Feedforward}: For each layer $i$, compute $a_i$ and $d_i$
\item \textbf{Output error} $\delta_0$: Compute the vector $\delta_0$ using Eq. (\ref{eq:dlayer0})
\item \textbf{Backpropagate the error}: For each $i=1,2,...,N$, compute $\alpha_{i-1}$ and $\delta_i$ iteratively using Eq. (\ref{eq:alayeri1}) and (\ref{eq:dlayeri})
\item \textbf{Output}: Compute the gradients of the cost function using Eq. (\ref{eq:C0theta}) and (\ref{eq:C0z})
\end{enumerate}

In a standard gradient descent method, the gradients of each sample $\nabla C_s$ are computed and averaged to get $\nabla C$ based on Eq. (\ref{eq:cost}),
\begin{equation}
\nabla C = \dfrac{1}{2N_s}\sum_s \nabla C_s.
\end{equation}
To accelerate the training speed, the stochastic gradient descent (SGD) is used to train the material network after the gradient vector is obtained. Instead of computing and averaging the gradients over all the samples at each step, a small number $M$ of samples are randomly picked to estimate the gradient $\nabla C$. In this way, the original dataset is divided into several mini-batches which will be used in a sequence of learning steps. If we label the samples in a mini-batch by $s_1, s_2,...,s_M$, the gradient of cost function for the corresponding learning step can be approximated by
\begin{equation}
\nabla C \approx \dfrac{1}{2M}\sum_{i=1}^{M} \nabla C_{s_i}.
\end{equation}
Writing out the gradient decent updating rule in terms of the components, we have
\begin{equation}
z_N^j {}'=z_N^j - \eta\dfrac{\partial C}{\partial z_N^j}
\end{equation}
and 
\begin{equation}
\theta^k_i {}'=\theta^k_i- \eta\dfrac{\partial C}{\partial \theta^k_i},
\end{equation}
where $\eta$ is a positive parameter known as the learning rate. Here, an epoch is defined as each time the algorithm has processed all the mini-batches and seen all the samples in the original dataset. In practice, after an epoch is completed, the dataset will be randomly shuffled and prepared for the next epoch to minimize the sample bias. 

All fitting parameters are initialized randomly following a uniform distribution at the beginning of the SGD algorithm,
\begin{equation}\label{eq:initialfit}
z_N^j{}^{(0)}\sim U(0.2,0.8)\quad\text{and}\quad \theta^k_i{}^{(0)}\sim U(-\pi/2,\pi/2).
\end{equation}
In order to reduce the influence of the constraint term $L(z)$ at early training stage, the hyper-parameter $\xi$ is chosen to be
\begin{equation}
\xi = E(z_N^j{}^{(0)})2^{N-1} = 2^{N-2}.
\end{equation} 
This will help to avoid unwanted early deactivation of the material network. Additionally, the learning rate $\eta$ is an important hyper-parameter that determines the performance of SGD. In the paper, we have utilized a Bold driver algorithm to dynamically adapting the learning rate, and it has effectively improved the training speed in our numerical study. In future, more sophisticated algorithms, like annealing, can be used to tune the learning rate. 

\subsection{Model compression and parameter reduction} \label{sec:compression}
\begin{figure}[!htb]
	\centering
	\includegraphics[clip=true,trim = 1.5cm 6cm 1.5cm 3cm, width = \textwidth]{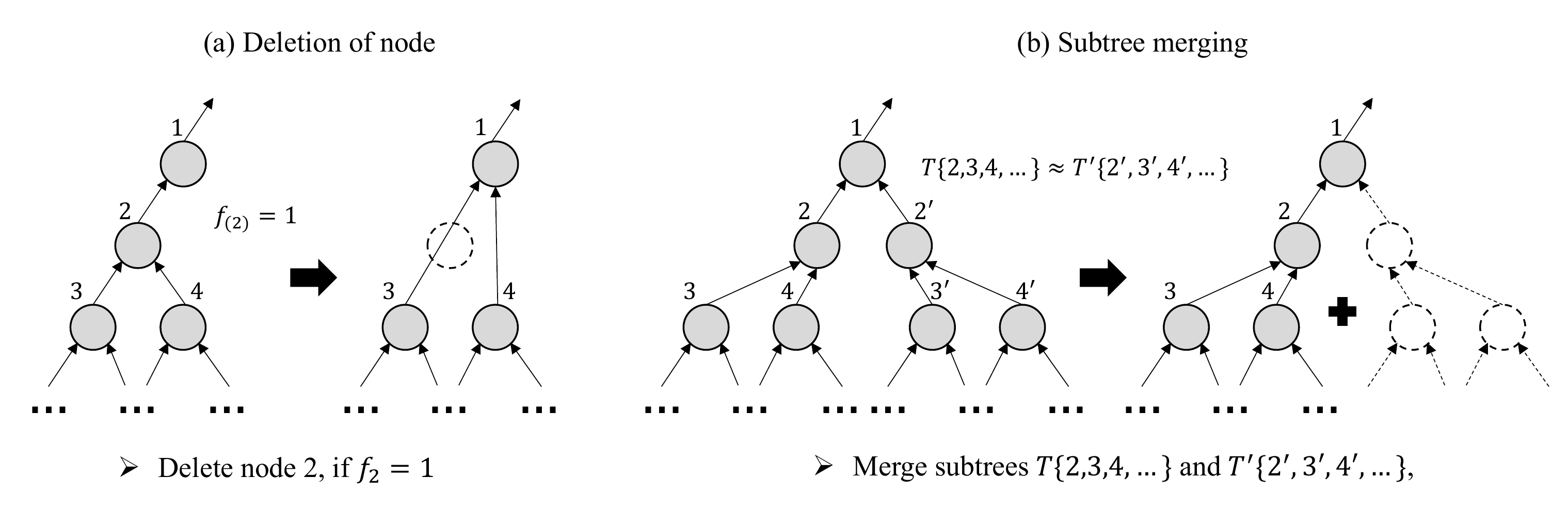}
	\caption{Illustrations of (a) deletion of node and (b) subtree merging for network compression. The merging operation is performed after reordering of the network. }
	\label{fig:compression}
\end{figure}

The speed and convergence rate of the network training process can be improved by removing the redundancy in the network, which is called model compression. On the other hand, a less complex network is always beneficial for the extrapolation process in the online stage. As discussed in Section \ref{sec:architects}, the rectified linear unit enables automatical deactivation of nodes during the training; hence it already has a function of network compression. Moreover, two additional approaches are also introduced for further model compression: 1)deletion of the parent node with only one child node; 2) subtree merging based on the similarity search. Illustrations of these two model compression operations are shown in Fig. \ref{fig:compression}.

In Fig. \ref{fig:compression} (a), node 2 is deleted as its volume fraction is equal to 1. Meanwhile, the rotation angle of node 1 needs to be updated to avoid a sudden jump in the cost function,
\begin{equation}
\theta_{(1)}^{new}=\theta_{(1)}^{old}+\theta_{(2)}^{old}
\end{equation}
As shown in Fig. \ref{fig:compression} (b), merging of the subtree structures is based on the similarity search. The comparison of the two subtrees, $T\{2,3,4,...\}$ and $T'\{2',3',4',...\}$, are performed between all descendant layers of their root nodes $2$ and $2'$. The differences are evaluated by
\begin{equation}
\delta_{T-T'}^f=\max_{m=3}\left(|f_{(m)}-f_{(m')}|\right) \quad \text{and} \quad 
\delta_{T-T'}^\theta=\max_{m=2}\left(\dfrac{1}{\pi}|\text{mod}{(\theta_{(m)}-\theta_{(m')},\pi)}|\right),
\end{equation}
where the index $(m)$ denotes the order of nodes in the subtree. If the differences of $\delta_{T-T'}^f$ and $\delta_{T-T'}^\theta$ are both below the tolerances, it is said that 
\begin{equation*}
T\{2,3,4,...\}\approx T'\{2',3',4',...\},
\end{equation*}
and $T'\{2',3',4',...\}$ will be merged left to compress the network and reduce the number of fitting parameters. Afterwards, the parameters in the new subtree $T\{2,3,4,...\}$ become
\begin{equation}
w^{new}_{(m)}=w^{old}_{(m)}+w^{old}_{(m')},\quad \theta^{new}_{(m)}=\dfrac{1}{2}(\theta^{old}_{(m)}+\theta^{old}_{(m')})\quad\text{for } m\geq2
\end{equation}
Note that a deletion operation will follow right after the merging operation since $f_{(2)}$ becomes 1 upon the merging of the two subtrees in Fig. \ref{fig:compression} (b).

The merging operation should be performed on an ordered material network. At any layer $i\in[1,N-1]$ of an order material network, the following condition should be satisfied,
\begin{equation*}
w_i^{2k-1}\geq w_i^{2k}  \quad \forall k \in [1,2^{i-1}].
\end{equation*}
Each time before the similarity search and subtree merging, the whole material network will be reordered based on the weighting functions. To save the training time, the network compression operations are performed every 10 epochs in our study.

\section{Prediction and extrapolation}\label{sec:online}
\subsection{Nonlinear small-strain plasticity}\label{sec:onlineplas}
Since the material network can capture the essential topological structure of an RVE, it can be used for online prediction of nonlinear plasticity, more than just linear elasticity. Each node at the bottom layer $N$ can be regarded as an individual material node with independent degrees of freedom (DOFs), precisely, the infinitesimal strain $\boldsymbol{\varepsilon}^j_N$.  As a result, the total number of DOFs $N_{dof}$ in the material network at the beginning of training is proportional to $2^N$. On the other hand, the ReLU activation function and several compression algorithms introduced in Section \ref{sec:compression} are used to decrease the model complexity during the training. For the final trained network, the number of DOFs becomes proportional to the number of active/remaining nodes at the bottom layer, denoted by $N_a$,
\begin{equation}
N_{dof}\propto N_{a}.
\end{equation}

In the bottom layer, the Cauchy stress at the $j$-th node is denoted by $\boldsymbol{\sigma}^j_N$. At each loading step, its stress-strain relation can be written as
\begin{equation}
\Delta\boldsymbol{\varepsilon}^{j}_{N}=\textbf{D}^{j}_{N}\Delta\boldsymbol{\sigma}^{j}_{N}+\delta\boldsymbol{\varepsilon}^{j}_{N}.
\end{equation}
The compliance matrix and incremental stress are given by the local constitutive law, 
\begin{equation}
\Delta\boldsymbol{\sigma}^{j}_{N}=\Delta\boldsymbol{\sigma}^{j}_{N}(\Delta\boldsymbol{\varepsilon}^{j}_{N},\boldsymbol{\varepsilon}^{j}_{N},\boldsymbol{\sigma}^{j}_{N},\boldsymbol{\beta}^{j}_{N},...)
\end{equation}
and
\begin{equation}
\textbf{D}^{j}_{N}=\textbf{D}^{j}_{N}(\Delta\boldsymbol{\varepsilon}^{j}_{N},\boldsymbol{\varepsilon}^{j}_{N},\boldsymbol{\sigma}^{j}_{N},\boldsymbol{\beta}^{j}_{N},...),
\end{equation}
where $\boldsymbol{\beta}^{j}_N$ is the vector of history-dependent internal variables. Once the compliance matrix and incremental stress are obtained, we can compute the residual strain in the bottom layer by 
\begin{equation}
\delta\boldsymbol{\varepsilon}^{j}_{N}=\Delta\boldsymbol{\varepsilon}^{j}_{N}-\textbf{D}^{j}_{N}\Delta\boldsymbol{\sigma}^{j}_{N}.
\end{equation}
Note that the residual strain comes from the material non-linearity. In other words, if all the materials are linear elastic, there will be no residual strain and the homogenization procedure will be same as the one described for the training process.
 
In a network with material non-linearity, both of the compliance matrix $\textbf{D}$ and residual strain $\delta\boldsymbol{\varepsilon}$ are feed forward from the bottom layer to the output layer . In Fig. \ref{fig:block}, the stress-strain relations for the $2k$-th and $(2k-1)$-th nodes on layer $i+1$ at each loading step can be written as
\begin{equation}\label{eq:cl1}
\Delta\boldsymbol{\varepsilon}^{2k-1}_{i+1}=\textbf{D}^{2k-1}_{i+1}\Delta\boldsymbol{\sigma}^{2k-1}_{i+1}+\delta\boldsymbol{\varepsilon}^{2k-1}_{i+1} \quad\text{and}\quad
\Delta\boldsymbol{\varepsilon}^{2k}_{i+1}=\textbf{D}^{2k}_{i+1}\Delta\boldsymbol{\sigma}^{2k}_{i+1}+\delta\boldsymbol{\varepsilon}^{2k}_{i+1},
\end{equation}
and the stress-strain relation of their parent node at layer $i$ is
\begin{equation}\label{eq:cl2}
\Delta\boldsymbol{\varepsilon}^{k}_{i}=\textbf{D}^{k}_{i}\Delta\boldsymbol{\sigma}^{k}_{i}+\delta\boldsymbol{\varepsilon}^{k}_{i}.
\end{equation}
In the forward homogenization process, the compliance matrix and residual strain at the parent node are calculated from the ones in its child nodes,
\begin{equation}\label{eq:p}
\textbf{D}^k_i=\textbf{p}\left(\textbf{D}^{2k-1}_{i+1},\textbf{D}^{2k}_{i+1}\right)
\end{equation}
and 
\begin{equation}\label{eq:q}
\delta\boldsymbol{\varepsilon}^k_i=\textbf{q}\left(\textbf{D}^{2k-1}_{i+1},\textbf{D}^{2k}_{i+1},\delta\boldsymbol{\varepsilon}^{2k-1}_{i+1},\delta\boldsymbol{\varepsilon}^{2k}_{i+1}\right).
\end{equation}
Again, due to the simplicity of the two-layer building block, analytical forms of function $\textbf{p}$ and $\textbf{q}$ can be derived. Since the function $\textbf{p}$ for compliance matrices takes the same form as Eq. (\ref{neuron})  in Section \ref{sec:buildingblock}, we will only provide the form of function $\textbf{q}$ for residual strain here. For the generic two-layer structure illustrated in Fig. \ref{fig:twolayer}, we have
\begin{equation}\label{eq:qb}
\delta\bar{\boldsymbol{\varepsilon}}=\textbf{q}\left(\textbf{D}^1,\textbf{D}^2,\delta\boldsymbol{\varepsilon}^1,\delta\boldsymbol{\varepsilon}^2\right),
\end{equation}
where $\delta\bar{\boldsymbol{\varepsilon}}$ is the overall residual strain at the parent node and $\delta\boldsymbol{\varepsilon}^1,\delta\boldsymbol{\varepsilon}^2$ are the residual strains at the two child nodes. Similarly,  the function $\textbf{q}$ contains a homogenization operation and a rotation operation. The homogenization operation gives the homogenized residual strain $\delta\bar{\boldsymbol{\varepsilon}}^r$ of the two-layer structure,
\begin{equation}\label{eq:rehom}
\delta\bar{\boldsymbol{\varepsilon}}^r=\delta\bar{\boldsymbol{\varepsilon}}^r\left(\textbf{D}^1,\textbf{D}^2,\boldsymbol{\varepsilon}^1,\boldsymbol{\varepsilon}^2, f_1\right),
\end{equation}
and the rotation operation outputs the overall residual strain $\delta\bar{\boldsymbol{\varepsilon}}$ by
\begin{equation}
\delta\bar{\boldsymbol{\varepsilon}}=\textbf{R}(-\theta)\delta\bar{\boldsymbol{\varepsilon}}^r,
\end{equation} 
where $\textbf{R}(-\theta)$ is the rotation matrix defined in Eq. (\ref{eq:rotatem}). Based on the equilibrium condition and kinematic constraints of the two-layer structure, the analytical forms of $\delta\bar{\boldsymbol{\varepsilon}}^r$ in Eq. (\ref{eq:rehom}) can be derived as
\begin{equation}
\delta\bar{\varepsilon}_{11}^r=\dfrac{1}{\Gamma}(f_1D_{11}^2\delta\varepsilon^1_{11}+f_2D_{11}^1\delta\varepsilon^2_{11}),
\end{equation}
\begin{equation*}
\delta\bar{\varepsilon}_{22}^r=f_1\delta\varepsilon^1_{22}+f_2\delta\varepsilon^2_{22}-\dfrac{1}{\Gamma}f_1f_2(D_{12}^1-D_{12}^2)(\delta\varepsilon^1_{11}-\delta\varepsilon^2_{11}),
\end{equation*}
\begin{equation*}
\delta\bar{\varepsilon}_{12}^r=f_1\delta\varepsilon^1_{12}+f_2\delta\varepsilon^2_{12}-\dfrac{1}{\Gamma}f_1f_2(D_{13}^1-D_{13}^2)(\delta\varepsilon^1_{11}-\delta\varepsilon^2_{11}),
\end{equation*}
with 
\begin{equation*}
\Gamma=f_1 D_{11}^2+f_2 D_{11}^1\quad\text{and}\quad f_2=1-f_1
\end{equation*}
Finally, after the forward homogenization process, the stress-strain relation at the output layer $i=0$ is obtained,
\begin{equation}
\Delta\boldsymbol{\varepsilon}^{1}_{0}=\textbf{D}^{1}_{0}\Delta\boldsymbol{\sigma}^{1}_{0}+\delta\boldsymbol{\varepsilon}^{1}_{0},
\end{equation}
Macroscopic boundary conditions are applied on the output layer, and $\Delta\boldsymbol{\varepsilon}^{1}_{0}$ and $\Delta\boldsymbol{\sigma}^{1}_{0}$ represent the macroscopic strain and stress, respectively. In the de-homogenization process, the incremental strain and stress data is feed backward from the output layer to the bottom layer, using the homogenized constitutive law Eq. (\ref{eq:cl2}) at each node.

\begin{remark}
For nonlinear history-dependent materials, the internal variables are stored at each individual active node in the bottom layer. Once each material law is evaluated at the bottom layer (e.g. loading-unloading conditions), the information of compliance (or stiffness) tensor and residual strain (or stress) are propagated forward through the network to give the homogenized properties at the output layer.
\end{remark}

 Newton's method is used for solving the nonlinear system. A Newton's iteration consists of one forward homogenization process and one backward de-homogenization process,
\begin{equation*}
\text{layer } N \xrightarrow[\text{homogenization}]{\text{forward}} \text{layer } 0 \text{ (macroscale)} \xrightarrow[\text{de-homogenization}]{\text{backward}} \text{layer } N.
\end{equation*}
After each iteration, the new strain increment $\Delta\boldsymbol{\varepsilon}_N^{j (new)}$ at the bottom layer can be computed. The relative difference between the strain increments at the current and previous iterations will be computed and used for the convergence check. If convergence is not yet achieved, the compliance matrices and residual strains at the bottom layer will be recomputed based upon the new $\Delta\boldsymbol{\varepsilon}_N^{j (new)}$, and used for the next iteration. Upon convergence, the internal variables at each active node in the bottom layer will be updated, and the analysis moves on to the next loading step. Note that the residual strain $\delta\boldsymbol{\varepsilon}_N^j$ is not necessarily equal to zero for nonlinear materials. The same iterative procedure can be applied to finite-strain problem as will be discussed in Section \ref{sec:onlinefinite}. 

If we define the homogenization and de-homogenization processes in Fig. \ref{fig:block} as one unit operation, the total number of operations of the whole material network in one Newton's iteration is proportional to the number of DOFs,
\begin{equation}
N_{op}\propto N_{dof}\quad\text{or}\quad N_{op}\propto N_{a}
\end{equation}
This is advantageous over most existing homogenization methods, such as FEM ($N_{op}\propto (N_{dof})^{2.x}$), integral equation-based methods ($N_{op}\propto (N_{dof})^3$) \cite{liu2016self}. The computational cost of material network will be evaluated numerically in Section \ref{sec:plasticity}.

\subsection{Finite-strain problem}\label{sec:onlinefinite}
The online extrapolation procedure of the material network based on the two-layer building block can also be extended to finite-strain problems with large deformation. 
In finite-strain problems, we choose the deformation gradient $\textbf{F}$ and the first Piola-Kirchhoff stress $\textbf{P}$ as the strain and stress measures, respectively. Here, we defined the tangent elasticity tensor $\textbf{A}$ in the rate form as
\begin{equation}
\utilde{\dot{\textbf{P}}}=\utilde{\textbf{A}}:\utilde{\dot{\textbf{F}}} \quad\text{ or}\quad \utilde{\dot{P}}_{ij} = \utilde{A}_{ijkl}\utilde{\dot{F}}_{kl}
\end{equation}
and
\begin{equation}
\utilde{A}_{ijkl}=\utilde{C}^{SE}_{jnpl}\utilde{F}_{in}\utilde{F}_{kp}+\utilde{S}_{jl}\delta_{ki},
\end{equation}
where $\utilde{\textbf{C}}^{SE}$ is the tangent stiffness tensor defined upon the second Piola-Kirchhoff stress $\utilde{\textbf{S}}$ and Green strain $\utilde{\textbf{E}}$. The first elasticity tensor has major symmetry, $A_{ijkl}=A_{klij}$, but does not have minor symmetries. As a result, Voigt notation cannot be applied in this case, instead, we represent the deformation gradient and first Piola-Kirchhoff stress in 2D as
\begin{equation}
\textbf{F} = \{\utilde{F}_{11},\utilde{F}_{22},\utilde{F}_{12},\utilde{F}_{21}\}^T=\{F_{1},F_{2},F_{3},F_{4}\}^T,
\end{equation}
\begin{equation*}
\textbf{P} = \{\utilde{P}_{11},\utilde{P}_{22},\utilde{P}_{12},\utilde{P}_{21}\}^T=\{P_{1},P_{2},P_{3},P_{4}\}^T.
\end{equation*}
Furthermore, the stiffness matrix $\textbf{A}$ can be written as
\begin{equation}\label{eq:finiteA}
{\dot{\textbf{P}}}={\textbf{A}}{\dot{\textbf{F}}},\quad\textbf{A}=
\begin{Bmatrix}
A_{11}&A_{12}&A_{13}&A_{14}\\
&A_{22}&A_{23}&A_{24}\\
&&A_{33}&A_{34}\\
\text{sym}&&&A_{44}\\
\end{Bmatrix}.
\end{equation}
Due to major symmetry, the matrix $\textbf{A}$ is symmetric. It can be seen from Eq. (\ref{eq:finiteA}) that the stiffness matrix for finite strain now has 10 independent components.

For each loading step, the stress-strain relation at each node in the material network takes the following form,
\begin{equation}
\Delta\textbf{P} = \textbf{A}\Delta \textbf{F}+\delta\textbf{P},
\end{equation}
where $\Delta\textbf{P}$ and $\Delta \textbf{F}$ are the incremental stress and strain, and $\delta\textbf{P}$ is the residual first Piola-Kirchhoff stress. In the forward homogenization process, $\textbf{A}_N$ and $\delta\textbf{P}_N$ at the bottom layer $i=N$ are given by the material constitutive laws. Infos of stiffness tensor and residual stress are feed forward all the way to the output layer $i=0$ through the homogenization and rotation operations at each building block.

In the generic two-layer structure shown in Fig. \ref{fig:twolayer}, the homogenized stiffness tensor $\bar{\textbf{A}}^r$ before the rotation can be written as a function of the stiffness tensors of its two constituents ($\textbf{A}^1$, $\textbf{A}^2$) and their volume fractions ($f_1$ and $f_2=1-f_1$),
\begin{equation}\label{eq:arfinite}
\bar{\textbf{A}}^r=\bar{\textbf{A}}^r\left(\textbf{A}^1,\textbf{A}^2,f_1\right).
\end{equation}
The homogenization function of $\delta\textbf{P}$ takes the following form,
\begin{equation}\label{eq:prfinite}
\delta\bar{\textbf{P}}^r=\delta\bar{\textbf{P}}^r\left(\textbf{A}^1,\textbf{A}^2,\delta\textbf{P}^1,\delta\textbf{P}^2,f_1\right).
\end{equation}
Analytical forms of Eq. (\ref{eq:arfinite}) and (\ref{eq:prfinite}) can be derived based on the equilibrium condition,
\begin{equation}\label{eq:equicon}
P_{2}^1 = P_{2}^2, \quad P_{3}^1 = P_{3}^2,
\end{equation}
and the kinematic constraints,
\begin{equation}\label{eq:kinecon}
F_{1}^1 = F_{1}^2, \quad F_{4}^1 = F_{4}^2.
\end{equation}

Explicit forms of $\bar{\textbf{A}}^r$ and $\delta\bar{\textbf{P}}^r$ can be found in Eq. (\ref{eq:arexplicit}) and (\ref{eq:prexplicit}) in \ref{sec:a1} with derivations.
\begin{remark}
In finite strain problems, we use the $\textbf{P}$ and $\textbf{F}$ as the stress and strain measures to keep the rotational information, instead of the symmetric ones such as second Piola-Kirchhoff stress and Green strain. To satisfy the equilibrium and kinematic conditions at the interface of the 2D building block, we need to constrain at least 4 DOFs as shown in Eq. (\ref{eq:equicon}) and (\ref{eq:kinecon}).
\end{remark}

Meanwhile, the finite-strain rotation matrix $\textbf{R}^f$ for a given angle $\theta$ becomes
\begin{equation}
\textbf{R}^f(\theta)=
\begin{Bmatrix}
\cos^2\theta&\sin^2 \theta&\sin\theta\cos\theta&\sin\theta\cos\theta\\
\sin^2 \theta&\cos^2\theta&-\sin\theta\cos\theta&-\sin\theta\cos\theta\\
-\sin\theta\cos\theta&\sin\theta\cos\theta&\cos^2\theta&-\sin^2\theta\\
-\sin\theta\cos\theta&\sin\theta\cos\theta&-\sin^2\theta&\cos^2\theta\\
\end{Bmatrix}.
\end{equation}
In the rotation operation, the new stiffness matrix and residual stress can be computed by
\begin{equation}
\bar{\textbf{A}}=\textbf{R}^f(-\theta)\bar{\textbf{A}}^r\textbf{R}^f(\theta)\quad\text{and}\quad \delta\bar{\textbf{P}}=\textbf{R}^f(-\theta)\delta\bar{\textbf{P}}^r
\end{equation}

After solving the macroscopic boundary value problem at the output layer, data of incremental deformation gradient $\Delta\textbf{F}$ and first Piola-Kirchhoff stress $\Delta \textbf{P}$ is feed backward from the output layer to the bottom layer. Similar to the procedure for nonlinear plasticity, Newton's method is used for finding the converged solution at each loading step.  

\section{Evaluations}\label{sec:results}
\subsection{Network training for various RVE morphologies}
Material networks are trained for four different types of RVE morphologies under 2D plane strain conditions: a) uniform material with a single phase; b) matrix-inclusion material with circular inclusions embedded in the matrix; c) amorphous material from solving the Cahn-Hilliard equation \cite{cahn1958free} for spinodal decomposition; d) anisotropic material with phases penetrated in one direction. 
\begin{figure}[!htb]
	\centering
	\includegraphics[clip=true,trim = 4cm 6.5cm 3.5cm 5cm, width = 0.95\textwidth]{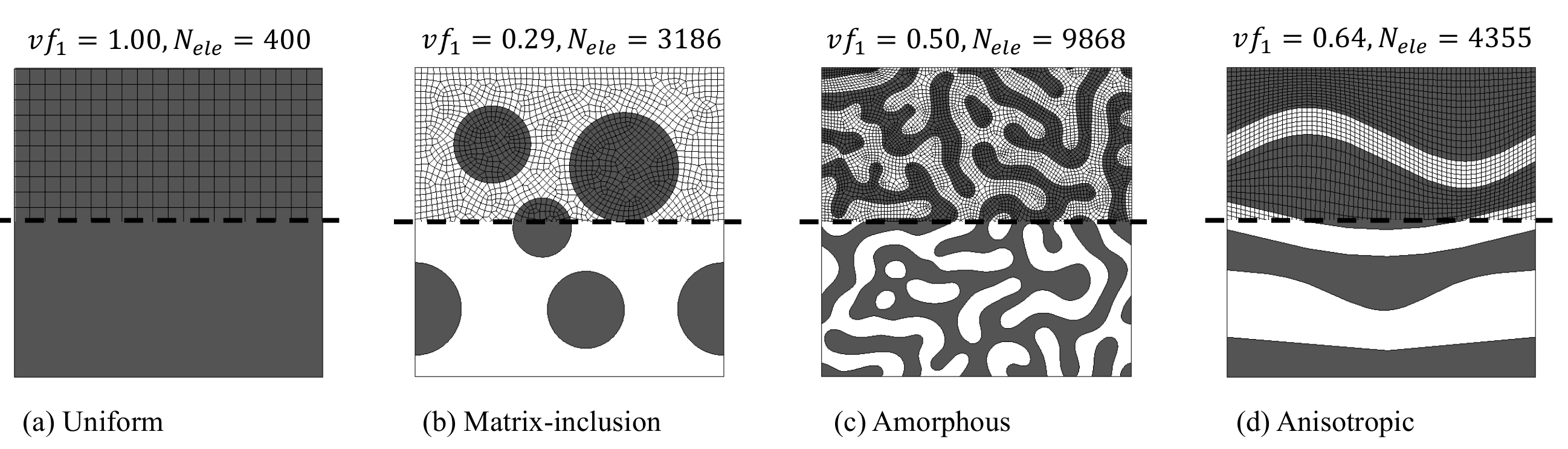}
	\caption{Geometry and finite element meshes (shown in the top half) of various two-phase RVEs evaluated in the work. Phase 1 is denoted by darker faces. The volume fraction of phase 1 material $vf_1$  and the number of elements in the FE mesh $N_{ele}$ are shown at the top of each RVE plot.}
	\label{fig:rves}
\end{figure}

The geometries and the corresponding finite element meshes of these four RVEs are provided in Fig. \ref{fig:rves}. Volume fraction of phase 1 material $vf_1$ in these four RVEs are 1.00, 0.29, 0.50 and 0.64. Number of elements in their FE meshes $N_{ele}$ are 400, 3186, 9868 and 4355. Direct numerical simulations (DNS) of RVEs are performed using the RVE package based on implicit finite element method in LS-DYNA\textregistered  $ $ with periodic boundary conditions. 

For each RVE, 200 training samples and 100 validation samples are generated, following the procedure introduced in Section \ref{sec:training}. In the SGD training process, the mini-batch size is chosen to be 20, so that there are 10 learning steps in each epoch. All the material networks are trained for 10000 epochs. The fitting parameters $z^j_N$ and $\theta_i^k$ are initialized randomly according to Eq. (\ref{eq:initialfit}). Note that the choice of initial fitting parameters may influence the topological structures of the trained network, and this initialization effect will be investigated in our future work.

Python libraries have been created for both network training and online extrapolation, and all the numerical examples including DNS are tested on one Intel\textregistered $ $ Xeon\textregistered $ $ CPU E5-2640 v4 2.40GHz processor. The computational time of each epoch in the SGD algorithm is approximately proportional to the number of active nodes in the bottom layer $N_a$. In our current study, the typical training times for $N=3$, 5 and 7 on one processor are 0.4, 2.1 and 9.1 hours, respectively. Several approaches can be potentially applied to further increase the training speed, such as code vectorization, hyper-parameter optimization, parallel programming and GPU computing.

\subsubsection{Training history of material network}
Performance of the training algorithm is evaluated by tracking the training and validation errors at each epoch. The (relative) error of a sample $s$ at the $t$-th epoch is defined as
\begin{figure}[!htb]
	\centering
	\subfigure[Uniform]{\includegraphics[clip=true,trim = 0.0cm 0.0cm 1.0cm 0.5cm,width=0.44\textwidth]{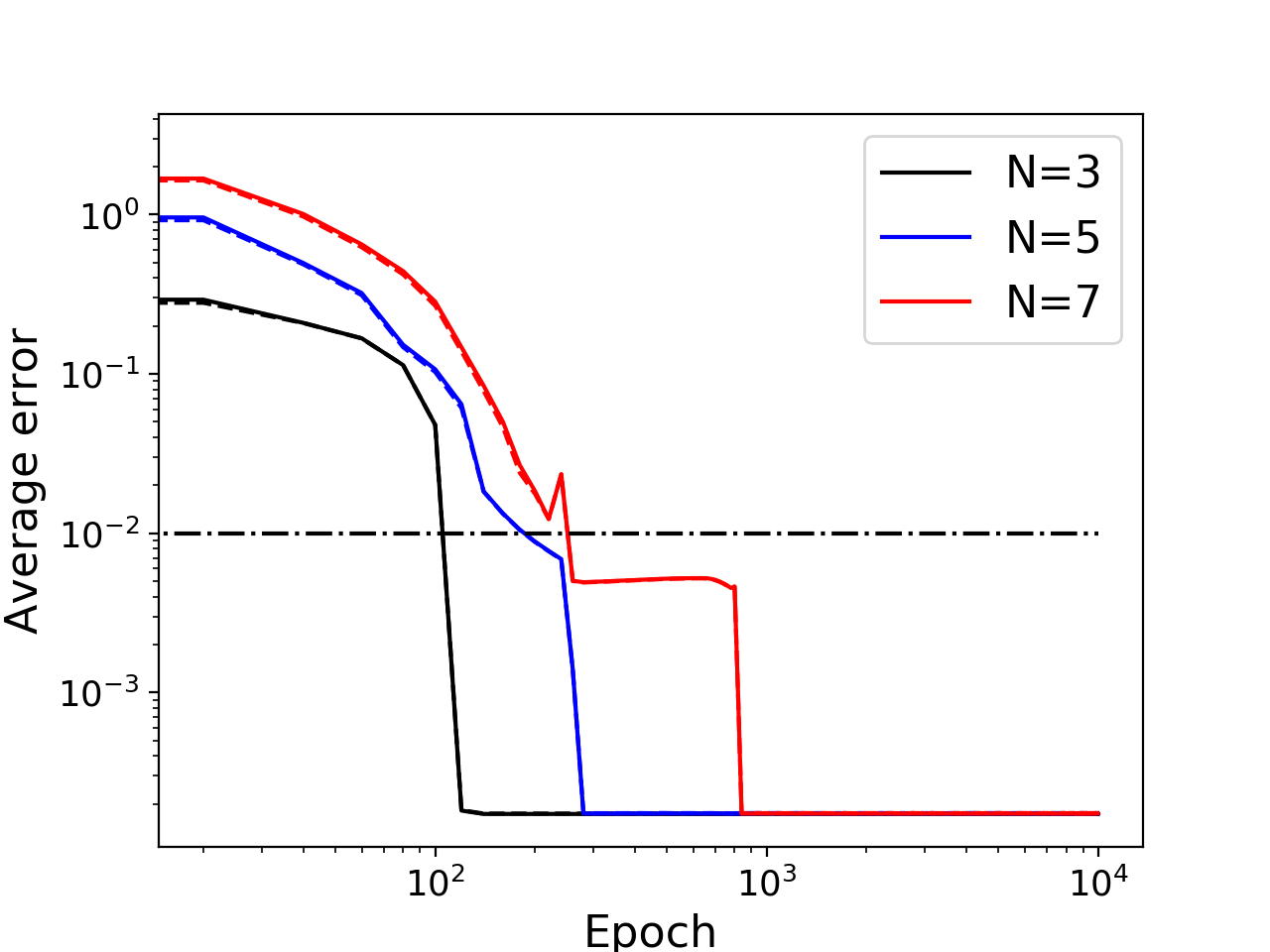}}
	\subfigure[Matrix-inclusion]{\includegraphics[clip=true,trim = 0.0cm 0.0cm 1.0cm 0.5cm,width=0.44\textwidth]{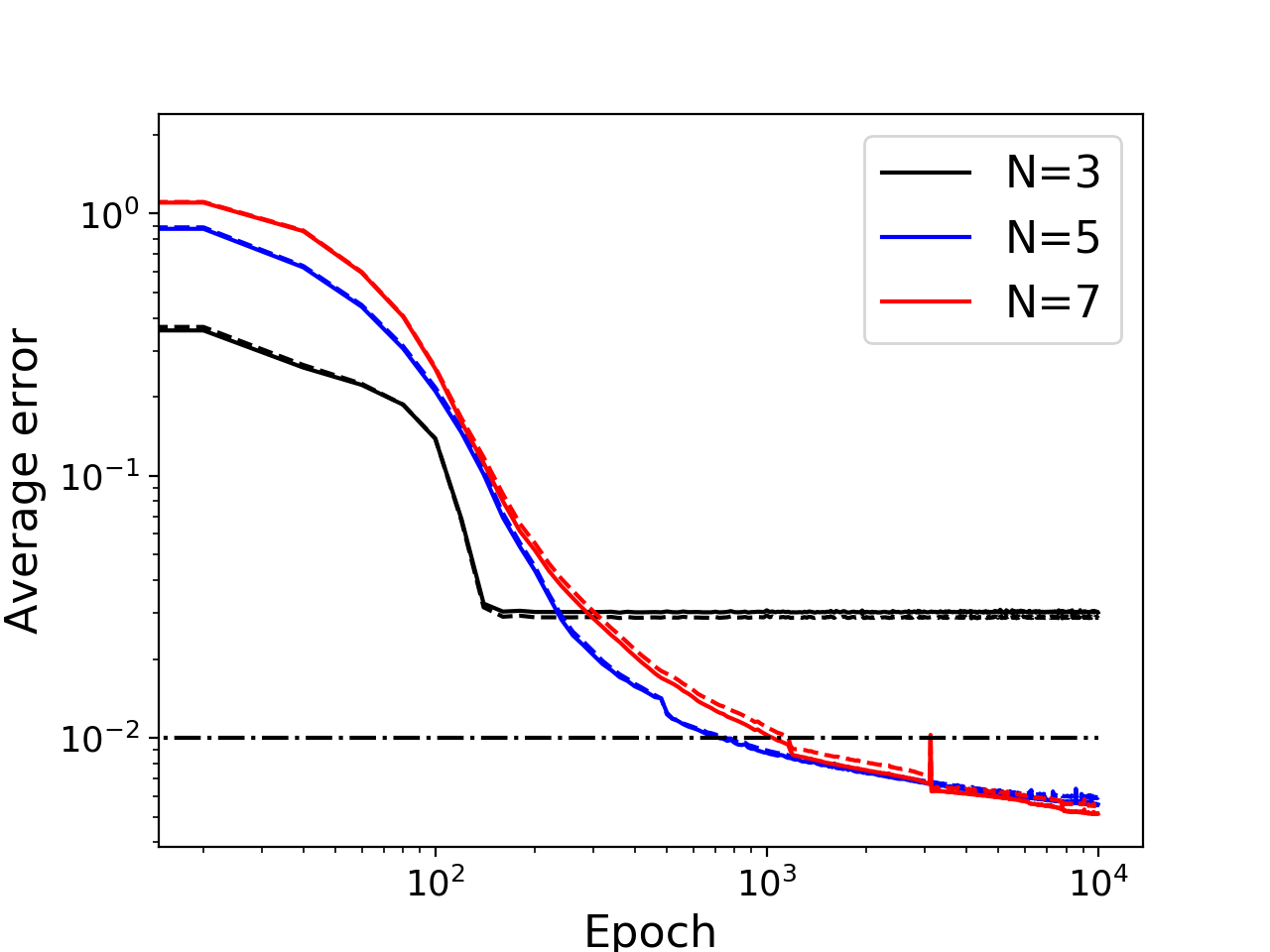}}
	\subfigure[Amorphous]{\includegraphics[clip=true,trim = 0.0cm 0.0cm 1.0cm 0.5cm,width=0.44\textwidth]{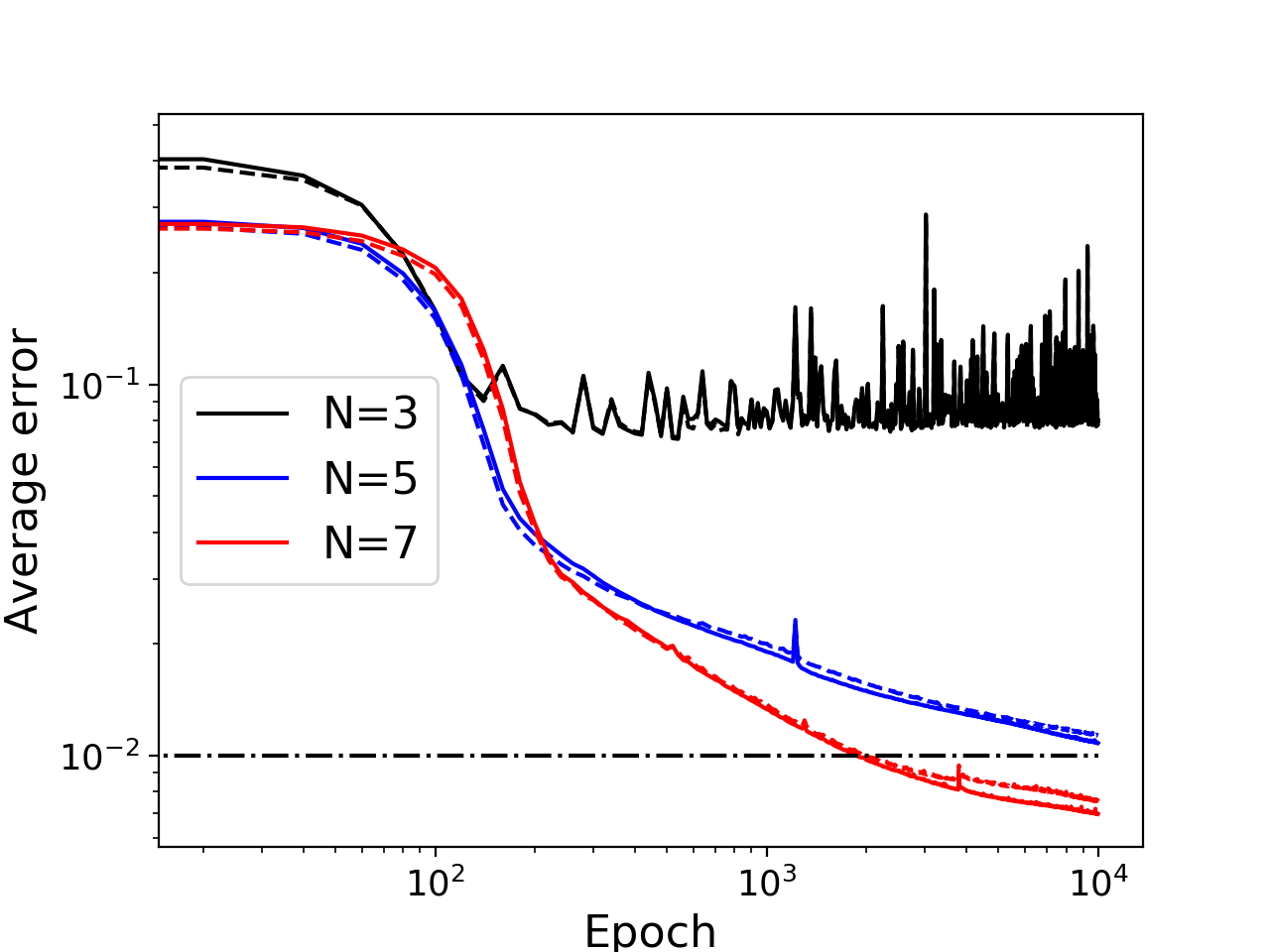}}
	\subfigure[Anisotropic]{\includegraphics[clip=true,trim = 0.0cm 0.0cm 1.0cm 0.5cm,width=0.44\textwidth]{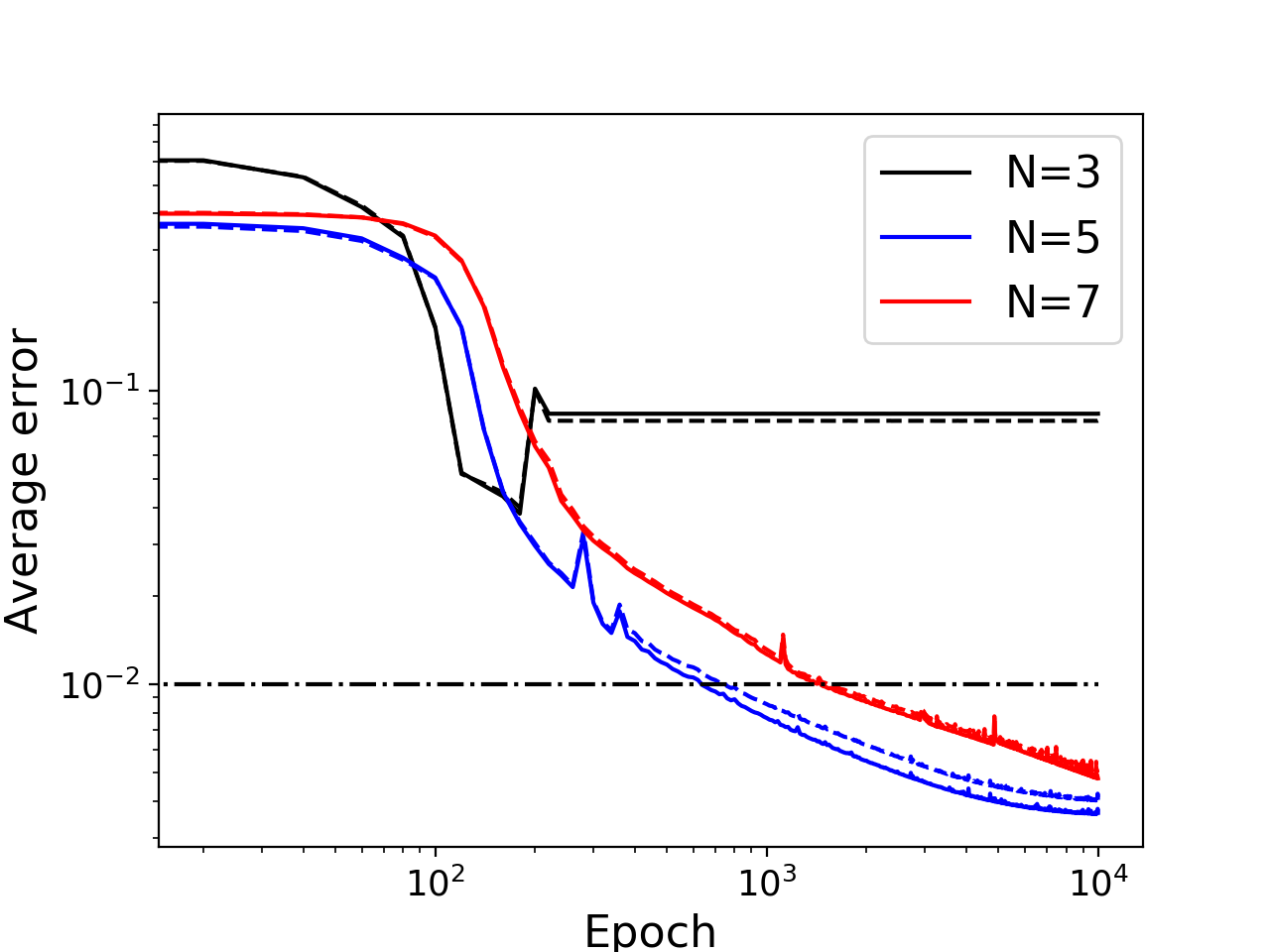}}
	\caption{Histories of average training errors (solid lines) and validation errors (dashed lines) for various RVEs. The dash-dot lines denote the average error equal to 1\%.}
	\label{fig:history}
\end{figure}
\begin{equation}\label{eq:relativeerror}
e_s^{(t)}=\dfrac{||\bar{\textbf{D}}^{dns}_s-\bar{\textbf{D}}^{mn(t)}_s||}{||\bar{\textbf{D}}^{dns}_s||},
\end{equation}
where $\bar{\textbf{D}}^{dns}_s$ and $\bar{\textbf{D}}^{mn(t)}_s$ are the compliance matrices predicted by DNS and the material network, respectively. The matrix norm $||...||$ has been defined in Eq. (\ref{eq:matrixnorm}). The average error of a dataset with $S$ samples at the t-th epoch is 
\begin{equation}
\bar{e}^{(t)}=\dfrac{1}{S}\sum_s e_s^{(t)}
\end{equation}
\begin{table}[!htb]
	\captionabove{Average errors of training datasets for various RVEs after 10000 epochs.} 
	\centering 
	\label{table:trlinear} 
	{\tabulinesep=1.5mm
		\begin{tabu}{c c c c c} 
			\hline 
			& uniform& matrix-inclusion & amorphous &  anisotropic \\
			\hline\hline
			$N=3$&0.02\%& 3.02\% & 7.86\%& 8.33\%\\ 
			\hline 
			$N=5$ &0.02\%& 0.55\% & 1.08\% & 0.37\% \\
			\hline
			$N=7$ &0.02\%& 0.51\% & 0.70\%& 0.48\% \\
			\hline
	\end{tabu}}
\end{table}
\begin{figure}[!htb]
	\centering
	\subfigure[Uniform]{\includegraphics[clip=true,trim = 0.0cm 0.0cm 1.0cm 0.5cm,width=0.44\textwidth]{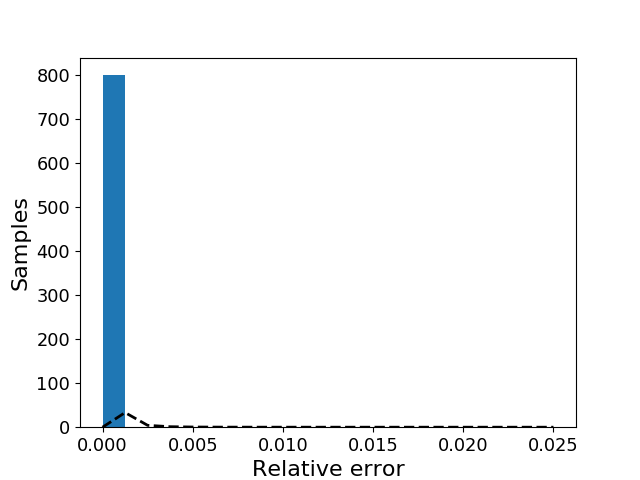}}
	\subfigure[Matrix-inclusion]{\includegraphics[clip=true,trim = 0.0cm 0.0cm 1.0cm 0.5cm,width=0.44\textwidth]{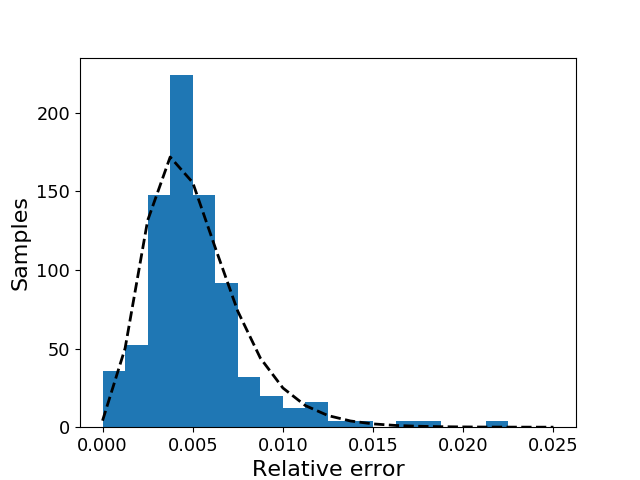}}
	\subfigure[Amorphous]{\includegraphics[clip=true,trim = 0.0cm 0.0cm 1.0cm 0.5cm,width=0.44\textwidth]{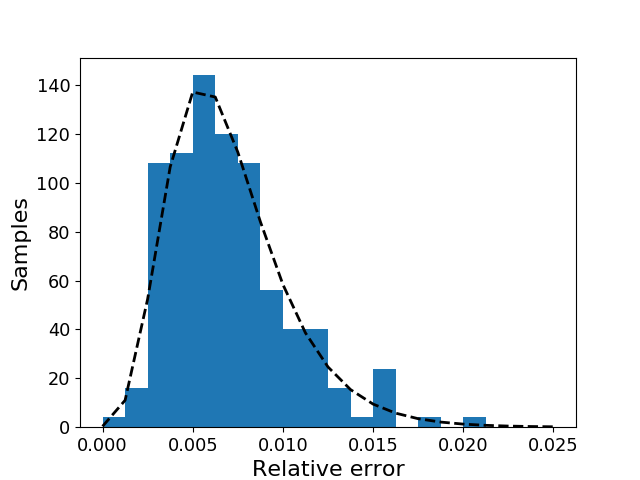}}
	\subfigure[Anisotropic]{\includegraphics[clip=true,trim = 0.0cm 0.0cm 1.0cm 0.5cm,width=0.44\textwidth]{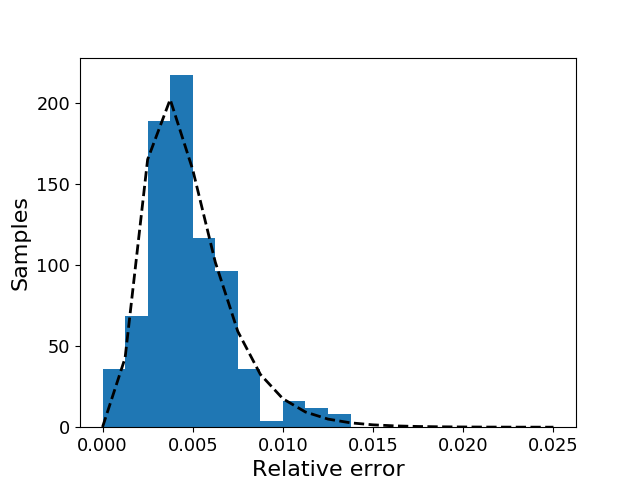}}
	\caption{Distributions of training errors for material networks with depth $N=7$ after 10000 epochs. The histograms are normalized.}
	\label{fig:trdistribution}
\end{figure}

Histories of average training and validation errors for material networks with different depths are provided in the Fig. \ref{fig:history}. The average errors of the training dataset for material networks with depth $N=$3, 5 and 7 are listed in Table \ref{table:trlinear}. It is observed that the average errors of all the RVEs were below 0.7\% after 10000 epochs of training when $N=7$. Actually, a good accuracy of 2\% was reached after 1000 epochs for $N=5$ and $7$ in all the cases.  Other than the uniform RVE, a network with $N=3$ is in general not deep enough to capture the RVE responses, herein, its average training error stopped decreasing at a relatively large value. Additionally, the difference between the training and validation errors was almost negligible, indicating that the hyper-parameters in the model are well tuned and no over-fitting has appeared.

\begin{remark}\label{rem:offline}
The training results depend on the initial parameters and the SGD algorithm converges to a band of critical points which are local minimal close to the "global optimum", as studied by Choromanska et al. \cite{choromanska2015loss} for multilayer neural networks. In addition, multiple realizations of the network with different sets of initial parameters can be trained separately, and the one with the minimum training or testing errors will be selected as the database for the RVE.
\end{remark}

Fig. \ref{fig:trdistribution} provides the distributions of the training error after 10000 epochs for network depth $N=7$. The figure shows that the maximum relative errors in the training datasets are below 2.5\% for all the RVEs. It can be concluded that a material network with sufficient depth ($N\geq5$) can represent the DNS model accurately, while with much less DOFs. Meanwhile, the proposed material network based on a simple two-layer building block can be effectively trained by SGD with backpropagation algorithms and network compression methods. 

\begin{table}[!htb]
	\captionabove{Number of active nodes in the bottom layer ($N_a$) after 10000 epochs of training.} 
	\centering 
	\label{table:activenodes} 
	{\tabulinesep=1.5mm
		\begin{tabu}{c c c c c | c} 
			\hline 
			& uniform& matrix-inclusion & amorphous &  anisotropic & \textit{initial $N_a$} \\
			\hline\hline
			$N=3$&1& 4 & 6& 2 & 8\\ 
			\hline 
			$N=5$ &1& 16 & 28 & 16 & 32\\
			\hline
			$N=7$ &1& 44 & 86& 62 & 128\\
			\hline
	\end{tabu}}
\end{table}
Numbers of active nodes in the bottom layer ($N_a$) of the trained network after 10000 epochs are listed in Table \ref{table:activenodes}. With the ReLU activation function and merging operation for network compression, $N_a$ gradually decreases during the training. The final $N_a$ of a trained network with a given depth $N$ depends on the RVE morphology. In our case, the amorphous RVE needs the most nodes (or DOFs) to minimize the training error. For uniform RVE, material networks with different depths all degenerated to the same structure with only one active node $N_a=1$. In fact, this structure is equivalent to the DNS model. Therefore, their relative errors are only about 0.02\%, which should come from the numerical error in the DNS calculation.

\subsubsection{Topologies of trained material network}
During the training, the topology of the network was kept varying to minimize the cost function. Here we use the so-called treemap to track and display the hierarchical structure of the material network using a set of nested rectangles. In a treemap, each rectangle represents an active node in the bottom layer of the material network, and its area is determined by the weighting function $w_N$ at the corresponding node.
\begin{figure}[!htb]
	\centering
	\includegraphics[clip=true,trim = 2cm 3cm 2.5cm 0.5cm, width = 0.9\textwidth]{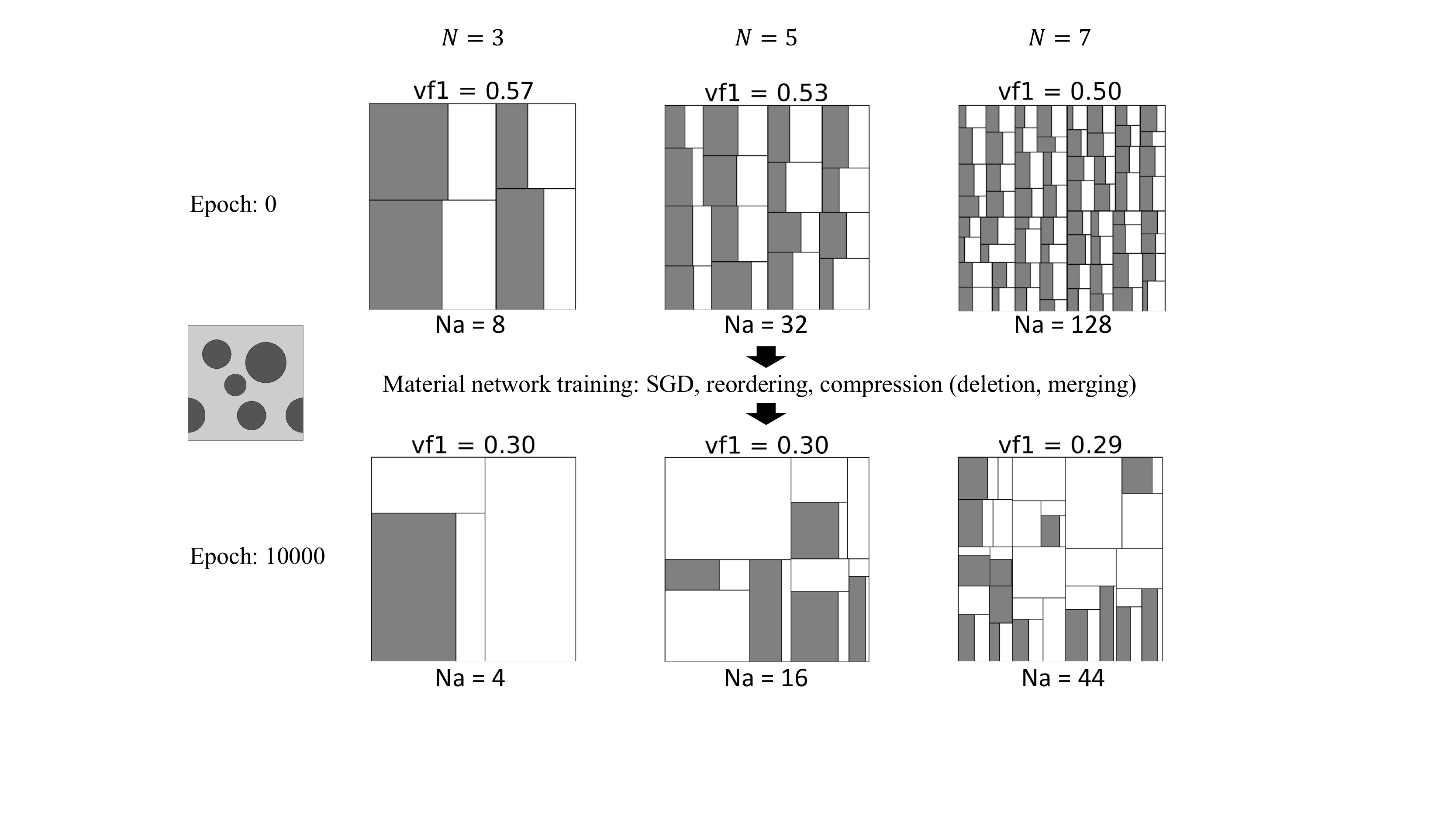}
	\caption{Treemaps of material network for the matrix-inclusion RVE at the beginning and the end of the training. Three network depths $N=3$, $5$, $7$ are considered. Volume fraction of phase 1 is provided at the top of each plot, while the number of active nodes at the bottom layer are shown at the bottom of each plot.}
	\label{fig:treemap-bf}
\end{figure}

\begin{figure}[!htb]
	\centering
	\includegraphics[clip=true,trim = 2cm 0cm 2.5cm 0cm, width = 0.9\textwidth]{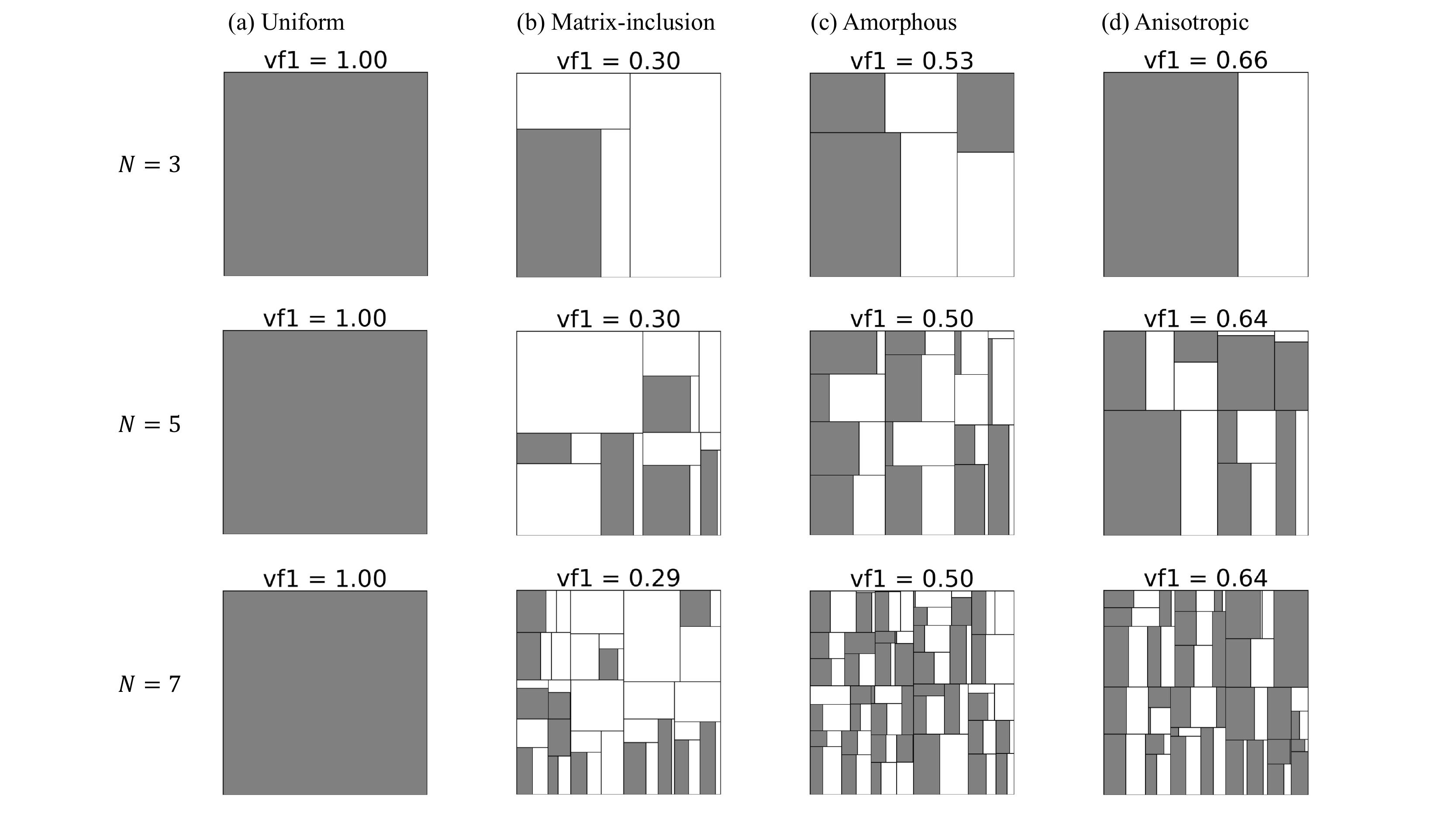}
	\caption{Treemaps of trained material network at the 10000-th epoch. Three network depths $N=3$, $5$, and $7$ are shown for each RVE. Volume fraction of phase 1 is provided at the top of each plot. }
	\label{fig:treemaps}
\end{figure}

Fig. \ref{fig:treemap-bf} presents the treemaps of material network for the matrix-inclusion RVE at the beginning and the end of the training, where the dark rectangles represent the phase 1 material. Since the weights of nodes at the bottom layer are assigned randomly for the initial network, the volume fraction of phase 1 is around 0.5 at the beginning of training. Note that the treemap only reflects the weight of each node, so that the rotation information at each node is not included in the plots.

Treemaps of trained material networks for all the RVEs after 10000 epochs are shown in Fig. \ref{fig:treemaps}. Three network depths $N=3$, $5$ and $7$ are considered. A material network with $N\geq5$ is capable of accurately learning the phase volume fraction $vf_1$ for all the RVEs. This indicates that the training dataset, which only contains the homogenized mechanical properties under different combinations of phases, can be used for extracting the geometric RVE information. It also suggests that the material network based on the two-layer building block can well represent the topological structure of the RVE. 

An intriguing feature of the material network  is that its topological structure is intrinsically parameterized. The rate of change or gradient of the overall material properties with respect to the geometric descriptors (e.g. volume fraction) and mechanical properties (e.g. modulus) of any node in the network can always be derived analytically. Thus, various gradient-based optimization methods could be easily employed to design the material both geometrically and mechanically. Potential applications to the area of material design will be investigated in our future work.

\subsection{Online extrapolation of trained network}
Based on the concept of material network and its machine learning algorithms, a reduced topological representation of the DNS model can be effectively mined from high-fidelity RVE training data. In this section, we will further investigate the material network's capability of online extrapolation to three different cases: 1) linear elasticity with high contrast of phase properties; 2) small-strain nonlinear plasticity; 3) finite-strain hyperelasticity under large deformations.

\subsubsection{Linear elasticity with high contrast of phase properties}\label{sec:elasticity}
The proposed material network is first extrapolated to linear elasticity problems with high contrast of phase properties. Comparing to the training dataset, wider ranges of material constants were used here. Both phases are assumed to be isotropic linear elastic. 

In this new testing dataset, ranges of elastic constants of phases 1 and 2 are
\begin{equation}
E^{p1}=1, \quad \nu^{p1} \in U[0.005, 0.495]
\end{equation}
and
\begin{equation}
\log_{10}(E^{p2})\in U[-3,3], \quad \nu^{p2} \in U[0.005, 0.495],
\end{equation}
where $U$ denotes the uniform distribution. The largest ratio of elastic moduli between the two phases is 1000. For each RVE, 100 testing samples are generated using Latin hypercube. The relative error of linear elastic compliance matrices is defined in Eq. (\ref{eq:relativeerror}).  
\begin{table}[ht]
	\captionabove{Average relative errors of the testing dataset on trained material networks.} 
	\centering 
	\label{table:exlinear} 
	{\tabulinesep=1.5mm
		\begin{tabu}{c c c c c} 
			\hline 
			& uniform& matrix-inclusion & amorphous &  anisotropic \\
			\hline\hline
			$N=3$&0.02\%& 2.32\% & 12.60\%& 9.93\%\\ 
			\hline 
			$N=5$ &0.02\%& 0.92\% & 2.34\% & 0.61\% \\
			\hline
			$N=7$ &0.02\%& 0.88\% & 1.70\%& 0.67\% \\
			\hline
	\end{tabu}}
\end{table}
\begin{figure}[!htb]
	\centering
	\subfigure[Uniform]{\includegraphics[clip=true,trim = 0.0cm 0.0cm 1.0cm 0.5cm,width=0.44\textwidth]{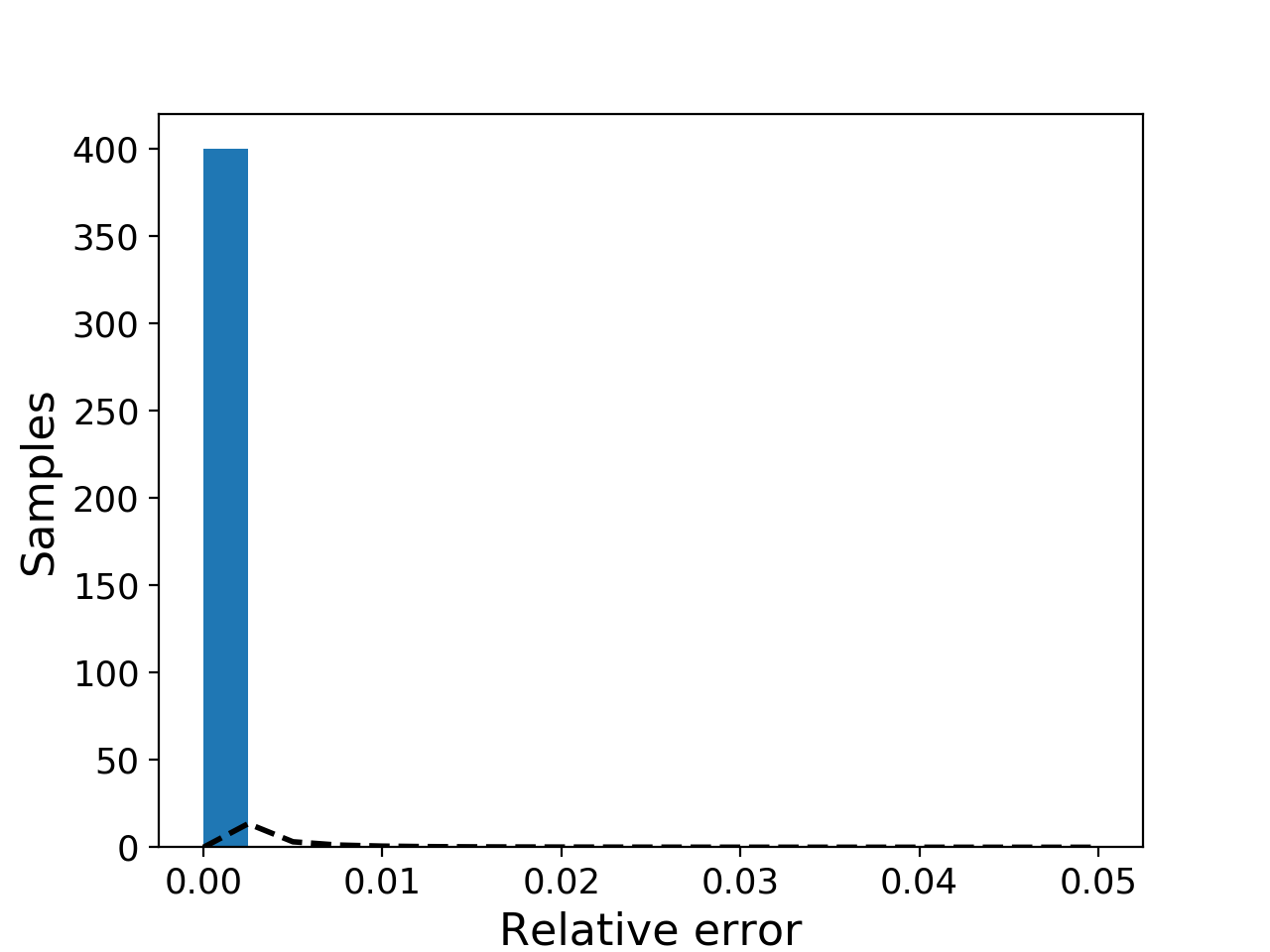}}
	\subfigure[Matrix-inclusion]{\includegraphics[clip=true,trim = 0.0cm 0.0cm 1.0cm 0.5cm,width=0.44\textwidth]{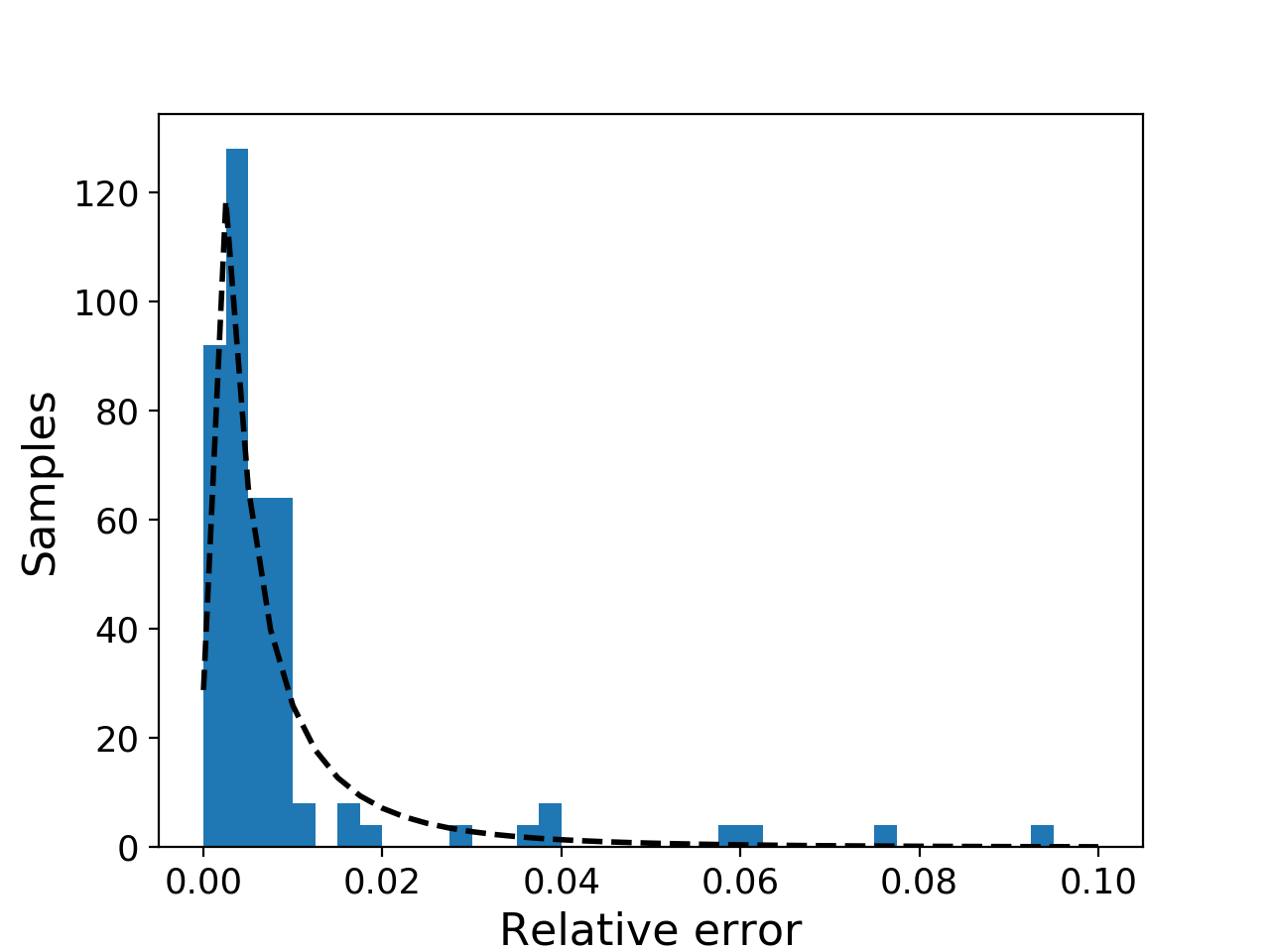}}
	\subfigure[Amorphous]{\includegraphics[clip=true,trim = 0.0cm 0.0cm 1.0cm 0.5cm,width=0.44\textwidth]{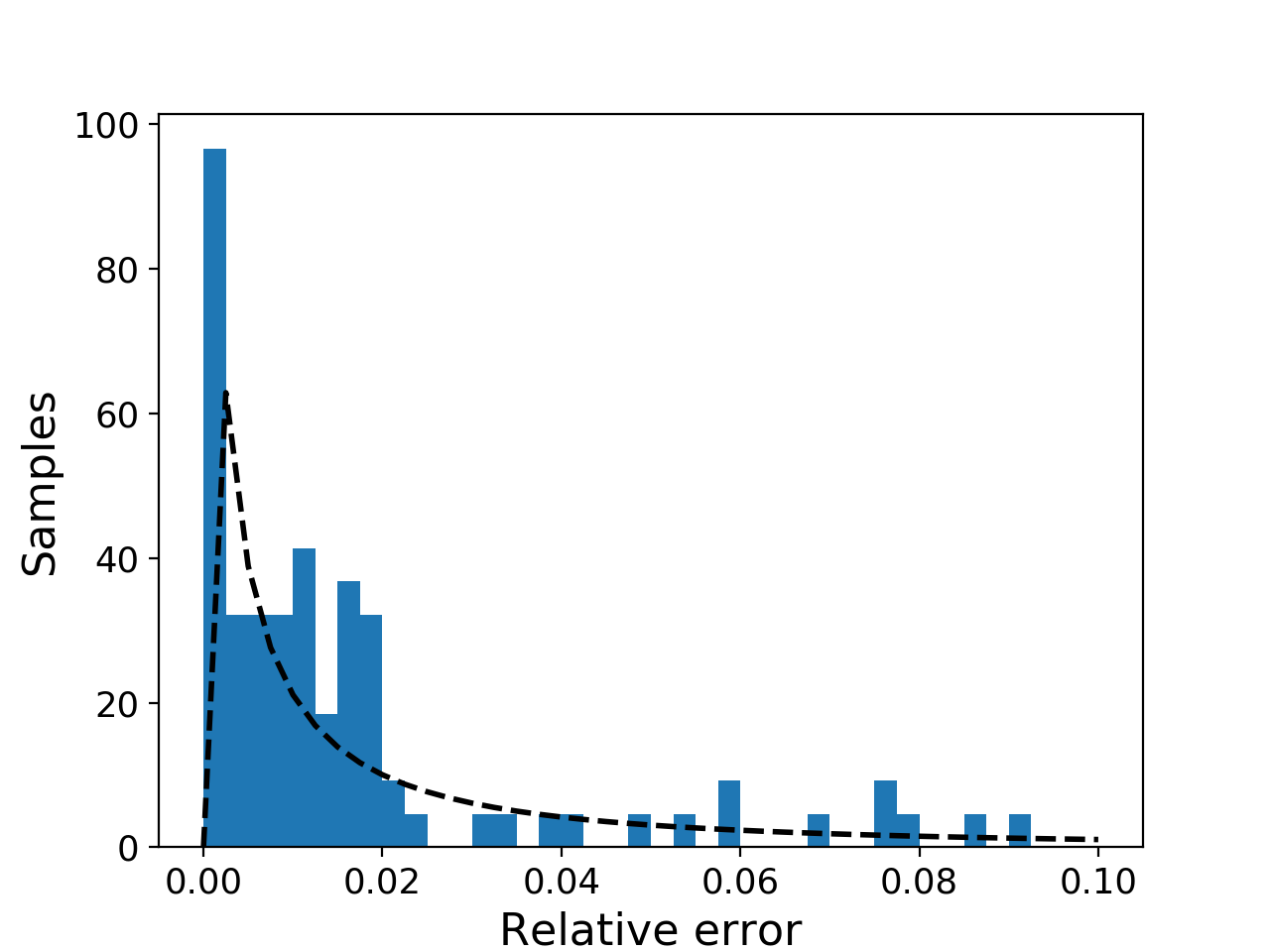}}
	\subfigure[Anisotropic]{\includegraphics[clip=true,trim = 0.0cm 0.0cm 1.0cm 0.5cm,width=0.44\textwidth]{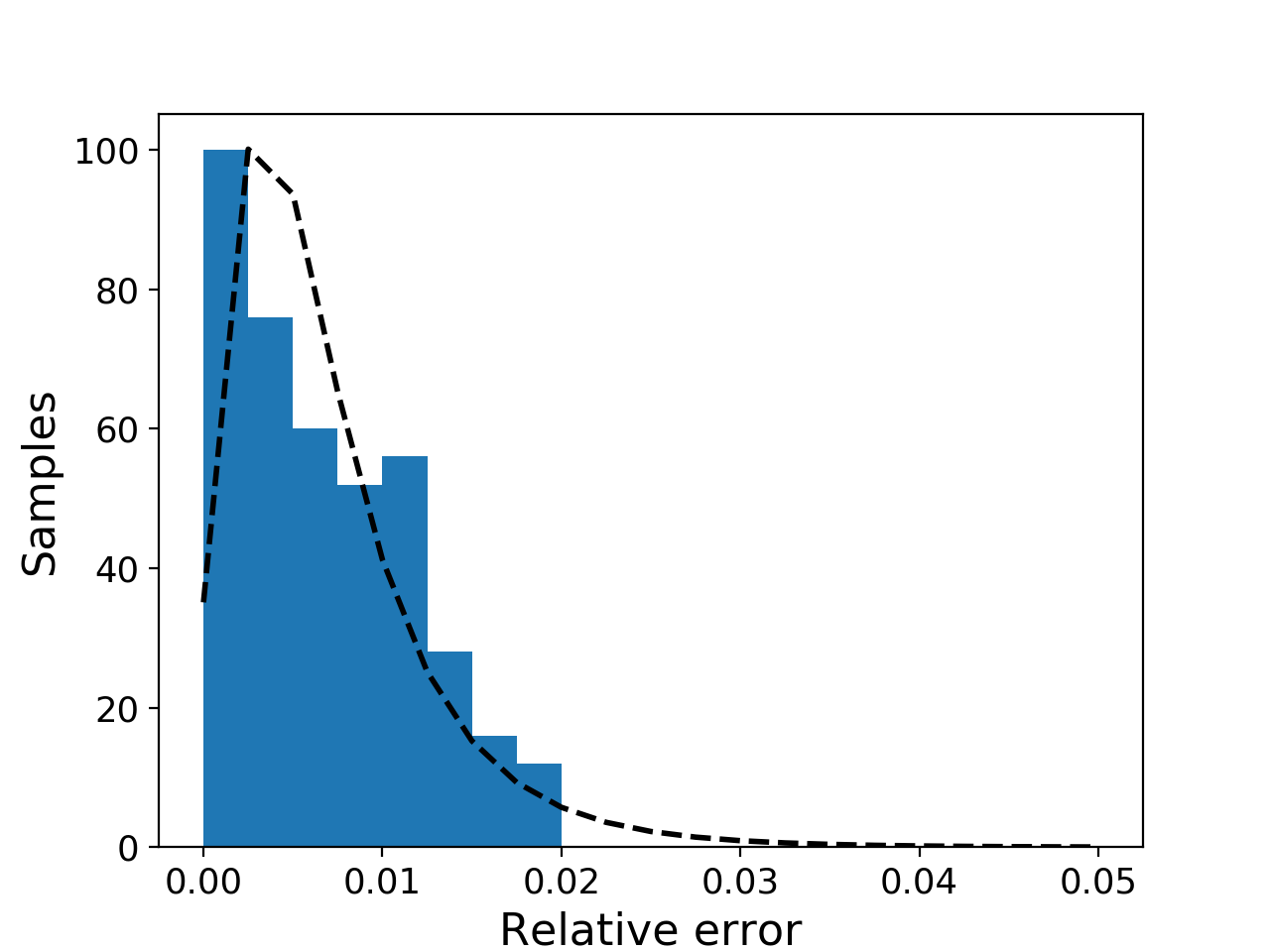}}
	\caption{Distributions of testing errors for trained networks with depth $N=7$. The histograms are normalized.}
	\label{fig:tedistribution}
\end{figure}

Average relative errors of the material networks with $N=3$, $N=5$ and $N=7$ are listed in Table \ref{table:exlinear} for all the RVEs. Distributions of the relative testing errors of material networks with $N=7$ are shown in Fig. \ref{fig:tedistribution}.  A log-normal density function is fitted to the histogram for each case. 

For the uniform RVE, the relative errors are negligible since all the networks have degenerated to the same one, that is equivalent to the DNS model. As expected, a material network with depth $N=3$ is not sufficient to maintain accurate predictions. For $N=7$, the average relative errors of the matrix-inclusion and anisotropic RVEs are below 1\%, while the one of the amorphous RVE reaches 1.7\%. As shown in Fig. \ref{fig:tedistribution}, the maximum relative error among the 100 testing samples of the anisotropic RVE is still below 2\%. However,  maximum errors of the matrix-inclusion and amorphous RVEs go up to 9.5\%, while the errors of their training and validation datasets are below 2.5\% as shown in  Fig. \ref{fig:trdistribution}. To reduce their prediction errors for models with high contrast of phase properties, it will be helpful to increase the sampling ranges of the training datasets.

\subsubsection{Small-strain nonlinear plasticity}\label{sec:plasticity}
Nonlinear elasto-plastic RVEs under small-strain assumption are studied in this section. The phase 2 material is considered to be elasto-plastic with an isotropic von Mises yield surface and piece-wise linear hardening law. Its elastic constants are
\begin{equation}
\quad E^{p2} = 100 \text{ GPa}, \nu^{p2}= 0.3
\end{equation}
\begin{figure}[!htb]
	\centering
	\subfigure[Matrix-inclusion, hard]{\includegraphics[clip=true,trim = 0.0cm 0.0cm 1.0cm 0.5cm,width=0.44\textwidth]{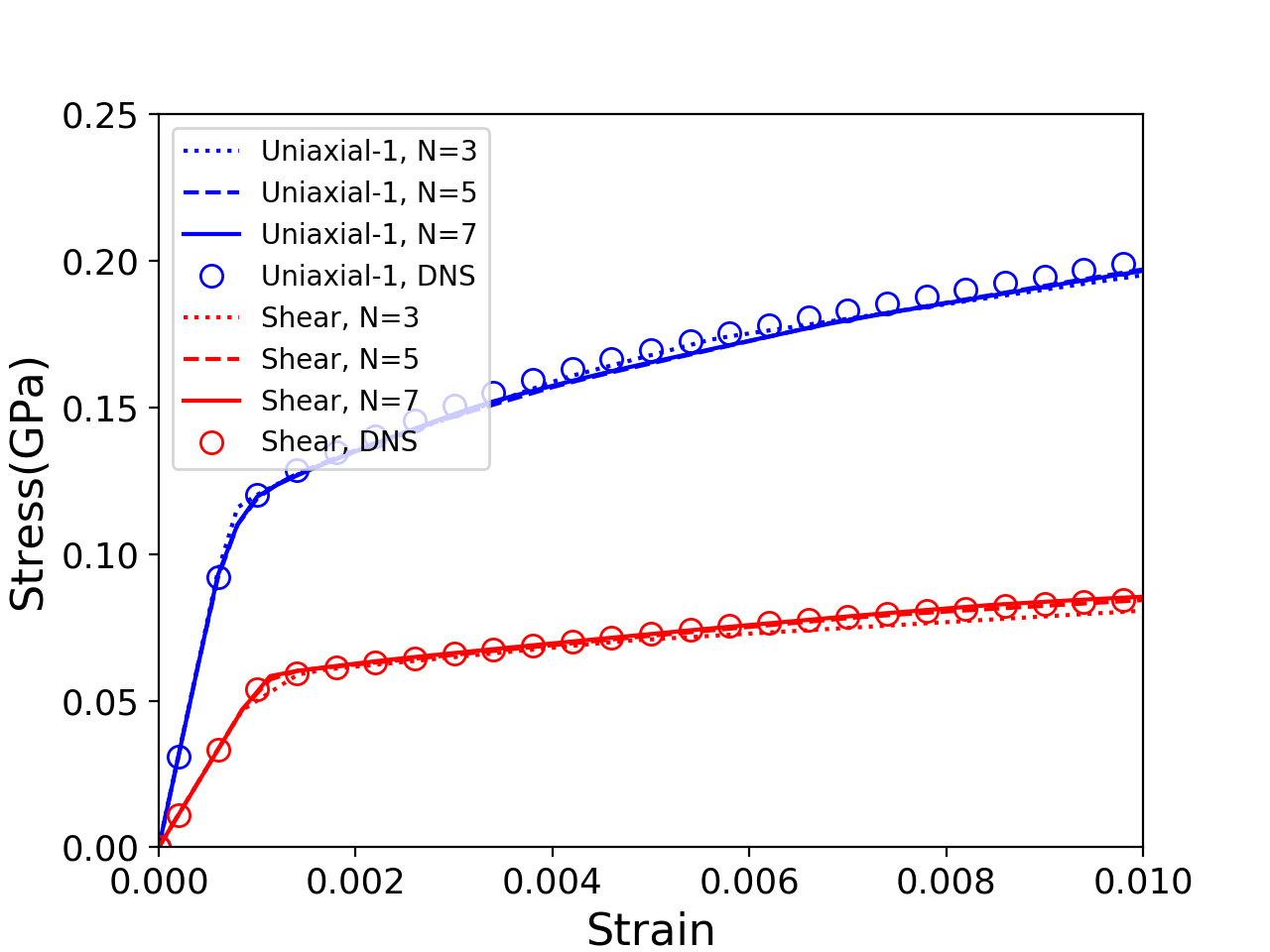}}
	\subfigure[Matrix-inclusion, soft]{\includegraphics[clip=true,trim = 0.0cm 0.0cm 1.0cm 0.5cm,width=0.44\textwidth]{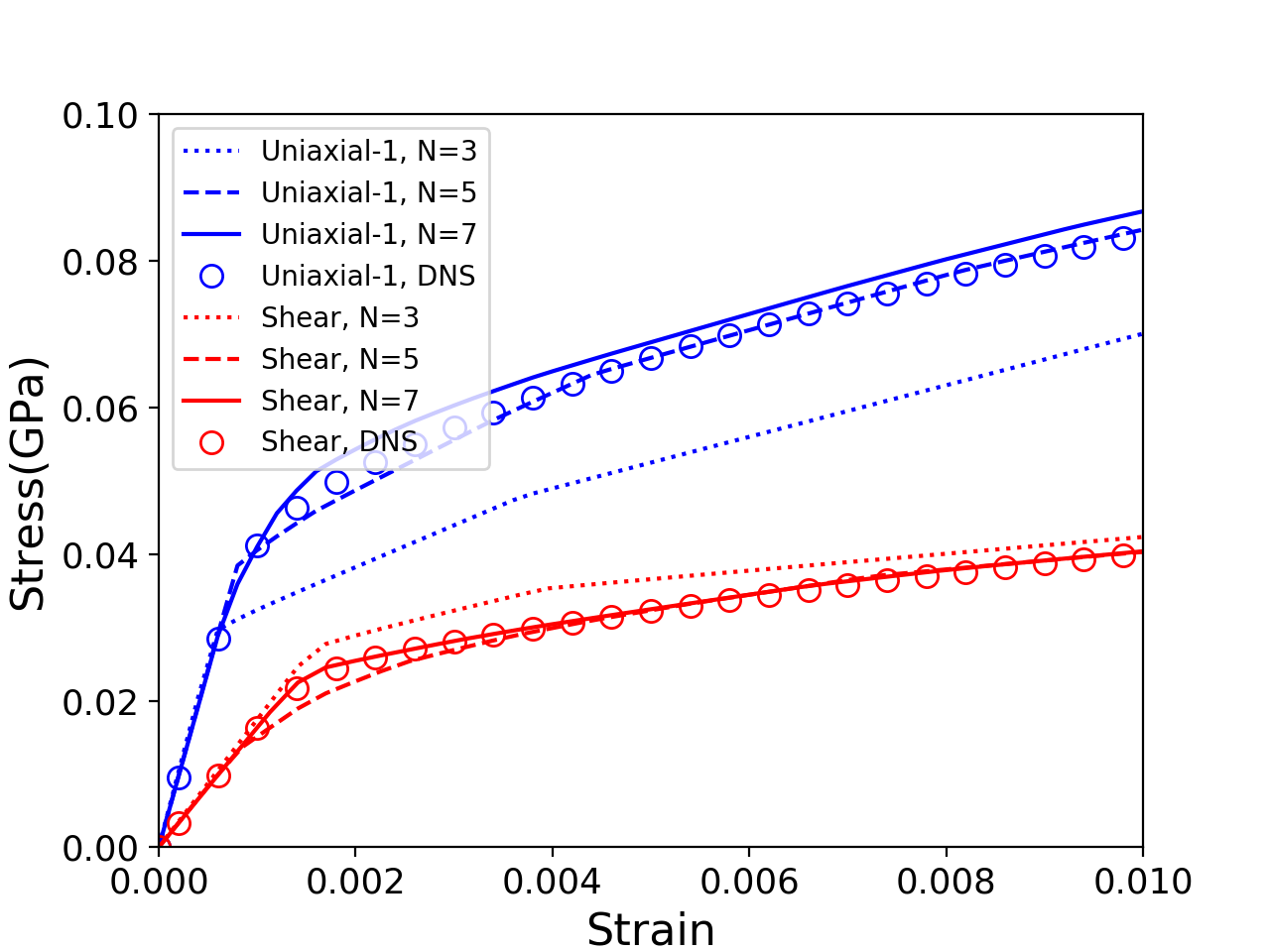}}
	\subfigure[Amorphous, hard]{\includegraphics[clip=true,trim = 0.0cm 0.0cm 1.0cm 0.5cm,width=0.44\textwidth]{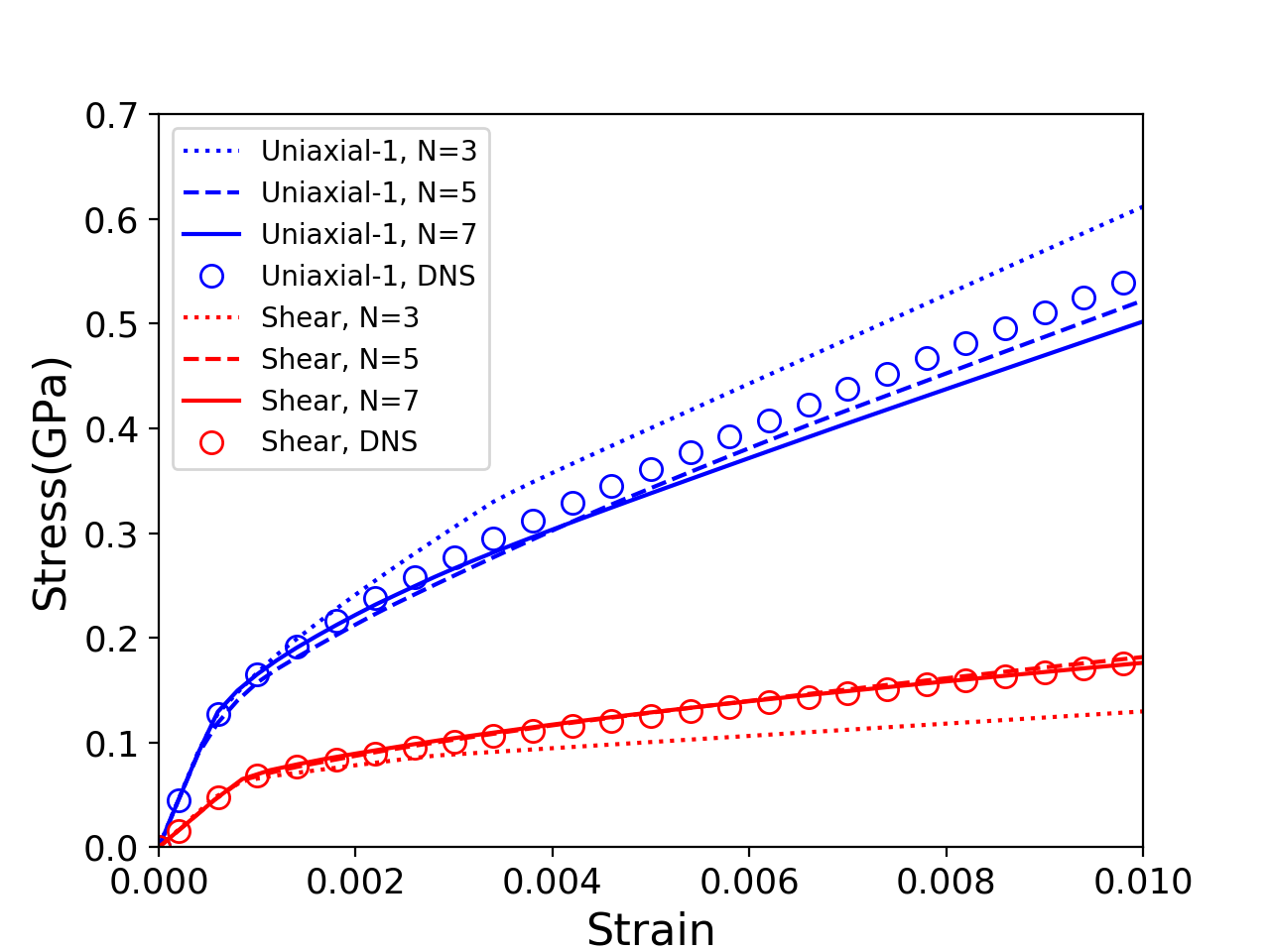}}
	\subfigure[Amorphous, soft]{\includegraphics[clip=true,trim = 0.0cm 0.0cm 1.0cm 0.5cm,width=0.44\textwidth]{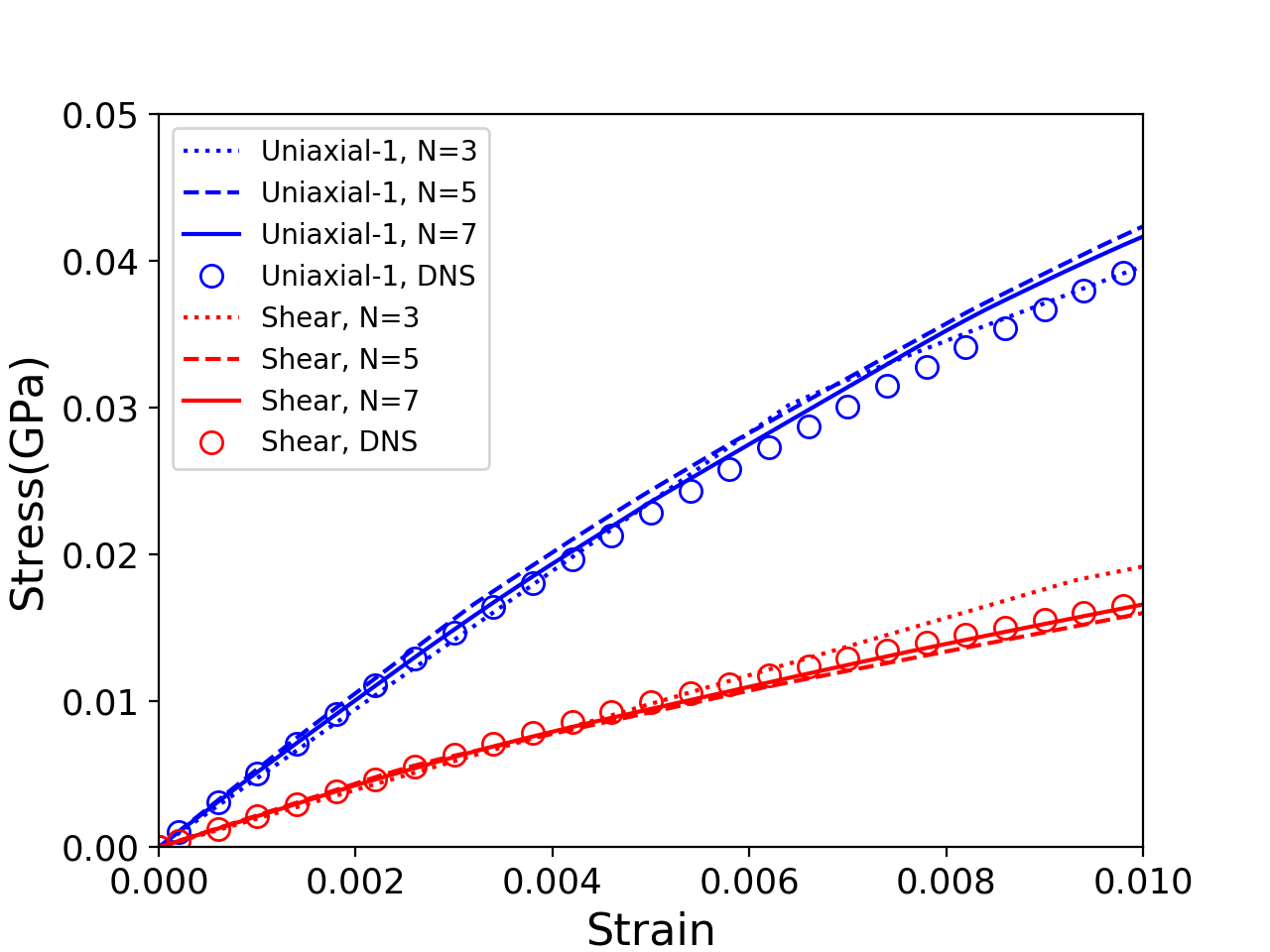}}
	\subfigure[Anisotropic, hard]{\includegraphics[clip=true,trim = 0.0cm 0.0cm 1.0cm 0.5cm,width=0.44\textwidth]{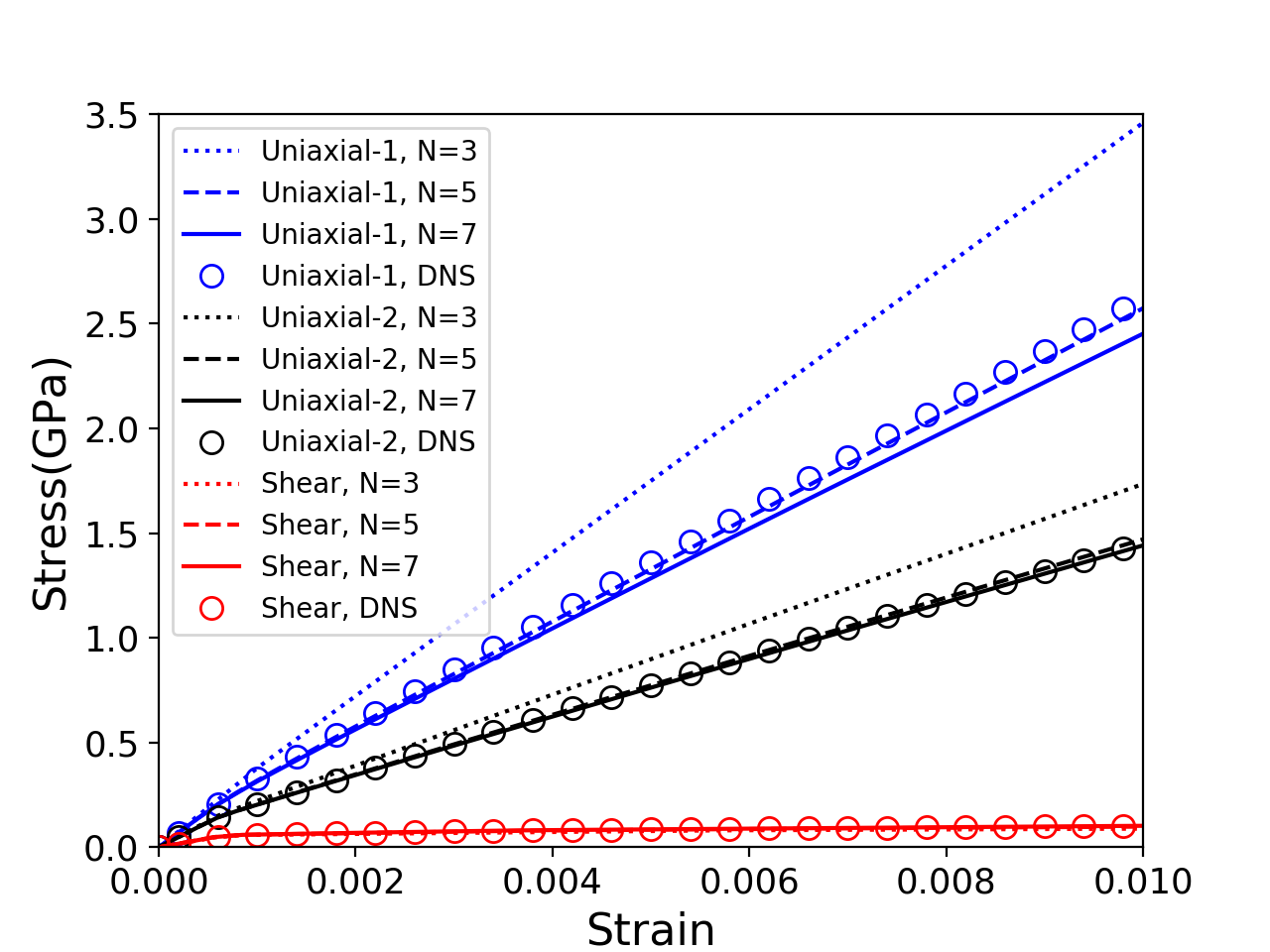}}
	\subfigure[Anisotropic, soft]{\includegraphics[clip=true,trim = 0.0cm 0.0cm 1.0cm 0.5cm,width=0.44\textwidth]{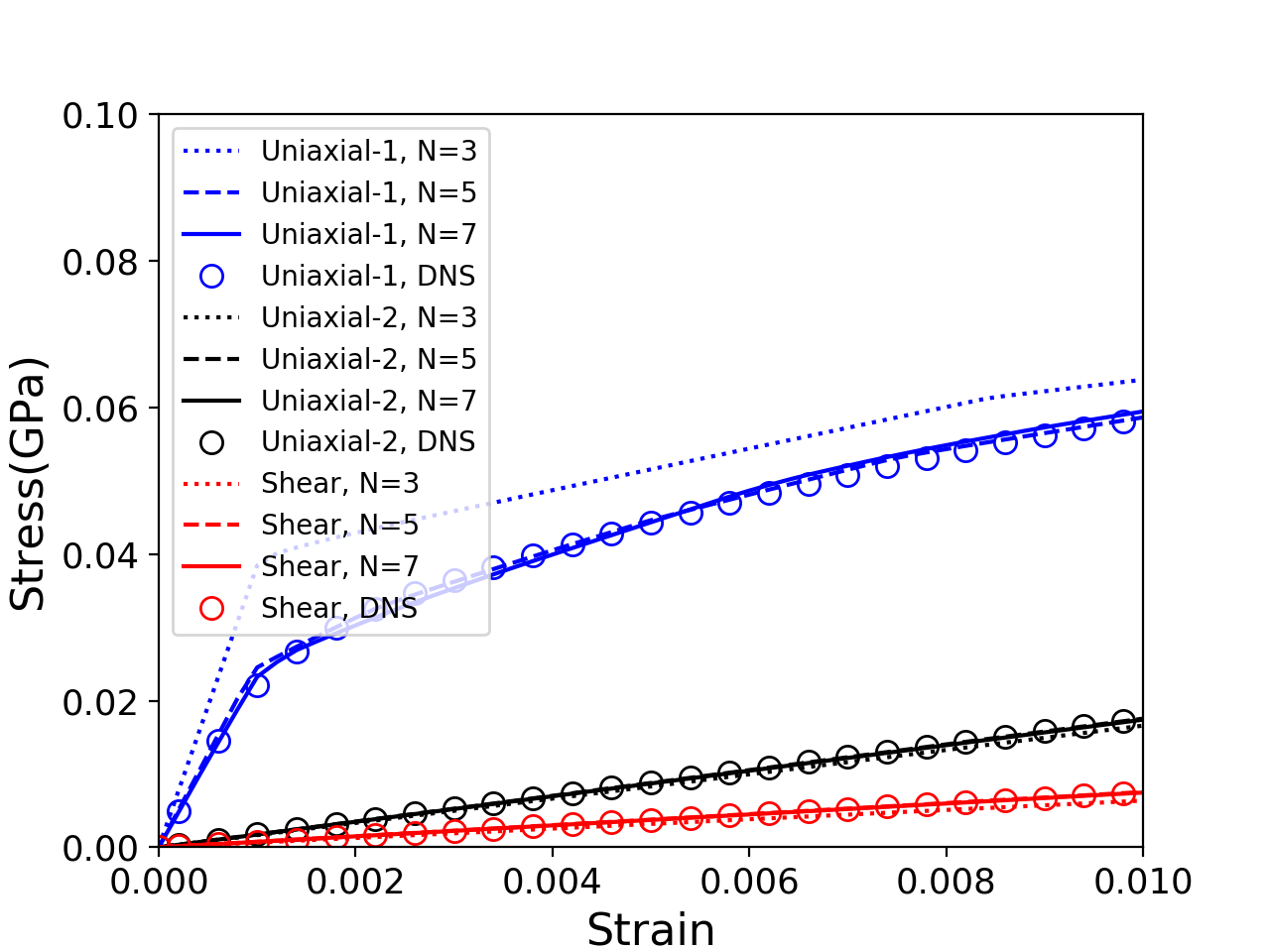}}
	\caption{Comparisons between material network and DNS for nonlinear small-strain plasticity for various RVEs under uniaxial tension and shear loading conditions. For the anisotropic RVE (e, f), uniaxial tension loadings are applied in two orthogonal directions. Both the hard (a, c, e) and soft (b, d, f) cases are considered. The network depths are $N=3$ (dotted), $5$ (dashed) and $7$ (solid). DNS results are marked by the circles ($\circ$). }
	\label{fig:plasticity}
\end{figure}  
The yield stress $\sigma_Y$ is determined by the hardening law as a function of the effective plastic strain $\bar{\varepsilon}^{pl}$, which is a monotonically increasing internal state variable of the plastic material during the deformation. The yielding stress $\sigma_Y^0$ is equal to 0.1 GPa. The hardening law is considered to be piecewise linear and isotropic,
\begin{equation}\label{eq:piecewise_hardening}
\sigma^Y_{p2}(\bar{\varepsilon}^{pl}_{p2})=
\begin{cases}
0.1 + 5\bar{\varepsilon}^{pl}_{p2} & \bar{\varepsilon}^{pl}_{p2} \in [0,0.008) \\
0.14 + 2\bar{\varepsilon}^{pl}_{p2} & \bar{\varepsilon}^{pl}_{p2} \in [0.008,\infty)
\end{cases}
\text{  GPa}.
\end{equation}
Phase 1 remains as a linear elastic material, but either hard or soft phases are considered by making the Young's modulus of phase 1 either harder or softer than phase 2. For these two cases, properties of phase 1 material are
\begin{equation}\label{eq:hard}
\quad E^{p1} = 500 \text{ GPa}, \nu^{p1} = 0.19 \text{ (hard)}\quad\text{and}\quad
\quad E^{p1} = 1 \text{ GPa}, \nu^{p1} = 0.19 \text{ (soft)}.
\end{equation}
\begin{figure}[!htb]
	\centering
	\subfigure[Illustration of loading-unloading path of $\varepsilon_{11}$]{\includegraphics[clip=true,trim = 9.5cm 4.5cm 9.5cm 4.0cm,width=0.44\textwidth]{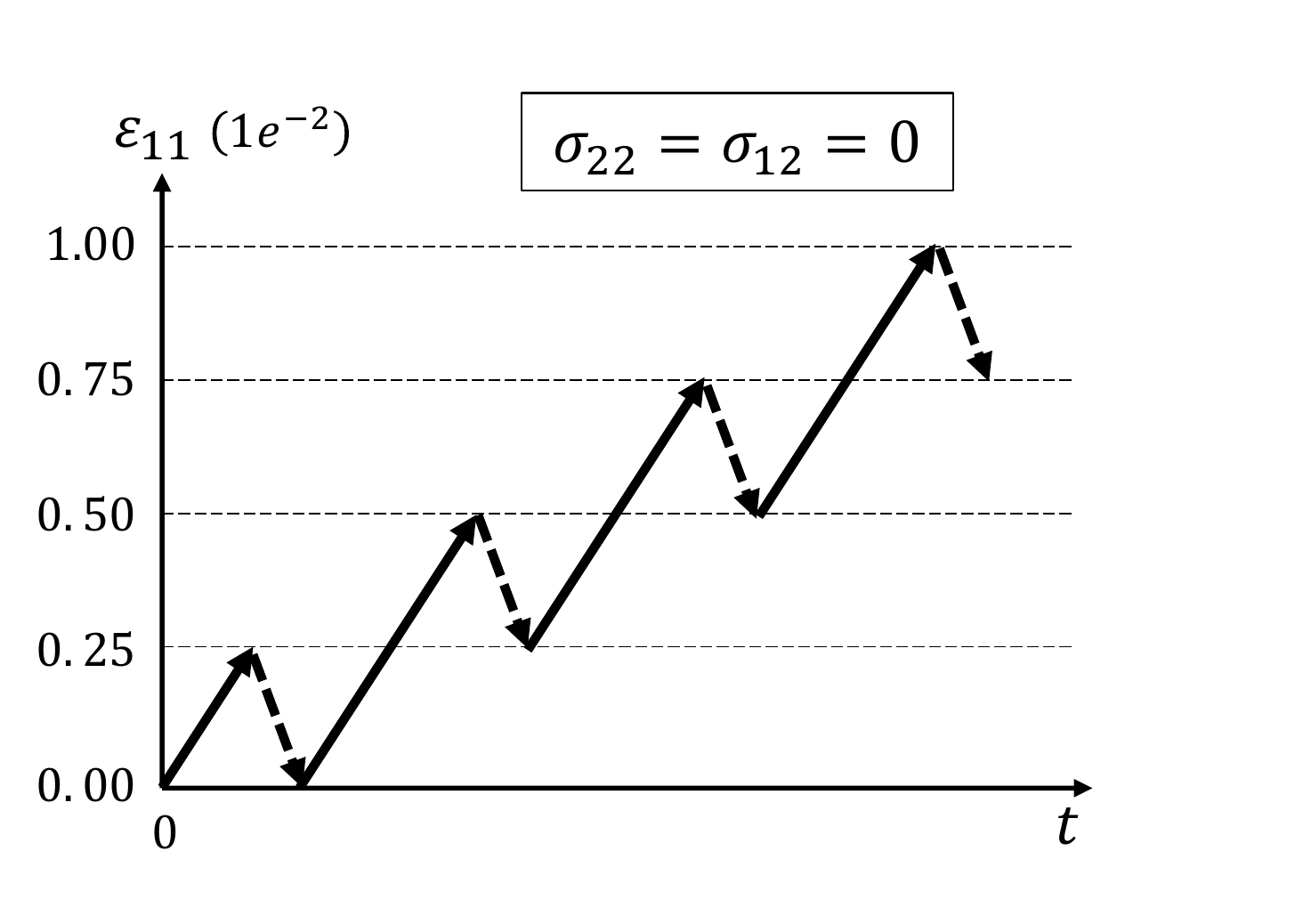}}
	\subfigure[Matrix-inclusion, hard]{\includegraphics[clip=true,trim = 0.0cm 0.0cm 1.0cm 0.5cm,width=0.44\textwidth]{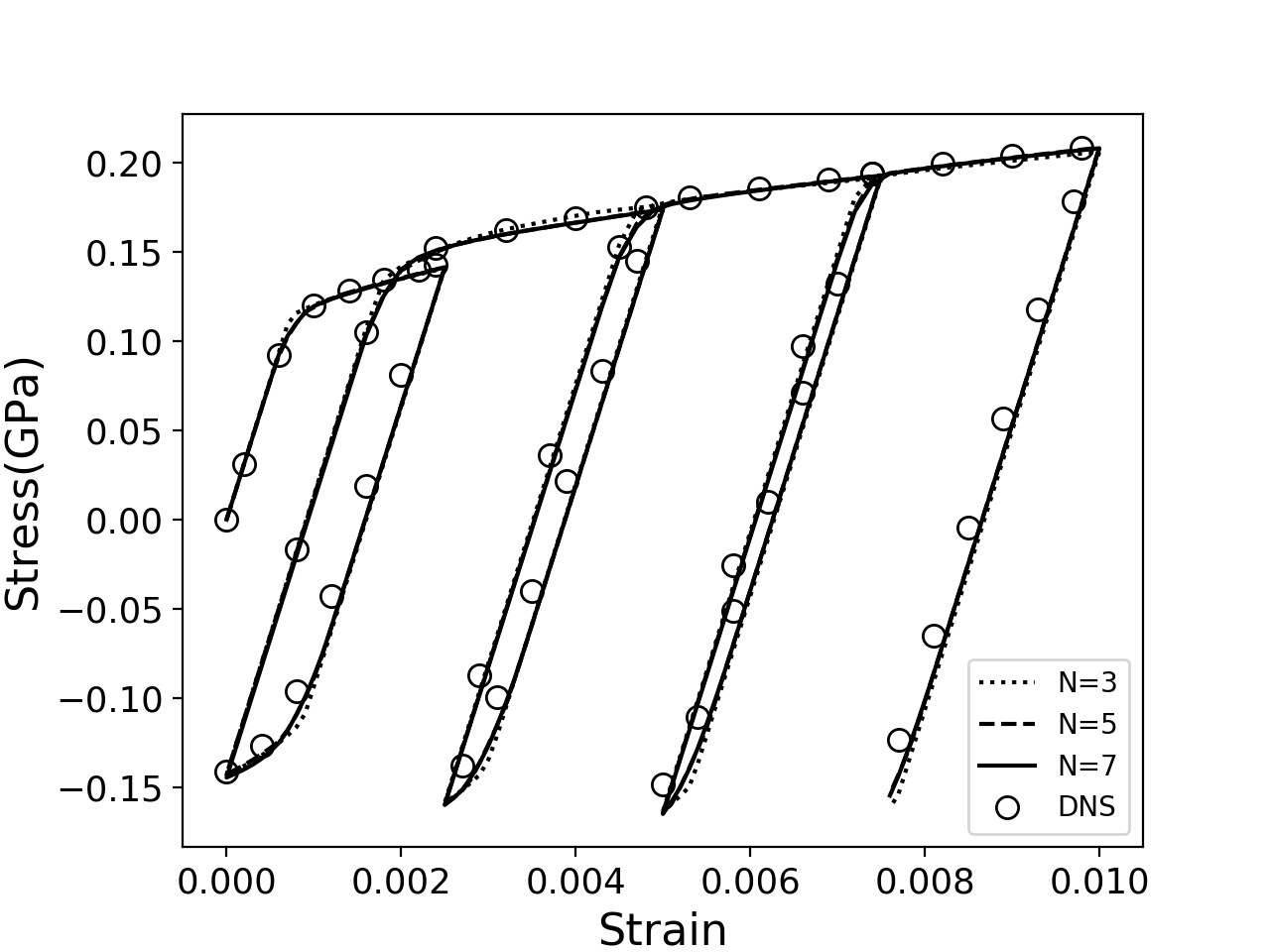}}
	\subfigure[Amorphous, hard]{\includegraphics[clip=true,trim = 0.0cm 0.0cm 1.0cm 0.5cm,width=0.44\textwidth]{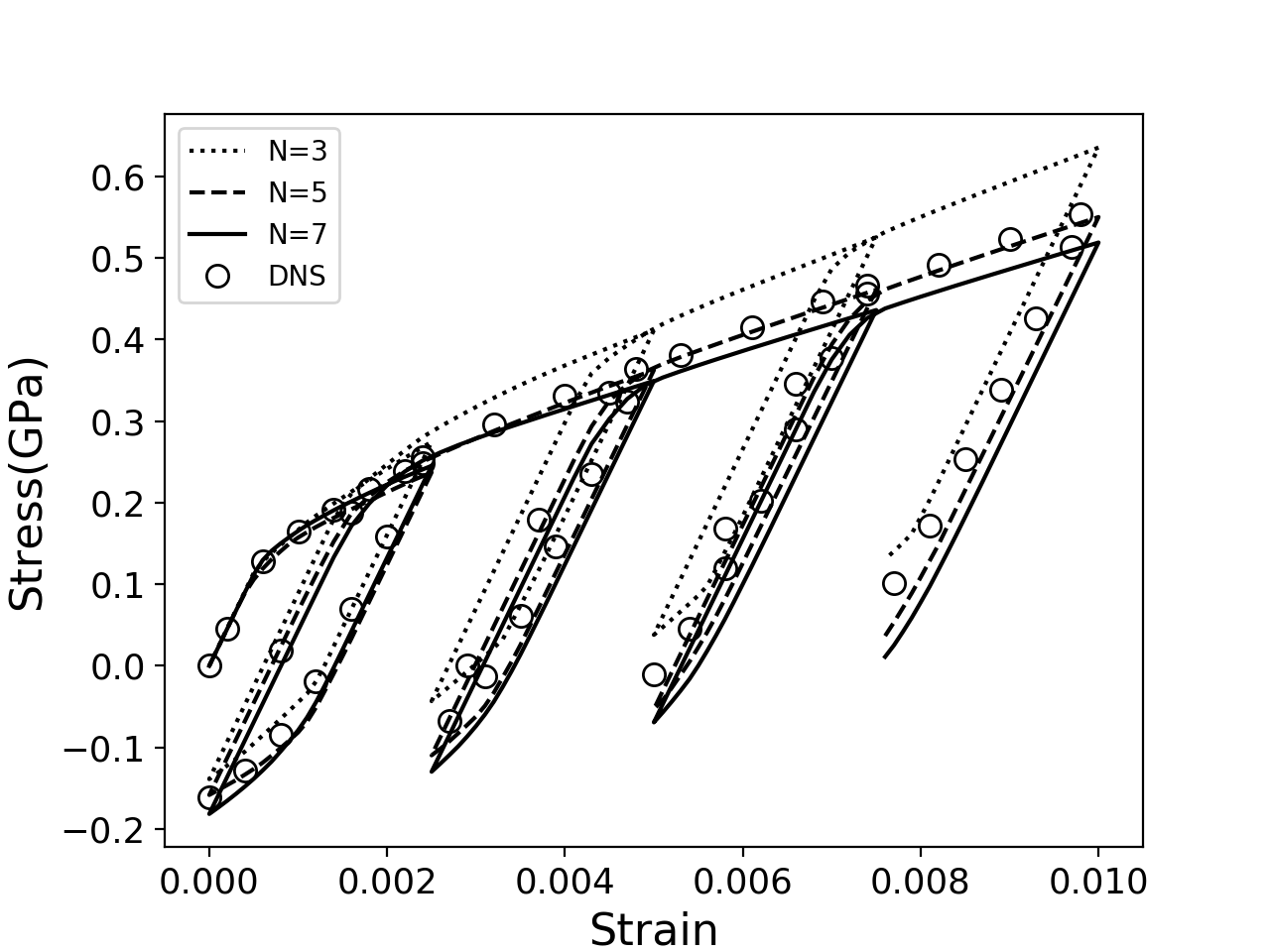}}
	\subfigure[Anisotropic, soft]{\includegraphics[clip=true,trim = 0.0cm 0.0cm 1.0cm 0.5cm,width=0.44\textwidth]{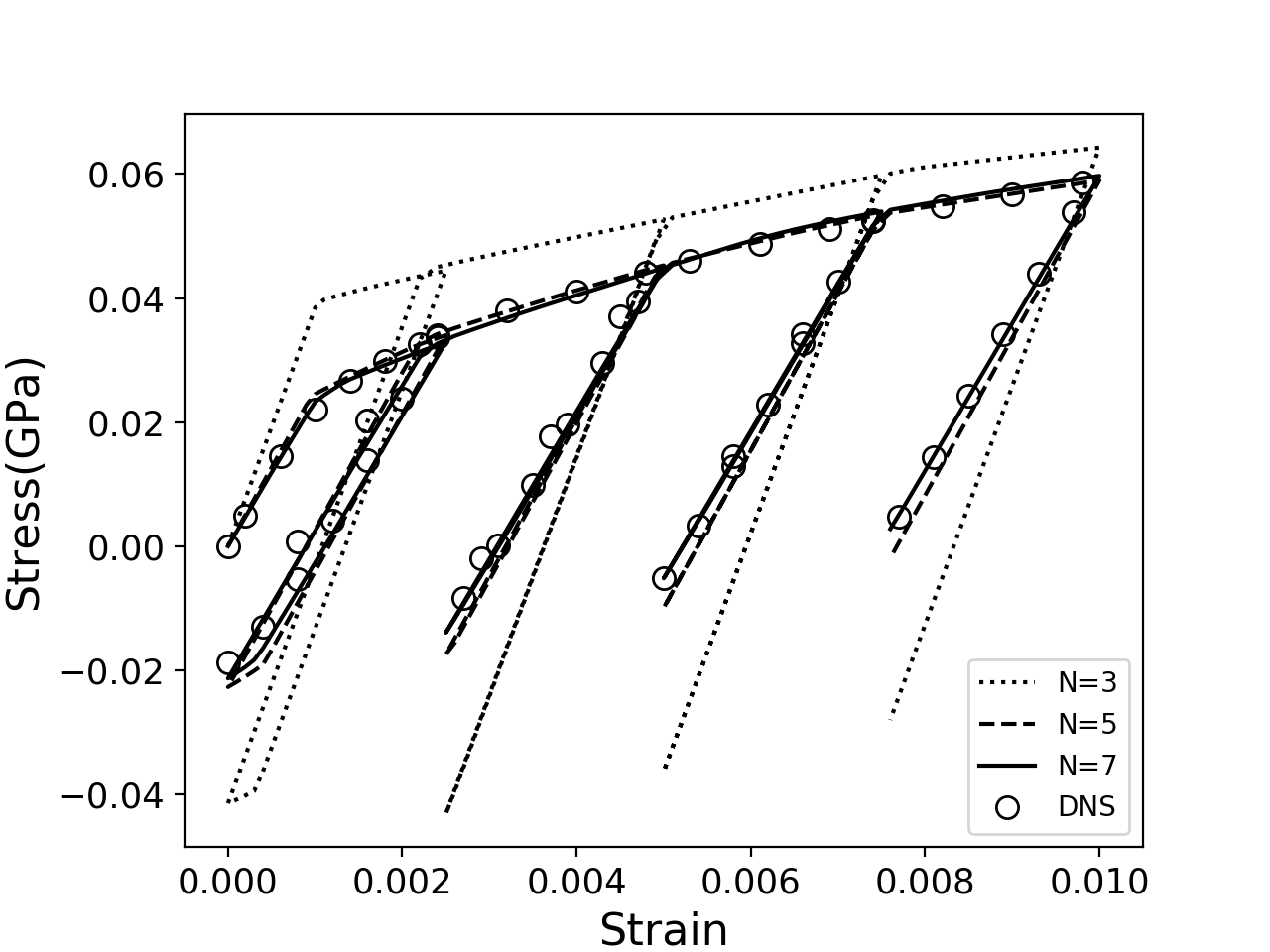}}
	\caption{Comparisons between material network and DNS under uniaxial loading-unloading path shown in (a). RVEs with apparent hardening regime are considered: matrix-inclusion and amorphous RVEs with hard phase 1, anisotropic RVE with soft phase 1. for The network depths are $N=3$ (dotted), $5$ (dashed) and $7$ (solid). DNS results are marked by the circles ($\circ$). }
	\label{fig:plasticity-unloading}
\end{figure} 

Fig. \ref{fig:plasticity} includes the stress-strain curves considering two different loading cases, uniaxial tension and pure shear, predicted by DNS and the material network. In the case of anisotropic RVE in Fig. \ref{fig:plasticity} (e) and (f), the uniaxial tension tests are performed in two orthogonal directions. By looking at the plots, we can conclude that the proposed method is capable of capturing the nonlinear plastic behavior for both hard and soft cases under different loading conditions, meanwhile, with significantly fewer degrees of freedom.  Note that it can also accurately capture anisotropic RVE behaviors, which can be very challenging for most micromechanics-based methods \cite{mura1987micromechanics,liu2016self}. Furthermore, the network should be deep enough to capture the nonlinear RVE behavior accurately, and in our case, $N=5$ is a good depth to start from.

Fig. \ref{fig:plasticity-unloading} (a) shows a uniaxial loading-unloading path which contains 4 loading and 4 unloading steps at different stress/strain levels. Three RVEs with apparent hardening regime are considered: matrix-inclusion RVE with hard phase 1, amorphous RVE with hard phase 1 and anisotropic RVE with soft phase 1. The stress-strain curves are provided in Fig. \ref{fig:plasticity-unloading} (b,c,d). For all the cases, the loading-unloading behavior can be well captured by the material networks for $N\geq5$.

\begin{figure}[!htb]
	\centering
	\subfigure[Illustration of complex loading path ($\varepsilon_{11}$, $\varepsilon_{12}$)]{\includegraphics[clip=true,trim = 9.5cm 4.5cm 9.5cm 4.0cm,width=0.44\textwidth]{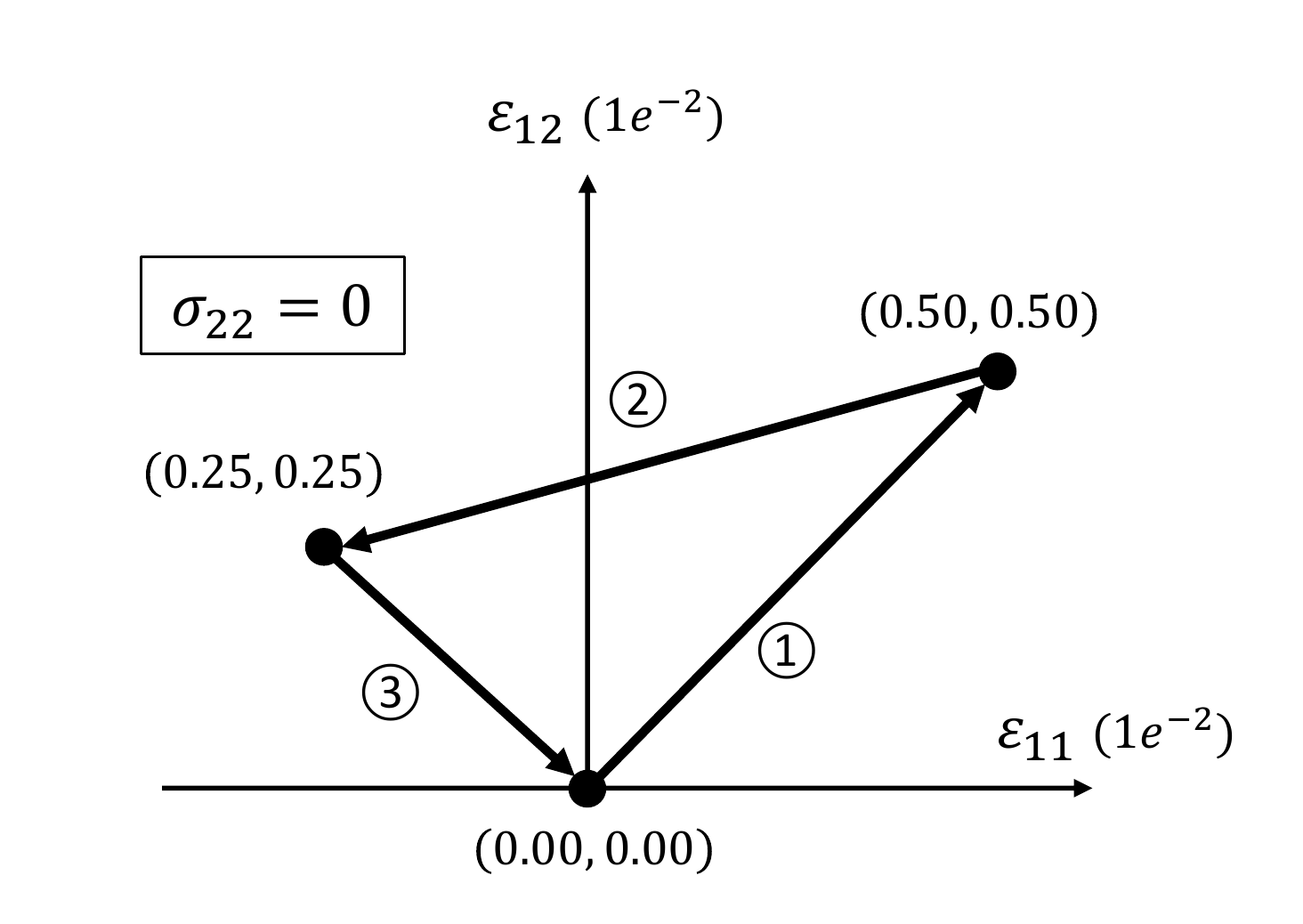}}
	\subfigure[Matrix-inclusion, hard]{\includegraphics[clip=true,trim = 0.0cm 0.0cm 1.0cm 0.5cm,width=0.44\textwidth]{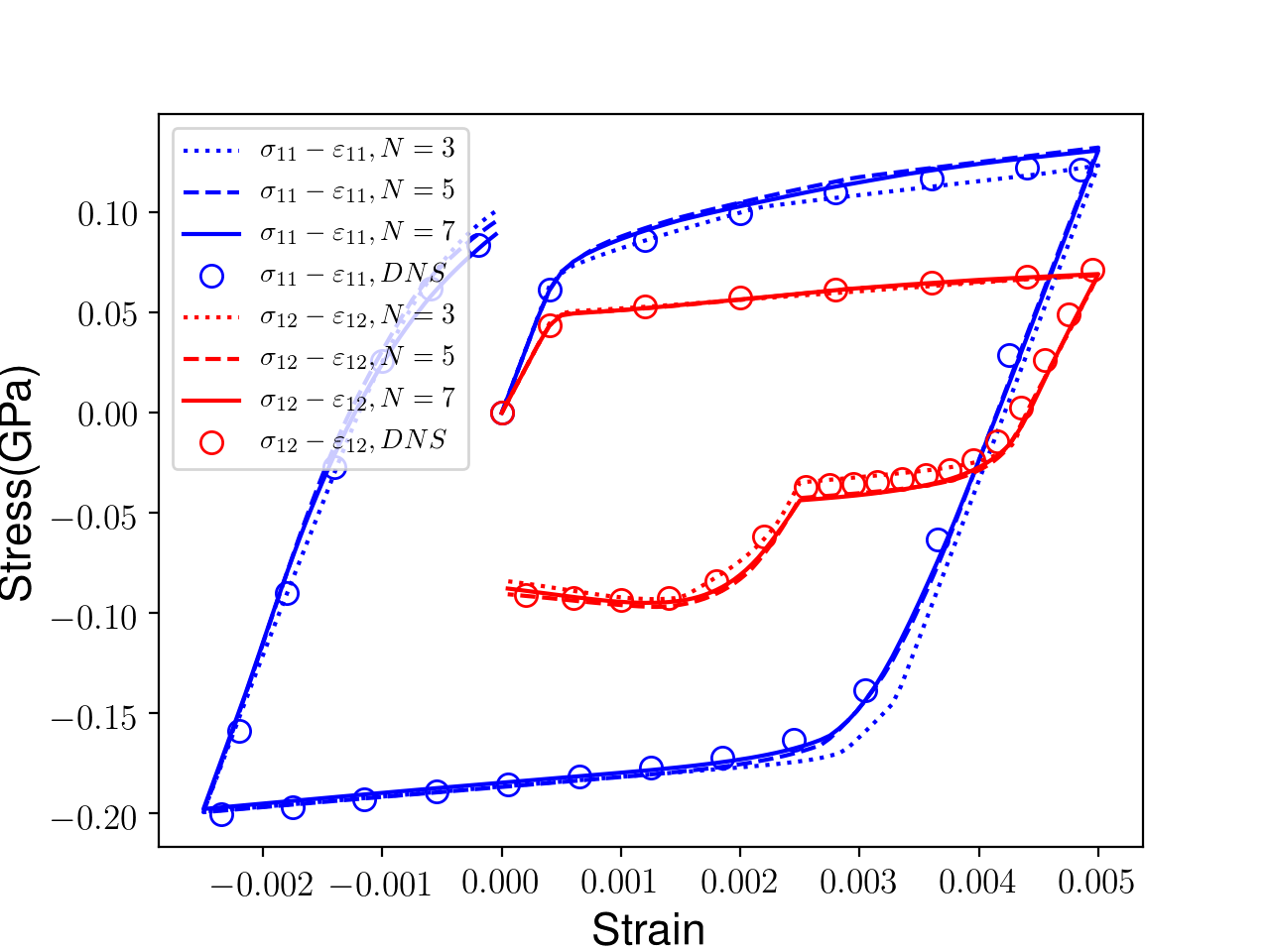}}
	\subfigure[Amorphous, hard]{\includegraphics[clip=true,trim = 0.0cm 0.0cm 1.0cm 0.5cm,width=0.44\textwidth]{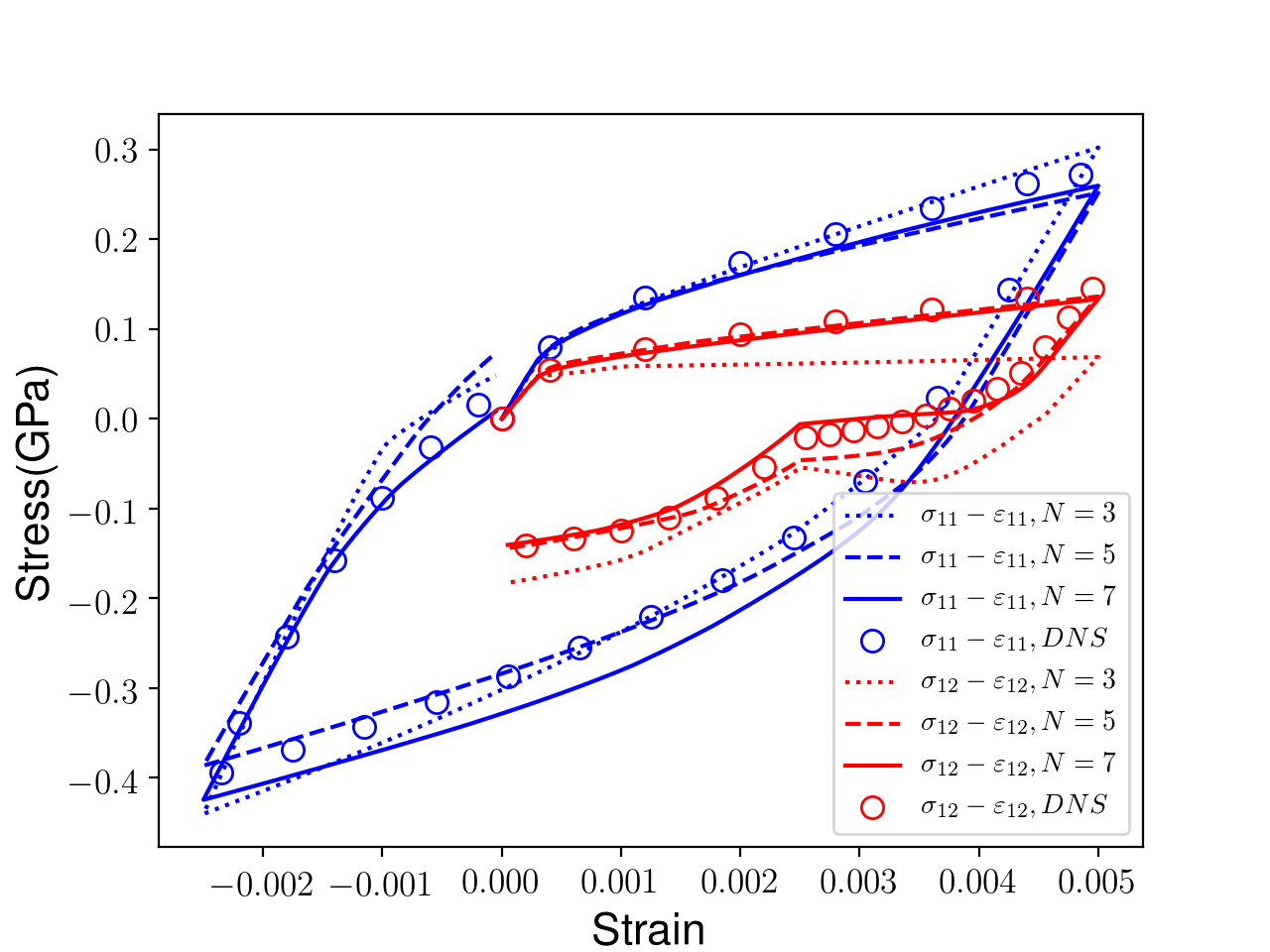}}
	\subfigure[Anisotropic, soft]{\includegraphics[clip=true,trim = 0.0cm 0.0cm 1.0cm 0.5cm,width=0.44\textwidth]{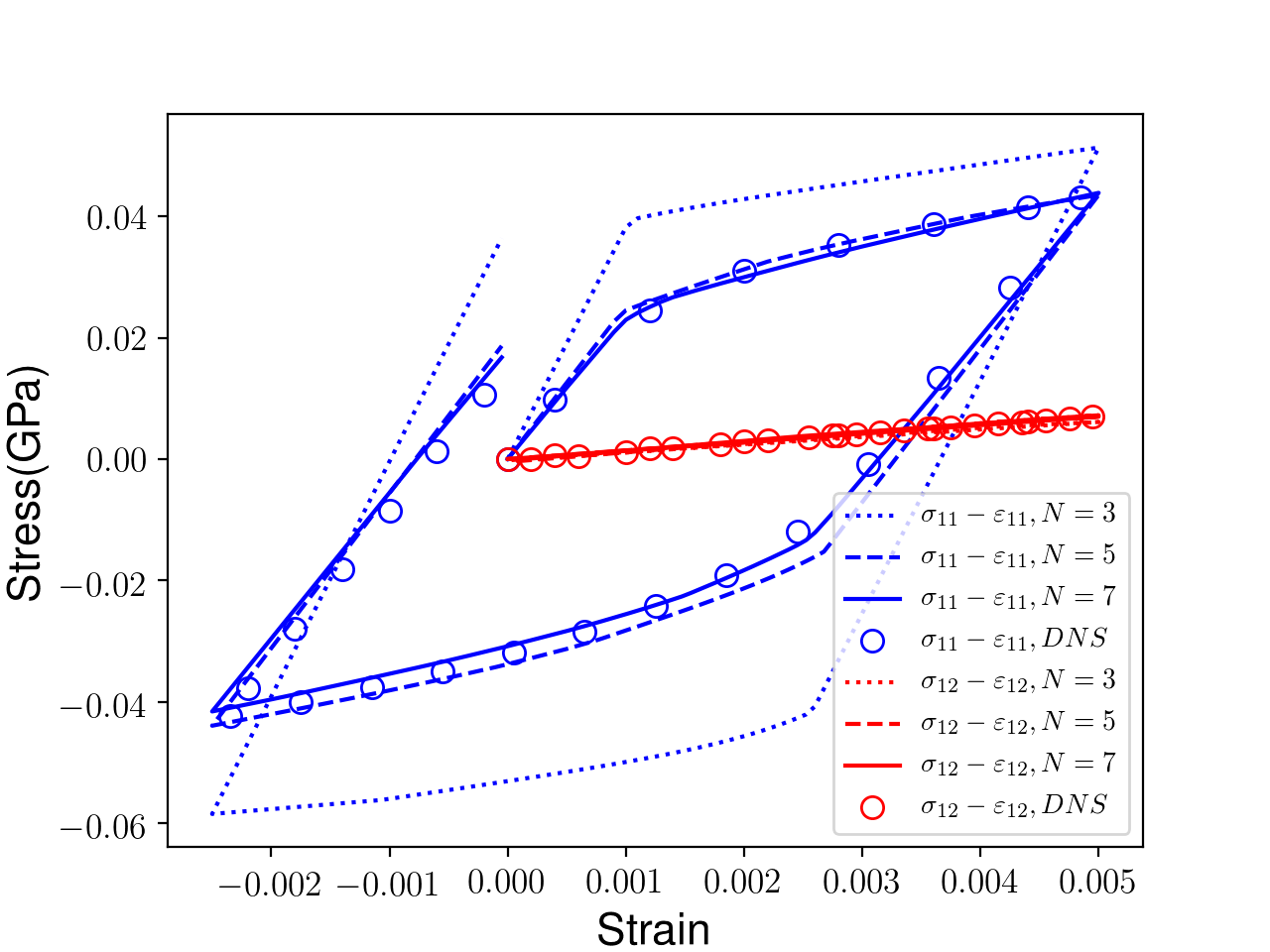}}
	\caption{Comparisons between material network and DNS under a complex loading path shown in (a). Constraints are applied on $\varepsilon_{11}$ and $\varepsilon_{12}$, and $\sigma_{22}=0$. RVEs with apparent hardening regime are considered: matrix-inclusion and amorphous RVEs with hard phase 1, anisotropic RVE with soft phase 1. The network depths are $N=3$ (dotted), $5$ (dashed) and $7$ (solid). DNS results are marked by the circles ($\circ$). }
	\label{fig:plasticity-complex}
\end{figure}

The material networks were also validated against a complex loading path for different RVEs. Macro-strain constraints are applied on $\varepsilon_{11}$ and $\varepsilon_{12}$, while $\sigma_{22}=0$. As shown in Fig. \ref{fig:plasticity-complex} (a), there are three steps within the loading path and the RVE eventually returns to the initial state ($\varepsilon_{11}=0$; 
$\varepsilon_{12}=0$). Due to plasticity, $\sigma_{22}$ does not necessarily vanish at the end of loading. The stress-strain curves $\sigma_{11}$-vs-$\varepsilon_{11}$ and
$\sigma_{12}$-vs-$\varepsilon_{12}$ given by the material networks and the DNS results are shown in Fig. \ref{fig:plasticity-complex} (b,c,d). Once again good agreement is observed for all three RVEs.

In theory, a deeper network can always be more accurate, since its solution space includes the one of a shallower network. However, with more fitting parameters,  the surface of the cost function of the deeper network has more local minimal, so that the training process will require more epochs to find the optimum representation of the RVE. For example, given 10000 epochs of training, a network with $N=7$ does not always provide a better prediction than the one with $N=5$ as we can see from Fig. \ref{fig:plasticity}, \ref{fig:plasticity-unloading} and \ref{fig:plasticity-complex}. Other than the method discussed in \textit{Remark} \ref{rem:offline} (multiple realizations), it is also possible to use different types of training datasets to improve the convergence of training algorithms. In practice, the optimum choice of network depth depends on one's devoted training time and desired accuracy.

\begin{figure}[!htb]
	\centering
	\includegraphics[clip=true,trim = 0.0cm 0.0cm .0cm 0.0cm,width=0.44\textwidth]{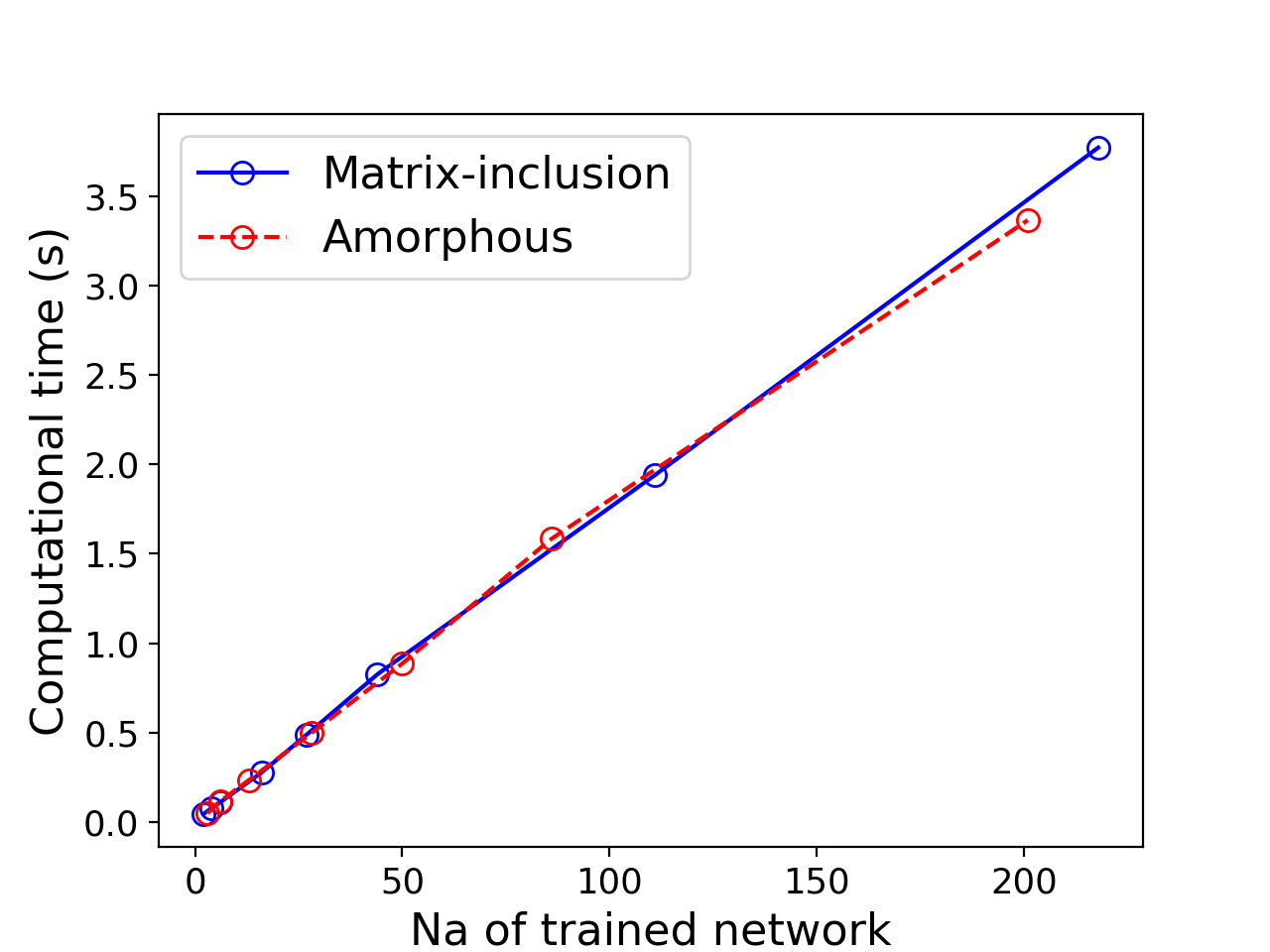}
	\caption{Computational time vs. number of active nodes in the bottom layer $N_a$ for nonlinear small-strain plasticity. Trained material networks for matrix-inclusion and amorphous RVEs are considered. 25 uniaxial loading steps are simulated with $e_{11}$ up to 0.01.}
	\label{fig:time}
\end{figure}

The computational times of material networks for the matrix-inclusion and amorphous RVEs are presented in Fig. \ref{fig:time} for different numbers of active nodes in the bottom layer. Uniaxial tension loading was considered, and there were 25 loading steps in each simulation. Typical FE simulations of the amorphous RVEs took about 79 s, while the trained material network took 0.12 s for $N_a=6$ $(N=3)$, 0.49 s for $N_a=28$ $(N=5)$, and 1.58 s for $N_a=86$ $(N=7)$. With the same $N_a$, the network for the matrix-inclusion has more active nodes belonging to the nonlinear plastic material, therefore, it costs more time in evaluating the local constitutive laws than the one of the amorphous RVE. For the 2D amorphous RVE, the trained network with $N=5$ is 150 times faster than the corresponding DNS, and we expect the improvement of efficiency from the proposed material network will be more significant for 3D RVEs.

It can be seen from the figure that the computational time is proportional to $N_a$ in the network, consistent of our statement in Section \ref{sec:onlineplas}. This is an important feature of the proposed material network based on hierarchical homogenization structure. As in most RVE homogenization techniques, the computational costs are proportional to $(N_{dof})^{2.x}$ (FEM), $N_{dof}\log(N_{dof})$ (FFT-based methods \cite{moulinec1998a}), or $(N_{dof})^3$ (SCA \cite{liu2016self}, TFA \cite{dvorak1992transformation}). Moreover, the material network is solved directly without the need for an iterative solver (e.g. the one used in the FFT-based method \cite{moulinec1998a}), herein, its robustness and reliability are well preserved.  All these features make the proposed material network a promising tool for concurrent multiscale simulations and material design.

\subsubsection{Finite-strain hyperelasticity under large deformations}\label{sec:onlinehyper}
In this section, the trained material networks are extrapolated to predict the responses of finite-strain hyperelastic RVEs under large deformation. To the authors' knowledge, finite-strain problems with large deformations are challenging for most existing model reduction methods. 
\begin{figure}[!htb]
	\centering
	\subfigure[Matrix-inclusion: $P_{11}-F_{11}$ plots]{\includegraphics[clip=true,trim = 0.0cm 0.0cm 1.0cm 0.5cm,width=0.44\textwidth]{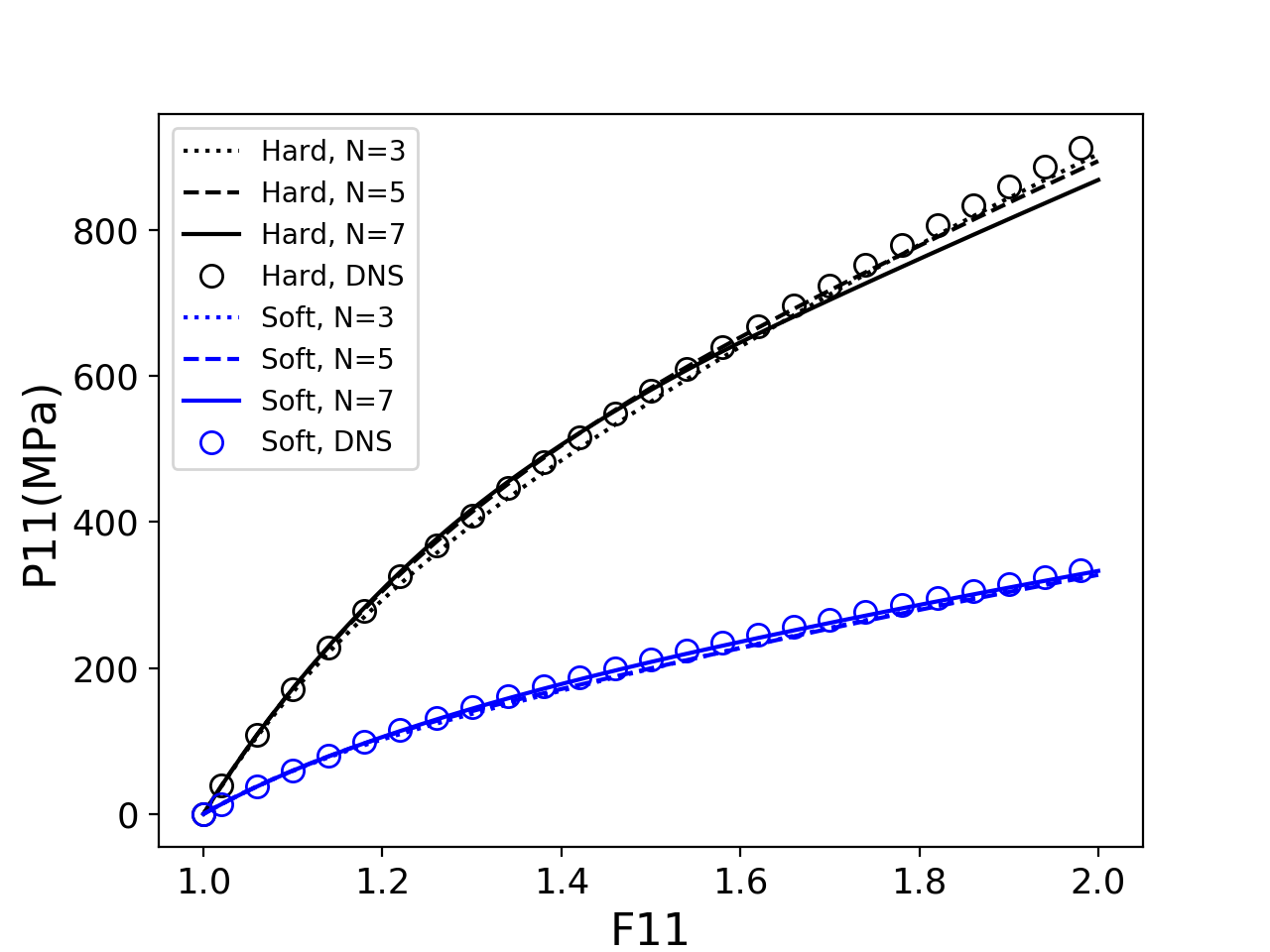}}
	\subfigure[Matrix-inclusion: deformed RVEs at $F_{11}=1.5$]{\includegraphics[clip=true,trim = 5.0cm 1.0cm 7.5cm 3.0cm,width=0.47\textwidth]{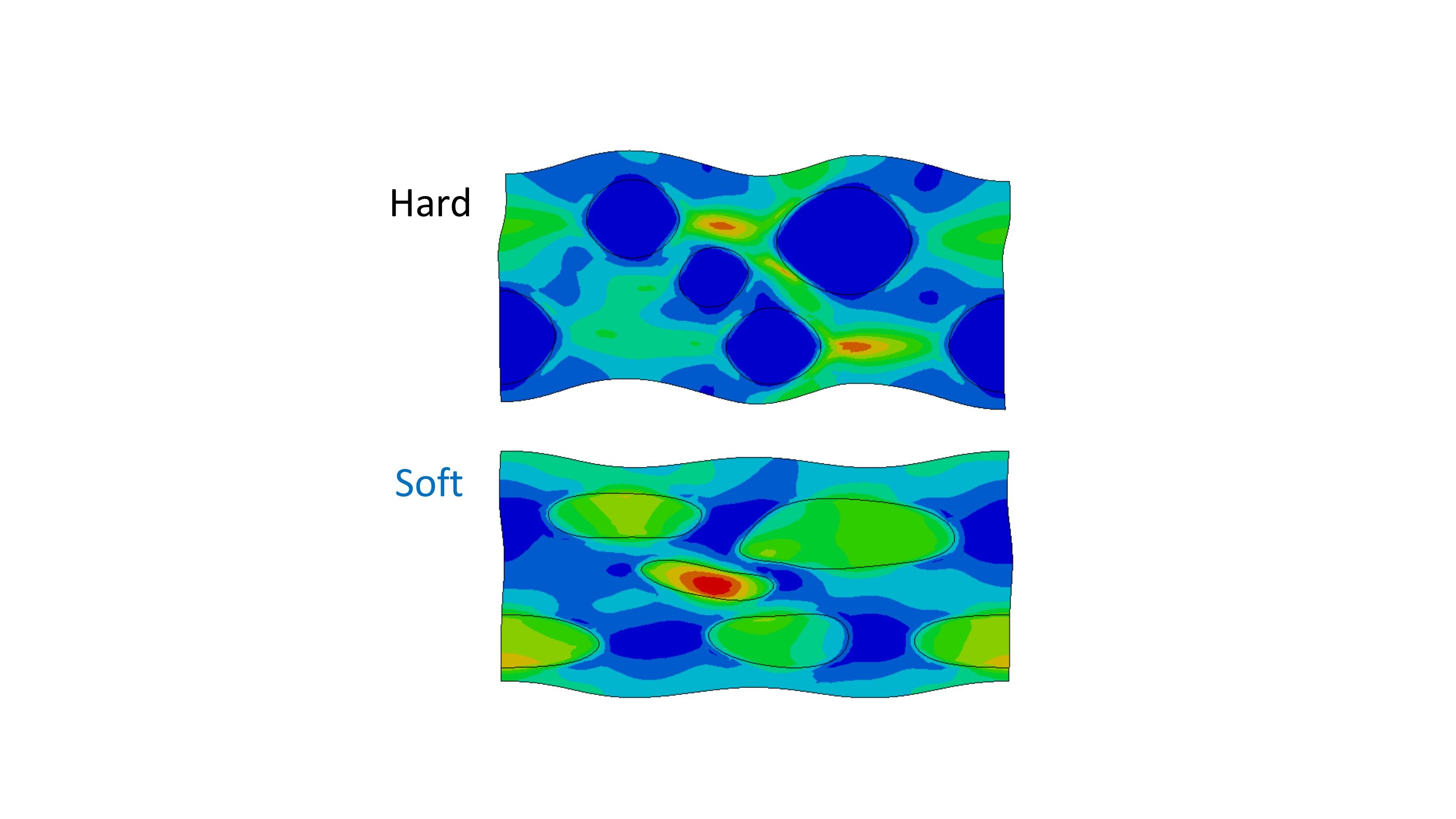}}
	\subfigure[Armophous: $P_{11}-F_{11}$ plots]{\includegraphics[clip=true,trim = 0.0cm 0.0cm 1.0cm 0.5cm,width=0.44\textwidth]{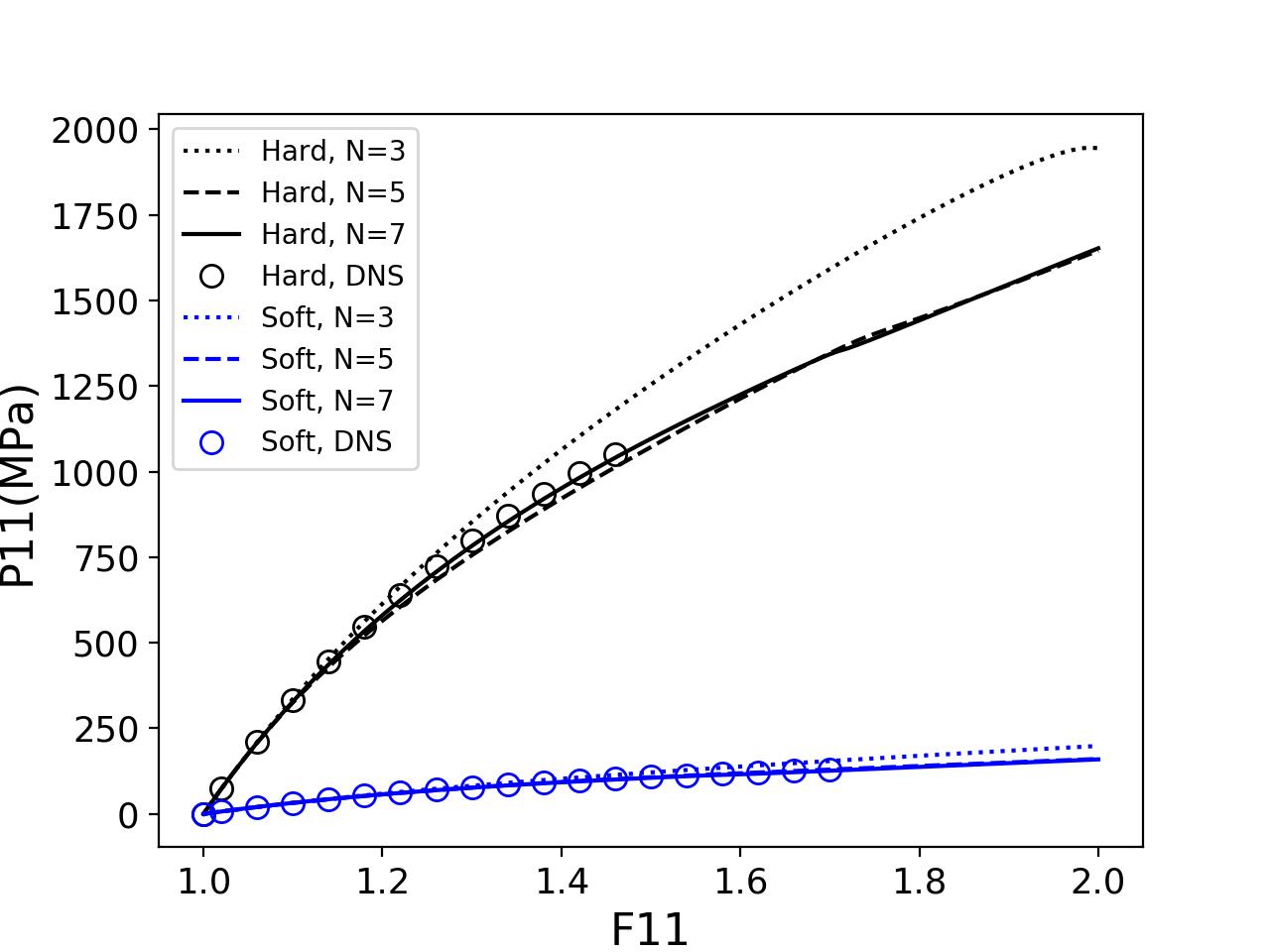}}
	\subfigure[Armophous: deformed RVEs at $F_{11}=1.5$]{\includegraphics[clip=true,trim = 5.0cm 1.0cm 7.5cm 3.0cm,width=0.47\textwidth]{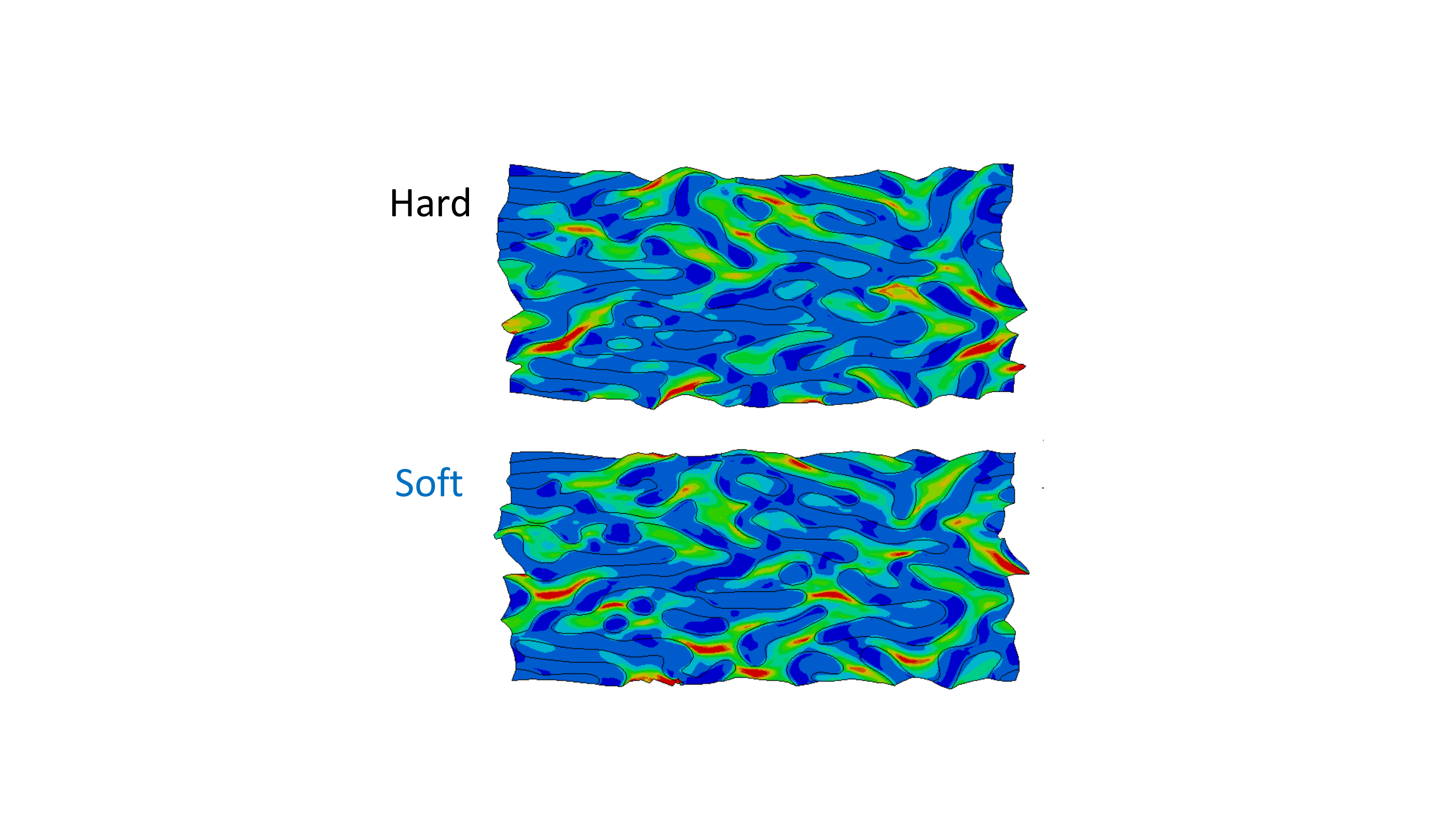}}
	\subfigure[Anisotropic: $P_{11}-F_{11}$ plots]{\includegraphics[clip=true,trim = 0.0cm 0.0cm 1.0cm 0.5cm,width=0.44\textwidth]{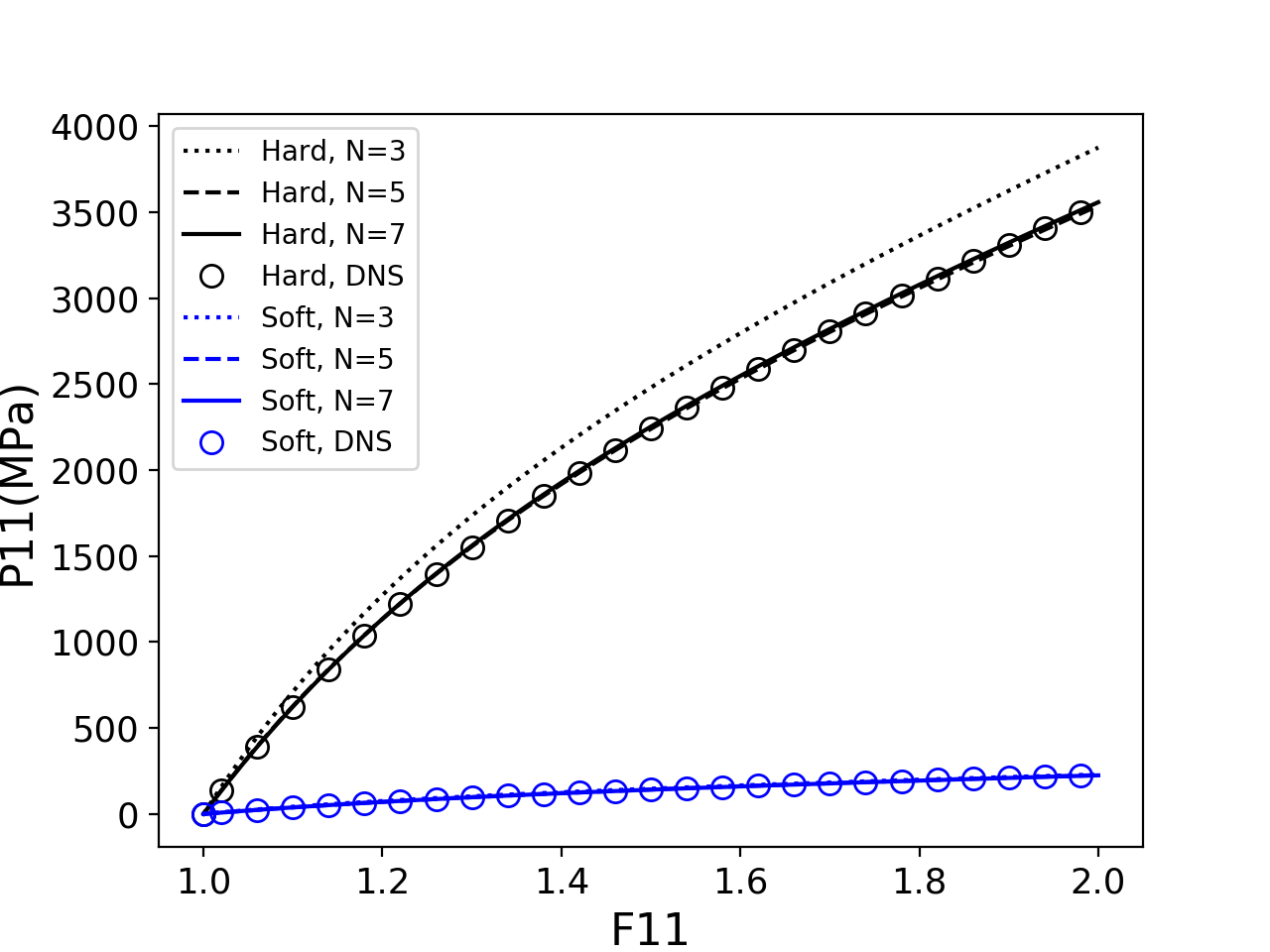}}
	\subfigure[Anisotropic: deformed RVEs at $F_{11}=1.5$]{\includegraphics[clip=true,trim = 5.0cm 1.0cm 7.5cm 3.0cm,width=0.47\textwidth]{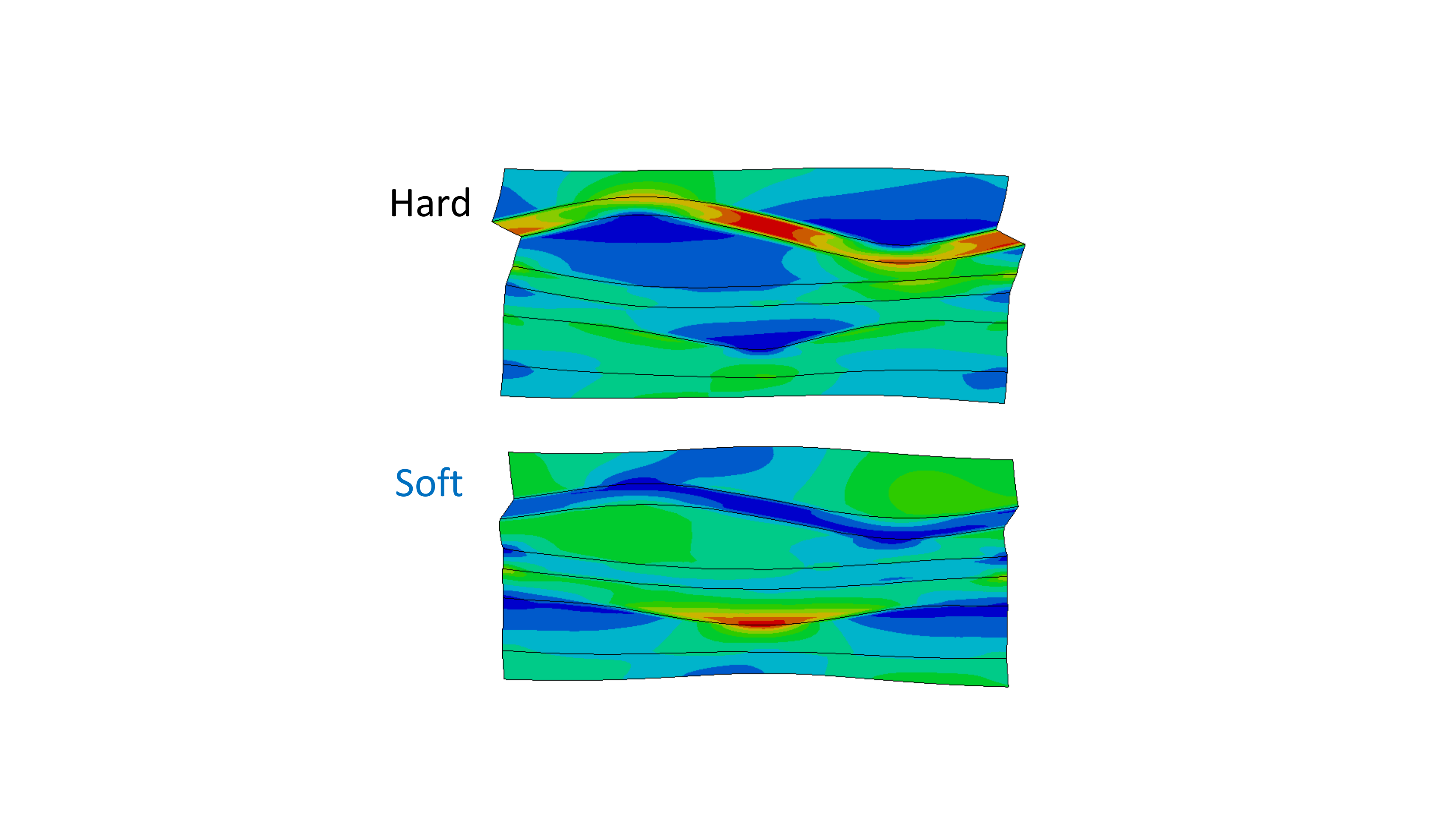}}
	\caption{Comparisons between material network and DNS for finite-strain hyperelasticity under uniaxial tension loading for various RVEs. Both hard and soft cases are considered. The $P_{11}$ vs. $F_{11}$ plots are shown on the left (a, c, e), and Green strain fields in the deformed RVEs at $F_{11}=1.5$ from DNS are shown on the right (b, d, f). The network depths are $N=3$ (dotted), $5$ (dashed) and $7$ (solid). DNS results are marked by the circles ($\circ$).}
	\label{fig:hyper}
\end{figure}
Analytical and semi-analytical micromechanics methods are no longer valid due to the failure of analytical solutions. Methods based on reduced basis also lose their accuracy when large geometric non-linearity is involved. Even DNS models may encounter some numerical difficulties due to severe material distortions. Nonetheless, we will demonstrate that the proposed material network with reduced DOFs is able to predict both global and local RVE responses accurately under large deformations.
\begin{figure}[!htb]
	\centering
	\subfigure[Matrix-inclusion: DNS model (3186 elements)]{\includegraphics[clip=true,trim = 0.0cm 0.0cm 1.0cm 0.5cm,width=0.44\textwidth]{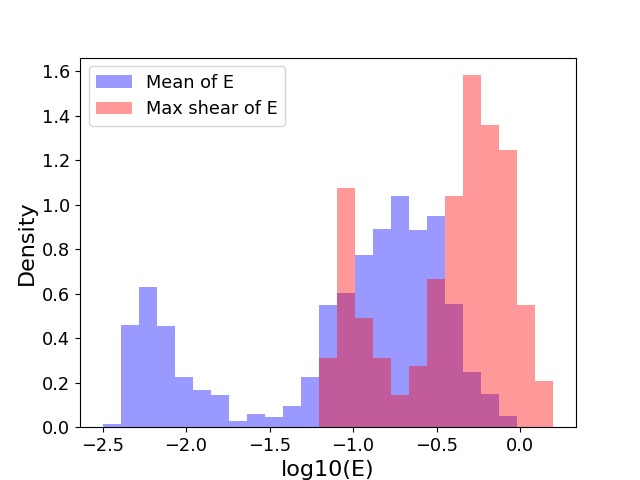}}
	\subfigure[Matrix-inclusion: $N=7$, $N_a=44$]{\includegraphics[clip=true,trim = 0.0cm 0.0cm 1.0cm 0.5cm,width=0.44\textwidth]{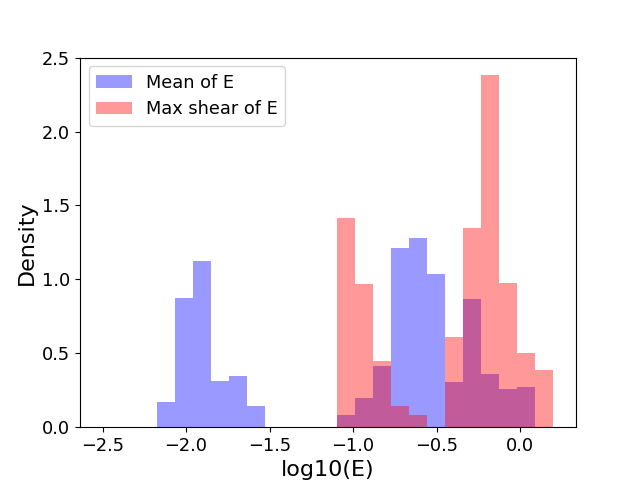}}
	\subfigure[Armophous: DNS model (9868 elements)]{\includegraphics[clip=true,trim = 0.0cm 0.0cm 1.0cm 0.5cm,width=0.44\textwidth]{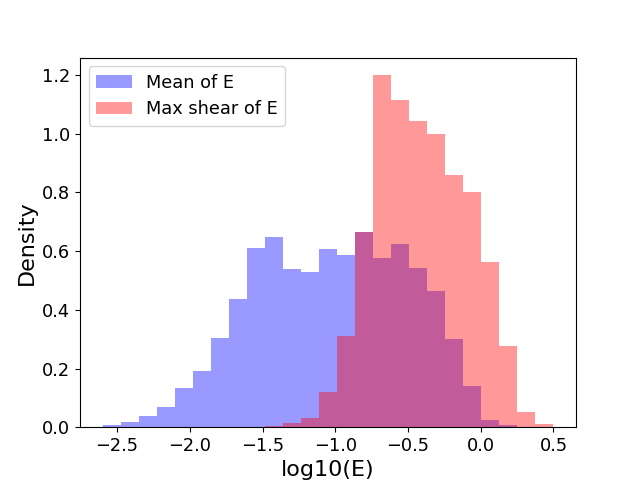}}
	\subfigure[Armophous: $N=7$, $N_a=86$]{\includegraphics[clip=true,trim = 0.0cm 0.0cm 1.0cm 0.5cm,width=0.44\textwidth]{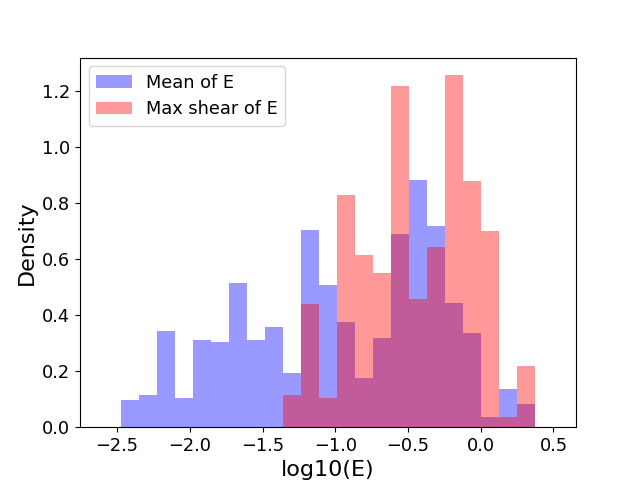}}
	\subfigure[Anisotropic: DNS model (4355 elements)]{\includegraphics[clip=true,trim = 0.0cm 0.0cm 1.0cm 0.5cm,width=0.44\textwidth]{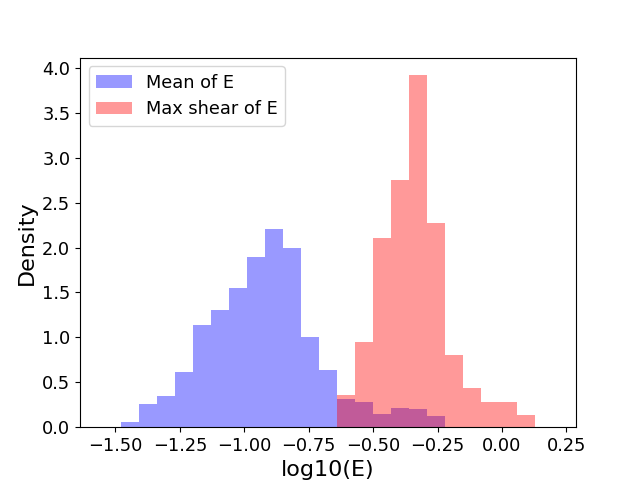}}
	\subfigure[Anisotropic: $N=7$, $N_a=62$]{\includegraphics[clip=true,trim = 0.0cm 0.0cm 1.0cm 0.5cm,width=0.44\textwidth]{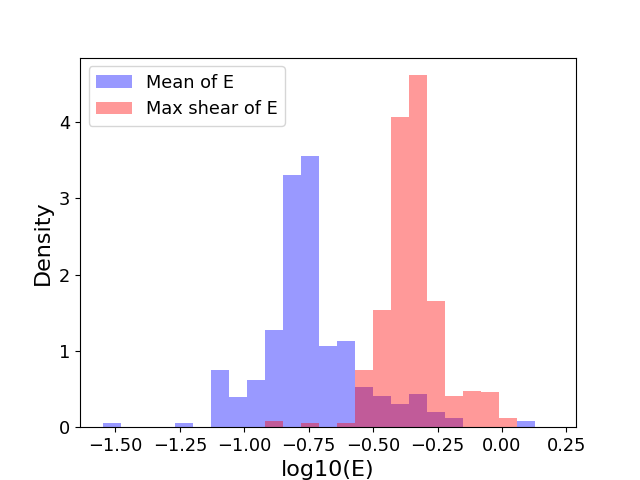}}
	\caption{Distributions of the mean Green strain and the max shear Green strain inside the RVE (hard) from DNS (left) and material networks with $N=7$ under uniaxial tension loading at $F_{11}=1.5$. The histograms are normalized. }
	\label{fig:field}
\end{figure}

Both phase 1 and phase 2 are considered to be Mooney-Rivlin hyperelastic materials. The strain energy density function of the Mooney-Rivlin material is defined as
\begin{equation}
W = A(I-3)+B(II-3)+C(III^{-2}-1) + D(III-1)^2.
\end{equation}
The independent material constants are $A$, $B$ and the Poisson ratio $\nu$, and $2(A+B)$ gives the shear modulus of linear elasticity. The other parameters in strain energy density function can be calculated by
\begin{equation}
C=0.5A+B,\quad D=\dfrac{A(5\nu-2)+B(11\nu-5)}{2(1-2\nu)},
\end{equation}
and $I,II,III$ are the invariants of right Cauchy-Green tensor $C=F^TF$.

Both phases are assumed to be nearly incompressible ($\nu\approx0.5$). The material constants of phase 2 are
\begin{equation}
A^{p2} = 100 \text{ MPa}, B^{p2}=50 \text{ MPa},\nu^{p2} = 0.49
\end{equation}
Phase 1 is considered to be either harder or softer than phase 2. For these two cases, the properties of phase 1 are
\begin{equation}
A^{p1} = 1000 \text{ MPa}, B^{p1}=500 \text{ MPa},\nu^{p1} = 0.49 \text{ (hard)}
\end{equation}
and
\begin{equation*}
A^{p1} = 10 \text{ MPa}, B^{p1}=5 \text{ MPa},\nu^{p1} = 0.49 \text{ (soft)}.
\end{equation*}

Fig. \ref{fig:hyper} provides comparisons of the $P_{11}-F_{11}$ curves obtained from DNS and from the material network under uniaxial tension loading, as well as plots of the deformed RVE at $F_11=1.5$. By default, each RVE is pulled to $F_{11}=2$, however, DNS for the amorphous RVE were terminated early due to severe distortion of elements during the analysis. Again, good agreements between the results from material networks and DNS are observed for all the cases.
\begin{figure}[!htb]
	\centering
	\includegraphics[clip=true,trim = 0.0cm 0.0cm .0cm 0.0cm,width=0.44\textwidth]{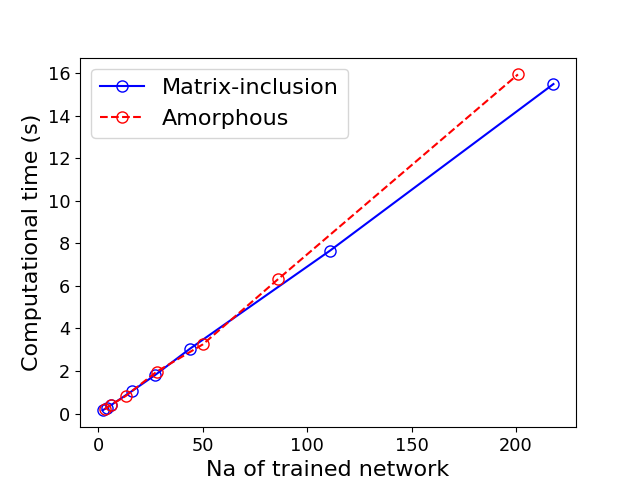}
	\caption{Computational time vs. number of active nodes in the bottom layer $N_a$ for nonlinear finite-strain hyperelasticity. Trained material networks for the matrix-inclusion and amorphous RVEs are considered. 50 uniaxial loading steps are simulated with $F_{11}$ up to 2.0.}
	\label{fig:timehyper}
\end{figure}

The distributions of local Green strains inside the RVEs predicted by DNS and material networks with $N=7$ under uniaxial tension loading at $F_{11}=1.5$ are shown in Fig. \ref{fig:field}. RVEs with hard phase 1 material were considered. For each case, the statistical characteristics of the local green strain in the material network agree well with the DNS results. The mean and max-shear of a Green strain $\textbf{E}=\{E_{11}, E_{22}, E_{12}\}$ are defined as,
\begin{equation}
E_{mean}=\dfrac{1}{2}(E_{11}+E_{22})
\end{equation}
and
\begin{equation}
E_{max-shear}=\sqrt{\dfrac{1}{4}(E_{11}-E_{22})^2+(E_{12})^2}
\end{equation}

The computational time of material networks for the matrix-inclusion and amorphous RVEs are presented in Fig. \ref{fig:timehyper} for different $N_a$. There were 50 loading steps in each simulation. A typical FE simulation of the matrix-inclusion RVE (3186 elements) took about 30 s, while the material network of took 1.05 s for $N_a=16 (N=5)$, and 3.06 s for $N_a=44 (N=7)$. Again, the computational time is still approximately proportional to $N_a$ in the network. However, a deeper network with more active nodes in its bottom layers usually has larger local deformations, herein, requires more newton's iterations to converge at each loading step. The material network is a simplified topological structure that reduces the number of DOFs and computational complexity of the DNS model, and the speed-up ratio will be more significant for DNS with a larger mesh and 3D problems.

\section{Conclusion}\label{sec:conclusion}
A new data-driven multiscale material modeling method called deep material network is developed for machine learning of RVE topologies and accelerated online predictions. Based on a simple two-layer building block with analytical homogenization solutions, the material network constructs a hierarchical structure that is capable of capturing complex RVE behavior. Optimization and training of the deep material network are enabled by the synergistic choices of various machine learning approaches, such as SGD with backpropagation algorithm, nodal deletion and subtree merging for model compression. Importantly, limitations encountered  in previous reduced order methods (e.g. extensive offline sampling, extra model assumption and calibration, danger of extrapolation, loss of physics) have been addressed by the proposed material network, with the following key features:
\begin{enumerate}
\item A novel RVE model reduction approach based on mechanistic building block and hierarchical topological structure;
\item A complete machine learning procedure for training the deep material network based on offline numerical DNS data or experimental testing data;
\item Efficient and accurate online predictions for challenging RVE homogenization problems, including nonlinear history-dependent plasticity and finite-strain hyperelasticity under large deformations.
\end{enumerate}

There are a vast number of research opportunities for improvement and extension of the deep material network, in terms of both theory development and applications. A few future directions are listed as below:
\begin{enumerate}
	\item Choices of proper data-sampling strategy and advanced machine learning techniques can help to optimize the material network, while a deeper understanding of the whole training process still requires more study. 
	\item Extension of the present 2D building block to 3D is of fundamental importance, and the possibility of including multiphysics effect can also be explored.
	\item Applications of the deep material network to industrial RVE problems, generally with more complex morphology and material laws, can be of significant interests, such as visco-hyperelastic polymer composite, carbon fiber reinforced polymer (CFRP),  and polycrystalline alloys modeled by crystal plasticity.
\end{enumerate}

\par Finally, this intelligent model reduction method provides a possibility to obtain accurate constitutive behaviors for accelerating the multi-scale concurrent computation in a large-scale heterogeneous structure. Additionally, it provides a way to avoid the numerical difficulty associated with large material distortion in conventional RVE approaches. Due to its efficiency and intrinsically parameterized structure, the proposed material network also offers a high application potential for multiscale material design.

\section*{Acknowledgement}
The authors would like to warmly thank Dr. John O. Hallquist of LSTC for his support to this research. The support from the Yokohama Rubber Co., LTD under the Yosemite project is also gratefully acknowledged.

\appendix
\section{Analytical solutions of 2D building block in finite strain}\label{sec:a1}
The analytical solutions in finite strain are derived based on the equilibrium condition and kinematic constraints at the interface between the two layers, which are listed below,
\begin{equation}\label{eq:ifcon}
P_{2}^1 = P_{2}^2, \quad P_{3}^1 = P_{3}^2, \quad
F_{1}^1 = F_{1}^2, \quad F_{4}^1 = F_{4}^2.
\end{equation}
First, assume an arbitrary overall deformation gradient after homogenization as
\begin{equation*}
\bar{\textbf{F}}^r=\{\bar{F}^r_1,\bar{F}^r_2,\bar{F}^r_3,\bar{F}^r_4\}^T,
\end{equation*}
Based on Eq. (\ref{eq:ifcon}) and definition of homogenization, we have
\begin{equation}\label{eq:ifcon1}
F_{1}^1 = F_{1}^2=\bar{F}^r_1, \quad F_{2}^2=\dfrac{1}{f_2}\bar{F}^r_2-\dfrac{f_1}{f_2}F_{2}^1,\quad F_{3}^2=\dfrac{1}{f_2}\bar{F}^r_3-\dfrac{f_1}{f_2}F_{3}^1,\quad F_{4}^1 = F_{4}^2=\bar{F}^r_4.
\end{equation}
with
\begin{equation*}
f_2 = 1-f_1.
\end{equation*}
Applying the constitutive laws (without residual stress)for both phases and substituting the above equation into the equilibrium condition yields
\begin{equation}\label{eq:ifcon2}
\begin{Bmatrix}
F_{2}^1\\
F_{3}^1\\
\end{Bmatrix}=\begin{Bmatrix}
\hat{A}_{22}&\hat{A}_{23}\\
\hat{A}_{32}&\hat{A}_{33}\\
\end{Bmatrix}^{-1}
\begin{Bmatrix}
f_2\Delta A_{12}&A_{22}^2&A_{23}^2&f_2\Delta A_{24}\\
f_2\Delta A_{13}&A_{32}^2&A_{33}^2&f_2\Delta A_{34}\\
\end{Bmatrix}\bar{\textbf{F}}^r=\textbf{s}^1_{2\times 4}\bar{\textbf{F}}^r,
\end{equation}
where
\begin{equation}
\hat{\textbf{A}}=f_2\textbf{A}^1+f_1\textbf{A}^2\quad\text{and}\quad \Delta\textbf{A} = \textbf{A}^2-\textbf{A}^1
\end{equation}
We can then combine Eq. (\ref{eq:ifcon1}) and (\ref{eq:ifcon2}) to define the concentration tensor for layer/material 1,
\begin{equation}
\textbf{F}^1 = \textbf{S}^1\bar{\textbf{F}}^r,\quad \text{with } S^1_{11}=S^1_{44}=1,\quad \textbf{S}^1_{([2,3],:)}=\textbf{s}^1_{2\times 4}
\end{equation}
The homogenized stiffness tensor $\bar{\textbf{A}}^r$ in Eq. (\ref{eq:arfinite}) can be derived as
\begin{equation}\label{eq:arexplicit}
\bar{\textbf{A}}^r = \textbf{A}^2-f_1\Delta\textbf{A}\textbf{S}^1.
\end{equation}
When residual stress is considered, we set the overall deformation gradient $\bar{\textbf{F}}^r=\textbf{0}$ and the homogenized residual stress $\delta\bar{\textbf{P}}^r$ in Eq. (\ref{eq:prfinite}) is equal to the homogenized stress in the RVE, which takes the following form,
\begin{equation}\label{eq:prexplicit}
\delta\bar{\textbf{P}}^r=f_1\delta\textbf{P}^1+f_2\delta\textbf{P}^2-f_1f_2\Delta\textbf{A}_{(:,[2,3])}
\begin{Bmatrix}
\hat{A}_{22}&\hat{A}_{23}\\
\hat{A}_{32}&\hat{A}_{33}\\
\end{Bmatrix}^{-1}\begin{Bmatrix}
\delta P_{2}^1\\
\delta P_{3}^1\\
\end{Bmatrix}
\end{equation}

Note that the same procedure can also be used to derive the analytical solutions of 3D two-layer building block in both small-strain and finite-strain formulation.

\bibliography{references_ROM}

\end{document}